\documentclass[longabstract,numberedappendix,twocolumn]{aastex63}
\usepackage{amsmath,amssymb,commath}
\usepackage{tikz}
\usetikzlibrary{matrix,positioning,decorations.pathreplacing}
\usepackage{lineno}
\usepackage{afterpage}
\usepackage{rotating}
\usepackage{graphicx}
\usepackage{color}
\usepackage[normalem]{ulem}
\usepackage{subfloat}
\usepackage{verbatim}
\usepackage{makecell}
\usepackage{enumitem}
\usepackage[ruled,linesnumbered]{algorithm2e}
\received{June 30, 2021}
\accepted{February 4, 2022}
\submitjournal{ApJ}

\shorttitle{SNSpec Predictor}
\shortauthors{Hu, Chen, \& Wang}

\graphicspath{{./}{}}
\begin{document}

\title{Spectroscopic Studies of Type Ia Supernovae Using LSTM Neural Networks}

\correspondingauthor{Lifan Wang}
\email{lifan@tamu.edu}

\author[0000-0001-7201-1938]{Lei Hu}
\affiliation{Purple Mountain Observatory \\
Nanjing 210023, People's Republic of China}
\affiliation{School of Astronomy and Space Sciences \\
University of Science and Technology of China \\
230029 Hefei People's Republic of China}

\author[0000-0003-3021-4897]{Xingzhuo Chen}
\author[0000-0001-7092-9374]{Lifan Wang}
\affiliation{George P. and Cynthia Woods Mitchell Institute for Fundamental Physics \& Astronomy, \\
Texas A. \& M. University, Department of Physics and Astronomy, 4242 TAMU, College Station, TX 77843, USA}

\begin{abstract}

We present a data-driven method based on long short-term memory (LSTM) neural networks to analyze spectral time series of Type Ia supernovae (SNe~Ia). The dataset includes 3091 spectra from 361 individual SNe~Ia. The method allows for accurate reconstruction of the spectral sequence of an SN Ia based on a single observed spectrum around maximum light. The precision of the spectral reconstruction increases with more spectral time coverages, but the significant benefit of multiple epoch data at around optical maximum is only evident for observations separated by more than a week. The method shows great power in extracting the spectral information of SNe~Ia, and suggests that the most critical information of an SN Ia can be derived from a single spectrum around the optical maximum. The algorithm we have developed is important for the planning of spectroscopic follow-up observations of future SN surveys with the LSST/Rubin and the WFIRST/Roman telescopes.

\end{abstract}


%%\keywords{editorials, notices --- 
%%miscellaneous --- catalogs --- surveys}


\section{Introduction} \label{sec:intro}

Type Ia supernovae (SNe~Ia) are among the most luminous phenomena in the transient universe which empowers accurate measurements of the history of expansion of the universe. It is believed that SNe~Ia result from the thermonuclear explosions of carbon/oxygen (C/O) white dwarfs \citep[WDs;][]{1960ApJ...132..565H} in binary systems. They are used as standardizable distance candles to probe the expansion history of the universe \citep{1998AJ....116.1009R,1999ApJ...517..565P,2019NatRP...2...10R} and constrain the properties of dark energy content of the universe \citep{Knop_2003,Kowalski_2008,Amanullah_2010,Suzuki_2012,c2014AA...568A..22B,2018ApJ...859..101S,2019ApJ...872L..30A}. 

Although the underlying progenitor systems and the physical pathways toward the explosions remain elusive, the diverse spectroscopic observations seem to epitomize, to a certain extent, intrinsic diversities among SNe~Ia \citep[e.g.,][]{2017suex.book.....B}. 
The ignition processes and explosion geometries may imprint their signatures on the observed optical spectra \citep{Wang:Wheeler:doi:10.1146/annurev.astro.46.060407.145139,Cikota:2019MNRAS.490..578C,Chen2020AIAI,Yang:2020ApJ...902...46Y}. The intrinsic brightness of SNe~Ia and their magnitude dispersion on the Hubble diagram seem to be correlated to the environment they explode \citep{Wang_1997,Uddin_2020ApJ...901..143U}. The viewing angle effect of asymmetric explosions has been proposed to be likely responsible for the diversity of the ejecta velocities measured from the spectral lines such as the Si~II $\lambda$6355 absorption line \citep{2010Natur.466...82M,MaundAsym2010ApJ...725L.167M}. The intrinsic diversity of SN~Ia phenomena implies that each individual SN~Ia bears discrete observational signatures, but the observational data set can be analyzed through data-driven approaches.

Normal SNe~Ia can be divided into two subclasses in terms of their expansion velocities at around peak brightness \citep{Benetti:HVG:2004MNRAS.348..261B,2009ApJ...699L.139W}. Correlations between the Si~II velocity and the host galaxy properties were identified in several studies \citep{2013Sci...340..170W,Wenxiong:2021ApJ...906...99L}.
A subset of SNe~Ia showed excess emission in the first few days after explosion \citep{2015Natur.521..328C,2016ApJ...820...92M,2017ApJ...845L..11H,2017Natur.550...80J,2018ApJ...863...90W,2019ApJ...870L...1D,2019ApJ...870...12L,2019ApJ...870...13S,2020ApJ...904...14W,2021ApJ...923L...8J,2021arXiv211008752L}.
An ultraviolet pulse was detected within 4 days of the explosion of iPTF~14atg \citep{2015Natur.521..328C}, in line with the theoretical predictions by models \citep{2010ApJ...708.1025K} involving ejecta-companion interaction in a single-degenerate scenario \citep[see][for a different view]{2016MNRAS.459.4428K}. 
The thermal radiation from the ejecta-companion star interaction can account for the early blackbody-dominated spectrum without prominent absorption features. 
Another example is MUSSES~1604D which was modeled in the context of the explosion of a WD triggered by a helium detonation \citep{2017Natur.550...80J}. MUSSES~1604D showed early excess in the red which contradicts the ejecta-companion star interaction but is consistent with the radiation due to surface radioactivity in a helium detonation scenario. The prominent titanium absorption trough around maximum light of MUSSES~1604D also corroborates with the existence of radioactive species synthesized by helium detonation. However, these diverse observational characteristics are usually compromised by the lack of extensive spectral and time coverage. New statistical tools need to be developed to accomplish comprehensive quantification of the observed features.

Optical spectra play an important role in SN cosmology \citep[e.g.,][]{Saunders2018SNEMO}. The classic standardization method is to describe the peak luminosity as a function of the light-curve width and the color at maximum light \citep{1993ApJ...413L.105P,1997ApJ...483..565P,1999ApJ...517..565P,1996ApJ...473...88R,c1998AA...331..815T,WangCMAGIC,Guy2005Salt1,Jha2007mlcs2k2,Taylor_2021MNRAS.504.4111T}. The intrinsic diversities among SN Ia properties beyond the stretch and color correction can introduce systematic errors to the measurement of cosmological parameters. It motivates additional terms to be included to better control potential systematic errors. 
For example, the mass of SN host galaxies has been recognized to be correlated to the Hubble residuals  \citep{2010ApJ...715..743K,2010MNRAS.406..782S}. Several studies have been devoted to the quantification of spectral features with applications to the standardization of SNe~Ia for distance calibrations \citep[e.g.,][]{Wagers:2010ApJ...711..711W,c2009AA...500L..17B,2009ApJ...699L.139W,2011ApJ...729...55F,c2011AA...529L...4C,2015ApJ...815...58F,2018ApJ...858..104Z,c2020AA...636A..46L}. 
The ejecta velocity can be used as an additional parameter to reduce the scatter of the Hubble residuals  \citep{2009ApJ...699L.139W,2013Sci...340..170W,2018ApJ...858..104Z}. \citet{2015ApJ...815...58F} developed an interesting approach to measure distances using spectroscopic twins of SNe~Ia. A data-driven approach to SN~Ia spectral analysis may also enable identifications of observational features that are most sensitive to the intrinsic luminosity of SNe~Ia and identify potential systematic errors in the cosmological applications of SNe~Ia.

Optical spectra are needed to perform K-corrections to derive the standardized magnitudes for cosmology. The K-corrections rely on the spectral energy distributions (SEDs) of the SNe~Ia with well-calibrated photometric and spectroscopic observations, which should cover a broad range of subtypes of SNe~Ia. Currently, empirical spectral models with low degrees of freedom \citep{Nugent2002Kcorr,c2007AA...466...11G,Hsiao2007Kcorrection,Jha2007mlcs2k2,Burns2011Snoopy} are employed for K-corrections, although more accurate models with more free parameters have been constructed \citep{Saunders2018SNEMO}. A data-driven approach to SN~Ia observations will naturally lead to spectral libraries that can be used for K-corrections of SN~Ia spectra.

The next-generation SN surveys with LSST \citep{2009arXiv0912.0201L} can discover a vast number of SNe, which makes the acquirement of spectral time series for these SNe extremely challenging. It is neither realistic nor cost-effective to trigger high-cadence spectroscopic follow-ups for the SNe discovered by the LSST. For the subset of posteriori SN Ia discoveries identified from photometric light curves, taking spectral data at multiple phases may become impossible for the majority of the SNe. However, in light of the rapidly growing spectroscopic dataset of nearby SNe, it is promising that a generative data-driven model for spectral inference via machine-learning techniques can be derived to mitigate these difficulties. 

There are some existing studies applying machine learning to transient studies. For example, the spectral types of the SNe can be classified based on their light-curve data \citep{Moller2016SNLSPhotClas,Muthukrishna2019Rapid,Takahashi2020HSC,Villar2020SuperRAENN}, and transients can be identified from the astronomical survey images \citep{Goldstein2015DECam,Mahabal2019ZTF,Gomez2020RNNImage}.
The light curves of SNe~Ia can be well modeled by functional principal component analysis (FPCA) \citep{He:2018ApJ...857..110H}, where it was shown remarkably that a set of FPCA eigenvectors that are independent of the photometric filters can be derived from the observed light curves of SNe~Ia. 
There are a few studies on the application of deep learning neural networks to the spectral data of SNe. For example, \citet{Muthukrishna2019Dash} used a convolution neural network (CNN) for automated SN type classification based on SN spectra.
Several other works \citep{Chen2020AIAI,Vogl2020typeII,Kerzendorf2021Dalek} applied a Gaussian process, principal component analysis (PCA), and deep learning neural networks to radiative transfer models of SNe. \citet{Sasdelli2016Dracula} used unsupervised learning algorithms to investigate the subtypes of SNe~Ia. 
\citet{Stahl2020deepSIP} developed neural networks to predict the photometric properties of SNe~Ia (phase and $\Delta m_{15}$) based on spectroscopic data. \citet{Saunders2018SNEMO} used PCA to find low dimensional representations of spectral sequences of 140 well-observed Type Ia SNe. 
\citet{Chen2020AIAI}, in particular, built an artificial intelligence assisted inversion (AIAI) of radiative transfer models and used that to link the observed SN spectra with theoretical models. 
The AIAI is able to retrieve the elemental abundances and density and temperature profiles from observed SN spectra. The AIAI approach has the potential for quantitatively coupling complex theoretical models with the ever-increasing amount of high-quality observational data.

This paper aims to build a data-driven model of the spectral time evolution of SNe~Ia using the long short-term memory (LSTM) neural networks \citep{LSTM}. Section~\ref{sec:samp} presents the data sample. Section~\ref{sec:preproc} shows the preprocessing of the data to bring the data set to a uniform standard for further processing using FPCA (Section~\ref{sec:fpca}). The neural network architecture and the assignment of statistical weights of the data are shown in Section~\ref{sec:lstm}. 
Section~\ref{sec:lstm-spectemp} shows the application of the neural network to the construction of spectral time sequences of SNe~Ia. Section~\ref{sec:lstm-12spec} presents the application of the neural network based on spectral data taken at a single epoch around optical maximum to the analyses of normal and high velocities (Section~\ref{ssec:onesepc-NVHV}) and the different subtypes of SNe~Ia (Section~\ref{ssec:onespec-subtypes}), and the application of the neural network based on spectral data taken at two epochs around optical maximum (Section~\ref{ssec:twospec}). 
Moreover, we explore the potential application of our method in predicting spectral phases in Section~\ref{ssec:unknownphase}.
Section~\ref{sec:discussions} gives the discussions and conclusions.

A {\tt\string Python} implementation of the method proposed in this paper is available on Github\footnote{\href{https://github.com/thomasvrussell/snail}{https://github.com/thomasvrussell/snail}}. The software allows users to apply the LSTM neural networks to their own observations and provide open access to the data sample presented in Section~\ref{sec:samp} and the spectral templates of 361 SNe~Ia constructed in Section~\ref{sec:lstm-spectemp}.

\section{Sample Selection} \label{sec:samp}

\begin{deluxetable*}{cccccccccc} 
    \tablenum{1}
    \tablewidth{0pt}
    \tabletypesize{\scriptsize}
    \tablecolumns{10}
    \tablecaption{\label{tab:tab1} Table of Supernovae in the spectral dataset}
    \setlength{\tabcolsep}{6pt}
    \setlength{\extrarowheight}{5pt}    

    \tablehead{
    \colhead{SN Name}      &
    \colhead{SN Subtype}      &
    \colhead{Redshift}      &
    \colhead{$\makecell[l]{\text{Number of}  \\  \text{Spectra}}$}      &
    \colhead{$\makecell[l]{\text{First}  \\  \text{Epoch}}$}      &
    \colhead{$\makecell[l]{\text{Last}  \\  \text{Epoch}}$}      &
    \colhead{$\makecell[lll]{\text{Spectrum}  \\  \text{Source}}$}      &
    \colhead{$\makecell[l]{\text{Redshift}  \\  \text{Reference}}$}      &
    \colhead{$\makecell[l]{\text{MJD}_{\text{max}}  \\  \text{Reference}}$}      &
    \colhead{$\makecell[l]{\text{Photometry}  \\  \text{Reference}}$}
    }
    
    \startdata
    SN~2012fr & Ia-norm & 0.005457 & 74 & -13.99 & +30.45 & WISeREP & 1 & 2 & 2 \\
    SN~2005cf & Ia-norm & 0.006461 & 73 & -12.40 & +29.29 & WISeREP, Kaepora, VLT & 1 & 3 & 4 \\
    SN~2011fe & Ia-norm & 0.000804 & 68 & -14.82 & +27.51 & WISeREP, Kaepora & 1 & 5 & 6, 7 \\
    SN~2002bo & Ia-norm & 0.004240 & 44 & -13.58 & +29.36 & WISeREP, Kaepora & 1 & 3 & 8 \\
    SN~2006X & Ia-norm & 0.005240 & 40 & -10.68 & +32.07 & WISeREP, Kaepora, VLT & 1 & 3 & 9 \\
    SN~1994D & Ia-norm & 0.002058 & 38 & -12.47 & +29.44 & Kaepora & 1 & 3 & 10, 11 \\
    SN~2007le & Ia-norm & 0.006721 & 36 & -10.63 & +23.33 & WISeREP, Kaepora & 1 & 3 & 9 \\
    SN~2003du & Ia-norm & 0.006384 & 35 & -12.70 & +32.98 & Kaepora & 1 & 3 & 12 \\
    SN~2004dt & Ia-norm & 0.019730 & 32 & -9.44 & +32.72 & WISeREP, Kaepora, VLT & 1 & 3 & 8 \\
    SN~2003cg & Ia-norm & 0.004130 & 32 & -7.48 & +26.67 & Kaepora & 1 & 3 & 8 \\
    SN~2001V & Ia-norm & 0.015018 & 27 & -13.20 & +27.97 & WISeREP, Kaepora & 1 & 3 & 13 \\
    SN~2007af & Ia-norm & 0.005464 & 27 & -5.60 & +32.07 & WISeREP, Kaepora & 1 & 3 & 9 \\
    SN~2002er & Ia-norm & 0.008569 & 26 & -9.94 & +32.66 & WISeREP, Kaepora & 1 & 3 & 8 \\
    \enddata
    \tablecomments{Reference: (1) \citet{2019MNRAS.486.5785S}; (2) \citet{2018ApJ...859...24C}; (3) \citet{2012AJ....143..126B}; (4) \citet{2009AAS...21331204W}; (5) \citet{2012ApJ...752L..26P}; (6) \citet{2019MNRAS.490.3882S}; (7) \citet{2013CoSka..43...94T}; (8) \citet{2013MNRAS.433.2240G}; (9) \citet{2011AJ....142..156S}; (10) \citet{1996AJ....112.2094G}; (11) \citet{1995AJ....109.2121R}; (12) \citet{c2007AA...469..645S}; (13) \citet{2009ApJ...700..331H}; (14) \citet{2015MNRAS.452.4307P}; (15) \citet{2006AJ....131..527J}; (16) \citet{c2018AA...611A..58G}; (17) \citet{2012PASP..124..668Y}; (18) \citet{2013MNRAS.436..222M}; (19) \citet{2015ApJS..219...13W}; (20) \citet{2014styd.confE.125B}; (21) \citet{cikota2018narrowing}; (22) \citet{c2014ApSS.354...89B}; (23) \citet{2017AJ....154..211K}; (24) \citet{c2014AA...561A.146S}; (25) \citet{2013ApJ...773...53F}; (26) \citet{2016MNRAS.457.1000S}; (27) \citet{2015MNRAS.453.3300A}; (28) \citet{2015ApJ...806..191Y}; (29) \citet{2012ApJS..200...12H}; (30) \citet{2012MNRAS.425.1789S}; (31) \citet{2002AAS...200.9503S}; (32) \citet{2013MNRAS.429.2228H}; (33) \citet{2017MNRAS.466.2436S}; (34) \citet{2018MNRAS.478.4575L}; (35) \citet{2016ApJ...826..144S}; (36) \citet{2016PASP..128c4501S}; (37) \citet{2017RNAAS...1...36K}; (38) \citet{2018AJ....155..201W}; (39) \citet{1995deun.book.....S}; (40) \citet{c1973AA....28..295A}; (41) \citet{c1975AA....45..429V}; (42) \citet{1998AJ....115..234L}; (43) \citet{2006AJ....131.1639K}; (44) \citet{2015ApJ...813...30S}; (45) \citet{1993AJ....105..301L}; (46) \citet{2019MNRAS.490..578C}; (47) \citet{2003AJ....125..166K}; (48) \citet{2015ApJS..220....9F}; (49) \citet{2004AJ....128.3034K}; (50) \citet{2017MNRAS.472.3437G}; (51) \citet{2009ApJS..183..109R}; (52) \citet{2011MNRAS.418..747M}; (53) \citet{1998MmSAI..69..245D}; (54) \citet{2016ApJ...820...92M}; (55) \citet{2014ApJ...782L..35Y}; (56) \citet{2015MNRAS.446.3895F}; (57) \citet{2018ApJ...863...90W}; (58) \citet{2017MNRAS.464.4476C}; (The full table is available in its entirety in a machine-readable form in the online journal. A portion is shown here for guidance regarding its form and content.)}
\end{deluxetable*}

We use the publicly available SN spectra from WISeREP \citep{2012PASP..124..668Y}, Kaepora \citep{2019MNRAS.486.5785S}, and a data set of high-quality Very Large Telescope (VLT) observations of SNe~Ia from the supernova polarimetry program \citep{Wang:Wheeler:doi:10.1146/annurev.astro.46.060407.145139,Cikota:2019MNRAS.490..578C,Yang:2020ApJ...902...46Y}. The SNe are selected with the following criteria:
\begin{itemize}
    \item The redshift of the host galaxy and the $B$-band maximum are accurately measured. 
    \item The SN is classified as one of the following five subtypes: Ia-norm, Ia-91T, Ia-91bg, Ia-99aa, and Iax.
    \item The spectrum of the SN is between -15 and 33 days relative to the $B$-band maximum, and the wavelength covers from 3800$-$7200 \AA\ in the rest frame.
    \item The SN has more than two distinct spectra.
\end{itemize}

The above distilling criteria lead to 3091 spectra from 361 SNe~Ia. 
Moreover, we collected the published $B$ and $V$ light curves of these SNe for spectrophotometric recalibration of the spectral data. In Table~\ref{tab:tab1}, we show the details of the selected SNe used in this study.

\begin{figure*}[ht!] 
    \centering
    \includegraphics[trim=0cm 1cm 0cm 2.2cm,clip=true,width=12.5cm]{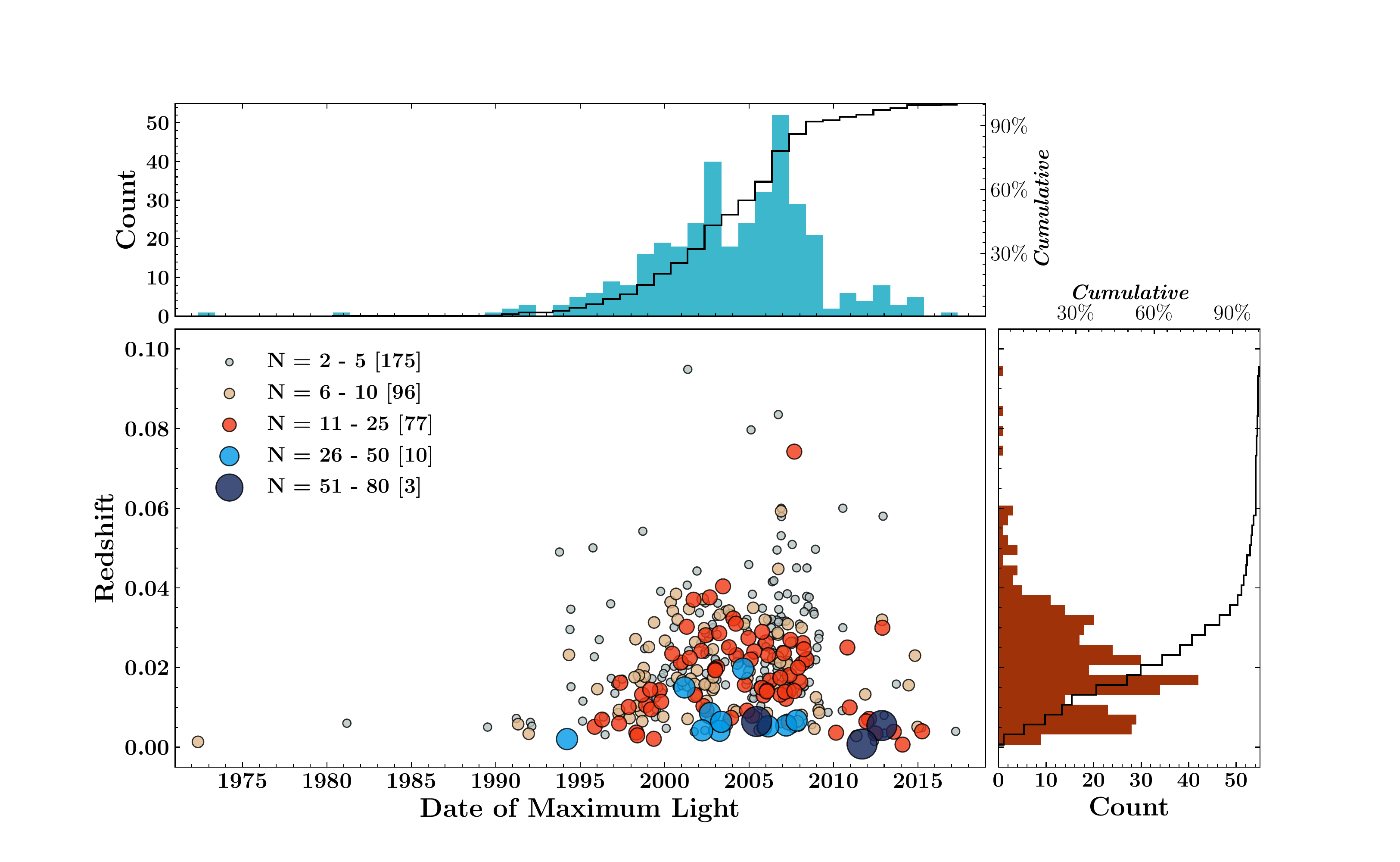}
    \caption{\label{fig:Distribution-SNe} The SN redshifts versus their times of $B$-band maximum for all the SNe~Ia in our dataset. These objects are divided into five subgroups according to the number of spectroscopic observations, where each subgroup is assigned a specific color and marker size, as shown in the upper left corner. The letter N in the legend stands for the number of spectra, and the numbers in the square brackets refer to the total counts of the subgroups. The attached panels at the top and on the right are the histograms of the times of maximum and host galaxy redshifts, respectively, with black curves showing their cumulative distributions.}
\end{figure*}

The characteristics of the SNe included in this work are shown in Figure~\ref{fig:Distribution-SNe}. The vast majority of them reached maximum light during 1995 and 2015, with 95\% of them originating from z $<$ 0.05, and nearly half of the SNe have $\leq$ five observations. About a quarter of the sample have temporal sampling with $> 10$ observations. A few nearby SNe, such as SN~2011fe, have $> 50$ spectroscopic measurements.

\begin{figure*}[ht!]
    \centering
    \includegraphics[trim=0cm 0.5cm 0cm 1.2cm,clip=true,width=12.5cm]{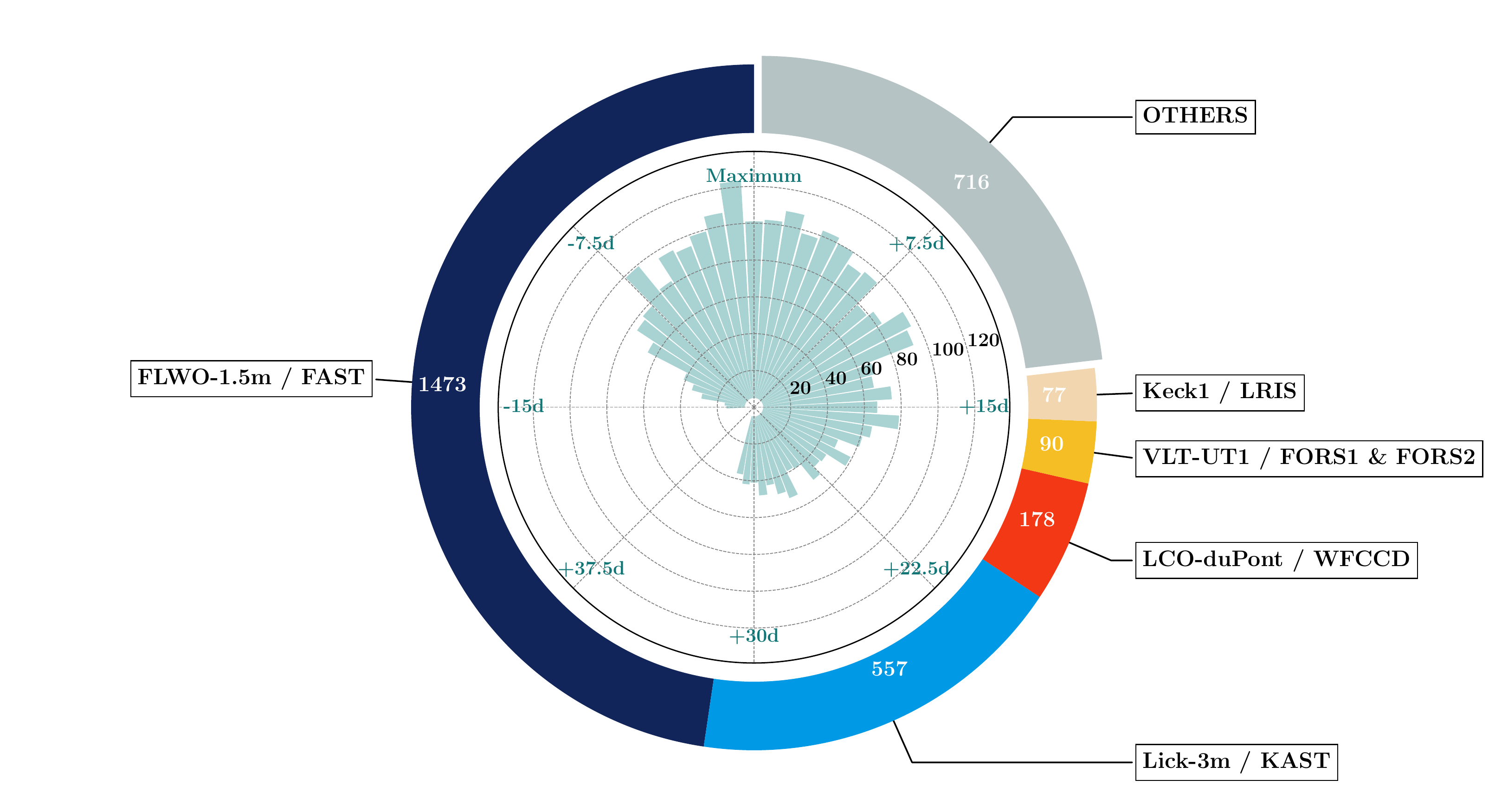}
    \caption{\label{fig:Distribution-Spec} \textit{Outer}: contributions of each major instrument to the total number of spectra in the dataset. \textit{Inner}: histogram of the number of individual SN spectra at each epoch in the dataset.}
\end{figure*}

The distribution of the source of the spectra in this compilation and the phase from B-band maximum are shown in Figure~\ref{fig:Distribution-Spec}. The sample consists of observations from four SN programs: the SN program from the Harvard-Smithsonian Center for Astrophysics \citep[CfA,][]{2012AJ....143..126B}, the Berkeley SuperNova Ia Program \citep[BSNIP,][]{2012MNRAS.425.1789S}, the Carnegie Supernova Program \citep[CSP,][]{2013ApJ...773...53F}, and the Supernova Polarimetry Program \citep{Wang:Wheeler:doi:10.1146/annurev.astro.46.060407.145139,Cikota:2019MNRAS.490..578C,Yang:2020ApJ...902...46Y}, thereby yielding $> 70\%$ spectra observed by FLWO 1.5m, Lick 3m, and LCO duPont telescopes. 
The VLT and Keck also provide a substantial portion of spectroscopic data, mostly with high signal-to-noises (S/Ns). One can also see in Figure~\ref{fig:Distribution-Spec} that spectroscopic time coverage peaks around maximum light, and there is a deficiency of SNe observed at the infancy stages ($< -10$ days).

In addition to the above dataset for the construction of a data-driven predictive model, we have also generated an auxiliary dataset by the following criteria:
\begin{itemize}
    \item The redshift of the host galaxy is accurately measured. 
    \item The SN is classified as an SN Ia. 
    \item The spectral wavelength covers 3800$-$7200 \AA\ in the rest frame.
\end{itemize}

These less restrictive criteria yield 8501 spectra from 3536 SNe~Ia. Throughout this paper, we will refer to the smaller dataset as ``the dataset" and use the term ``the {\it extended} dataset" to denote the larger dataset.

\section{Preprocessing of the data} \label{sec:preproc}

%% ***** Section: SAMPLES

The sources of the spectral data are heterogeneous and are often not well calibrated photometrically. 
We evaluated the noise levels of the spectral data (Section \ref{ssec:snr}), which were used to assign statistical weightings to each spectrum.
Moreover, the spectral data were processed through the following four steps to ensure the self-consistency and uniformity of the dataset: deredshifting (Section \ref{ssec:dered}), smoothing and resampling (Section \ref{ssec:smooth}), removal of telluric lines (Section \ref{ssec:telem}), and recalibration of spectral flux levels (Section \ref{ssec:colorcal}). 

\subsection{Estimation of the Spectral Noise} \label{ssec:snr}

The data are of diverse S/N, and their statistical weights need to be approximately accounted for in our studies.  
An approximate error spectrum for each original spectrum in the {\it extended} dataset is constructed following the method introduced in the Kaepora database \citep{2019MNRAS.486.5785S}. We further calculate the S/N of each spectrum in the wavelength range 4800 and 6200 \AA\ and will use this as a measure of the quality of each spectrum (see Section~\ref{sec:lstm} for more details on the assignments of statistical weights).

\subsection{Deredshifting} \label{ssec:dered}

The original spectra in the {\it extended} dataset were deredshifted from the observer frame to the rest frame using the host galaxy redshifts. 
In some cases, a correction had already been applied to the data we have downloaded. 
To avoid double correction, we visually inspected each spectrum to ensure that the wavelengths of the telluric absorption lines at $6867-6884$\AA\ and $7594-7621$\AA\ were consistent with the values expected in the observer's frame before deredshifting. 
We also compared the spectra with other spectra of the same SN at similar phases to the confirm that the spectra had not been deredshifted in their source database. 
We found 185 spectra in our dataset showing conspicuous redshift inconsistencies, and they were included in our analyses after removing the inconsistencies.

\subsection{Smooth and Rebinning} \label{ssec:smooth}

The next step was to apply a two-order Savitzky–Golay filter \citep{SG-SmoothFilter} to suppress the ubiquitous random noise in the SN spectra. Here we adopted a smoothing width of 1000 km $\text{s}^{-1}$, which is broader than the high-frequency noise but significantly smaller than the typical spectral absorption features. 
Each deredshifted spectrum from the {\it extended} dataset was resampled to a wavelength grid with a bin size of 2 \AA\ from 3800 and 7200 \AA\ by linear interpolation and normalized by dividing its average flux. For clarity, these processed spectra are hereafter referred to as the {\it homogenized spectral dataset} (HSD) which contains a total of 8501 spectra. Each {\it homogenized} spectrum is a 1700 dimensional array with unit mean.

\subsection{Removal of Telluric Features and Galaxy Emission Lines} \label{ssec:telem}

We removed the telluric absorptions at 7605 and 6869 \AA\ and all of the conspicuous hydrogen Balmer lines originating from the host galaxies at 6565 and 4861 \AA\ for each {\it homogenized} spectrum. 
The wavelength regions affected by the absorption and emission lines were filled with values from second-order B-spline interpolations. 
Note that the interpolation was not directly performed on the {\it homogenized} spectra but rather on the {\it homogenized} spectra already smoothed by an inverse-variance Gaussian smooth algorithm \citep{2006AJ....131.1648B,2019MNRAS.486.5785S} to ensure numerical stability.

\subsection{Spectral Flux Recalibration} \label{ssec:colorcal}

Both the continuum and spectral line components are important in our study, since our goal is to develop a model that enables predictions of both the spectral features and the overall spectral energy density distribution.
However, the spectral fluxes and colors integrated from SN spectra are usually inconsistent with those derived from broadband photometries. 
Such inconsistency is mainly due to the technical difficulties in spectrophotometry and can be conspicuous in many spectra observed at the same phases by different telescopes.

The flux scales of the preprocessed {\it homogenized} spectra were recalibrated to eliminate the flux scale inconsistency by enforcing the $B-V$ colors integrated from the spectra to agree with their corresponding photometric observations. 
For this to work, we must have a sample of SNe~Ia with excellent multicolor light-curve coverage. We went through a comprehensive literature search for the photometry of all of these SNe and collected published light curves of all of them, together with the specific filter bands at which they were observed. 
The sources of the light curves are shown in Table~\ref{tab:tab1}. For the data that are published in the natural system, the spectral flux level corrections were made with the appropriate transmission curves downloaded from the sources of the original data.

Fortunately, in general, the SNe with more spectral coverage had more complete photometric coverage. Over 80\% of the SNe with more than four optical spectra were found to have excellent light-curve coverages to allow for detailed template fits \citep{He:2018ApJ...857..110H} or interpolations to their light curves.
We used the Gaussian process with a radial-basis function (RBF) kernel \citep[\texttt{scikit-learn};][]{scikit-learn}, a model-independent interpolation method, to fit the $B$ and $V$ light curves and hence derive the $B-V$ colors at the epochs of the spectroscopic observations.
Subsequently, we should adjust the original spectrum so that its synthetic $B-V$ color can be in line with the derived photometric $B-V$ color, as the flux levels of the available spectra are usually poorly calibrated. 
One straightforward approach is to multiply each original spectrum by a monotonic flux scaling function with low degrees of freedom such that the adjusted spectrum will have a $B-V$ color that is consistent with the corresponding photometric measurement at the same epoch. 
The specific choice of the function can be somewhat arbitrary. In our work, we adopt the functional form of the CCM89 extinction law \citep{1989ApJ...345..245C} to perform the flux scaling, with the parameter $R_V$ fixed to 3.1, but leaving $E(B-V)$ as the only fitting parameter. All the adjusted spectra thus have $B-V$ colors that are identical to the values on their corresponding photometric color curves.

The {\it homogenized} spectra, starting from 3800 \AA\ in the rest frame, do not fully cover the entire effective wavelength range of the $B$ band.
The missing data in the {\it homogenized} spectra were set to zero first when calculating the $B$-band magnitudes. This introduces a systematic error to each $B-V$ color integrated from a spectrum that was corrected by employing a spectral template of SNe~Ia \citep{Hsiao2007Kcorrection} at the nearest phase. 
In doing so, the spectra of \cite{Hsiao2007Kcorrection} were truncated by setting the fluxes outside of 3800$-$7200 \AA\ to zero. The $B-V$ colors of both the Hsiao template (covering 1000$-$25000 \AA) and its truncates were calculated, and the differences were taken as their approximate systematic offsets. 
Note that this offset is not merely a function of the phase but also a function of the redshift of the SN and the specific transmission curves used for the corresponding photometric data.

The fit has zero degrees of freedom and results in a precise match of $B-V$ color between a color-calibrated spectrum and its corresponding photometric data. Such a treatment of the data may introduce systematic uncertainties that are difficult to quantify. We defined a quantity $|\Delta_{B-V}|$, which is the absolute value of the difference between the $B-V$ color measured on an uncalibrated spectrum and its corresponding observed photometric color to account for the amount of the color correction. Presumably, larger values of $|\Delta_{B-V}|$ imply higher levels of uncertainties. We used this quantity to set the statistical weightings of each spectrum in training the neural networks (see Section~\ref{sec:lstm}).

There are 801 spectra ($\sim$ 1/4) in the dataset for which the above $B-V$ calibration cannot be applied due to the lack of sufficient photometric coverage. This could be either due to unavailable photometric data or because the phases of the spectra are beyond the limited photometric data coverage. No photometric corrections were applied to these spectra, but their weightings are lowered in the training of the neural network (see Section~\ref{sec:lstm} for details).

Each spectrum was then renormalized by its average flux across the wavelength range of 3800-7200 \AA. This forms a new spectral dataset which we hereafter refer to as the {\it corrected spectral dataset} (CSD). There are a total of 3091 {\it corrected} spectra in the CSD, out of a total of 8501 spectra of the {\it extended} spectral dataset. 

\section{FPCA Parameterization} \label{sec:fpca}

The \citep[FPCA;][]{hall2006} was utilized to reduce the spectral dimensionality of the spectral data. 
It was applied to construct light-curve templates of SNe~Ia and build Hubble Diagrams using nearby, well-observed supernovae \citep{He:2018ApJ...857..110H}, and it was adopted by \citet{2020ApJ...890..177K} to parameterize the spectra of SNe~Ia. 
Here we follow a similar approach and use the $\texttt{fpca}$ package in the $R$ language \citep{Peng:FPCA} to solve for the optimal set of FPCA solutions.

The FPCA algorithm uses a series of orthogonal functions as the principal components and a linear combination of these principal components to reconstruct the input dataset. 
The function $\psi(\lambda)$ for the reconstruction can be written as
\begin{equation}
    \psi_i(\lambda)=\mu(\lambda)+\sum_{n=1}^{N} x_{i,n} \phi_n (\lambda),
\end{equation}
where $\mu(\lambda)$ is the average function, $\phi_n(\lambda)$ is the $n$th order component in the form of a function, $x_{i,n}$ is the $n$th order FPCA score for the $i$th spectrum.
To find the best FPCA score series $x_{i,n}$, the mean squared error (MSE) between the input data $Y_i(\lambda)$ and the fitting function $\psi(\lambda)$ is minimized
\begin{equation}
    \text{MSE}_i=\int_{\lambda_0}^{\lambda_1}( Y_i(\lambda)-\psi_{i}(\lambda))^2 d\lambda,
\end{equation}
where $\lambda_0$ and $\lambda_1$ are the lower and upper limits of the wavelength range. The principal components $\phi_n(\lambda)$ and the average function $\mu(\lambda)$ are solved for a given spectral dataset of size $N$.
The average function is
\begin{equation}
    \mu(\lambda)=\frac{1}{N}\sum_{n=1}^{N} Y_n(\lambda),
\end{equation}
and the principal components are solved by maximizing the variance of FPCA scores over the input spectral data set,
\begin{equation}
    \phi_i(\lambda)=argmax\left(\text{Var}(x_i) \right),
\end{equation}
where $\text{Var}(x_i)$ is the variance of all the $i$-th FPCA scores of the input data, $x_{i,1},\ x_{i,2},\ ..., x_{i,N}$. 
The solutions to the principal components are  subject to two additional conditions. 
First, the principal components are orthogonal:
\begin{equation}
    \int_{\lambda_0}^{\lambda_1} \phi_{i}(\lambda)\phi_{j}(\lambda) d\lambda=\delta_{ij}.
\end{equation}
Secondly, the variances are ordered with the order of principal components by
\begin{equation}
    \forall \ i,\ \text{Var}(x_{i})>\text{Var}(x_{i+1}).
\end{equation}

The solutions to the FPCA were derived from the more diverse {\it extended} dataset to achieve maximum generalizability.
Given the high computational cost (mostly RAM limitations) of $\texttt{fpca}$, 500 {\it homogenized} spectra were randomly drawn from the HSD to solve for the FPCA principal components, and each {\it corrected} spectrum in the CSD was decomposed into the resulting basis functions.
The spectral data were divided into two sections: a blue section covering the wavelength range from 3800 to 5500 \AA\ and a red section covering 5500 - 7200 \AA. At a spectral resolution of 2 \AA $\text{pixel}^{-1}$, each spectral section has a total of 850 pixels. Each spectral section was then subtracted by its own average flux and subsequently divided by its standard deviation.
We used 90 FPCA principal components for each spectral section. As a result, the full spectrum with two sections (1700 pixels) could be reconstructed by 180 FPCA scores and four additional factors accounting for the average fluxes and standard deviations of the two spectral sections. 
We denote the combined array that concatenates 180 FPCA scores and the four additional parameters as $[V_{\text{FPCA}}]$ (hereafter the FPCA-encoded array).
The FPCA parameterization for each spectral section was performed separately. This could introduce some artifacts at the wavelength boundary around 5500 \AA, but the effect was not significant enough to affect the analyses in this paper. 
Applying FPCA on the full range (5500$-$7200 \AA) of spectra without division over wavelength can eliminate these artifacts at the boundary. However, the corresponding RAM demand for solving FPCA with a satisfactory reconstruction accuracy exceeds the affordable level of our current computation platform.

%% ***** Section: FPCA
\begin{figure}[ht!]
    \centering
    \includegraphics[trim=0cm 0cm 0cm 0cm,clip=true,width=8.5cm]{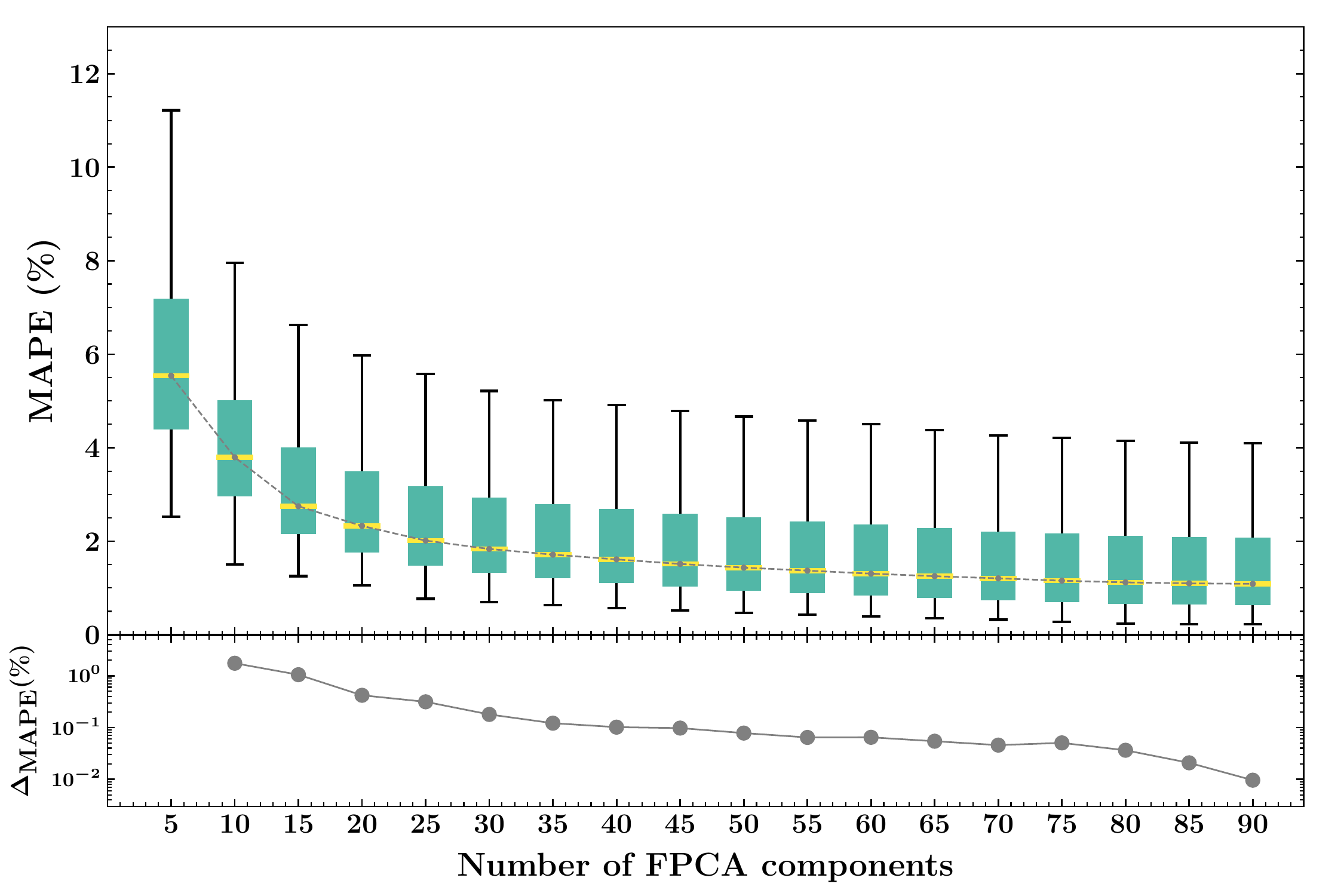}
    \caption{\label{fig:FPCA-ReconAccuracy} Accuracy of applying FPCA to the corrected spectra in the dataset. The upper panel gives the MAPE of FPCA reconstruction as a function of the number of used FPCA components. The box plot shows the distribution of MAPE for all of the corrected spectra. Note that the box is drawn from the first quartile (Q1) to third quartile (Q3), with a horizontal yellow line to denote the median; in addition, the lower (upper) whisker is at the lowest (highest) datum above $\text{Q1} - 1.5\times \text{IQR}$ (below $\text{Q3} + 1.5\times \text{IQR}$), where $\text{IQR} = \text{Q3} - \text{Q1}$. The dashed gray curve indicates a monotonically decreasing trend of the median MAPE as more FPCA components are employed. The lower attached panel shows the decrease of the median-level MAPE (i.e., overall accuracy improvement) for every five additional FPCA components used.}
\end{figure}

\begin{figure}[ht!]
    \centering
    \includegraphics[trim=0cm 1cm 1cm 2.6cm,clip=true,width=8cm]{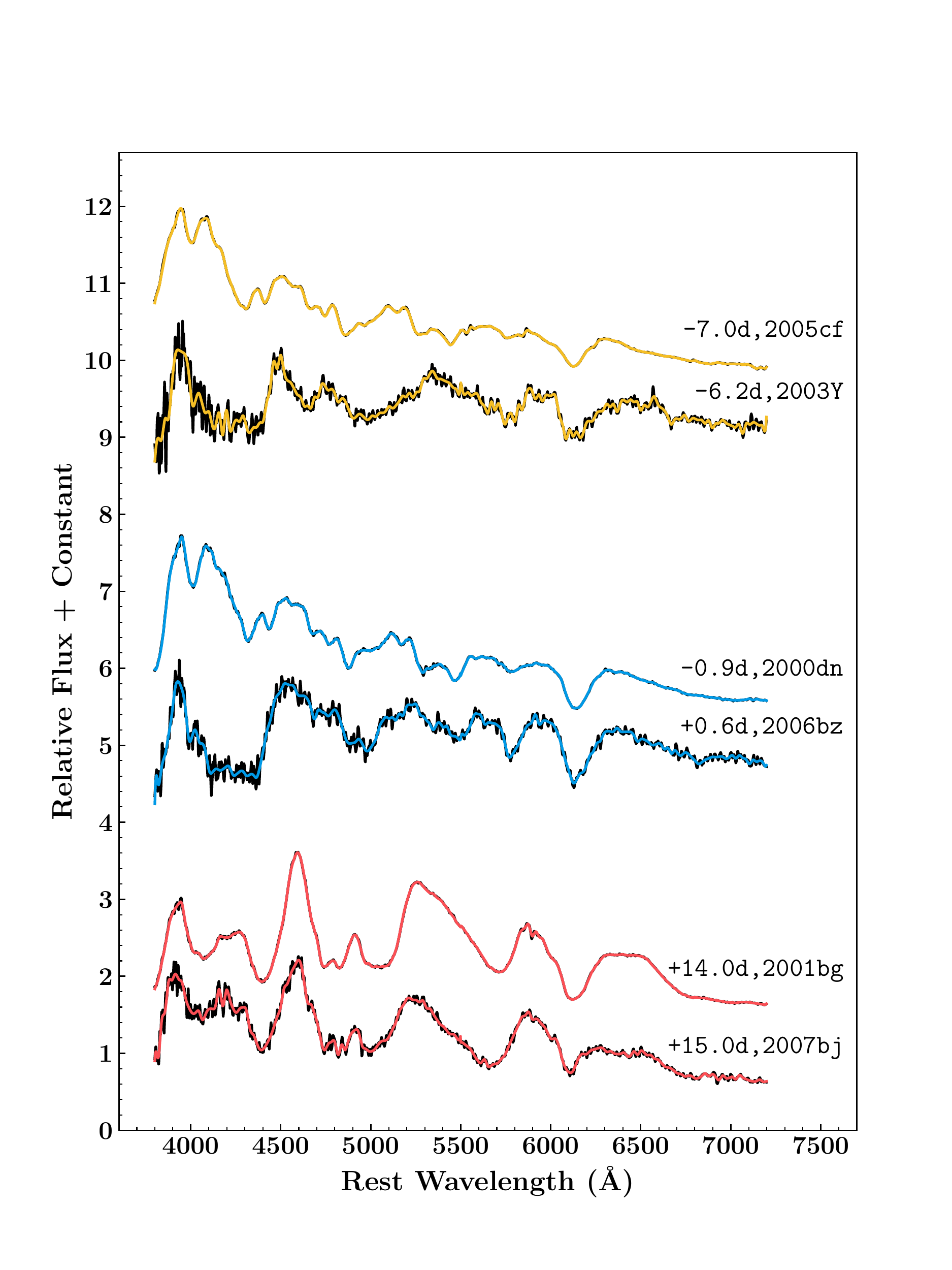}
    \caption{\label{fig:FPCA-ReconExamples} The FPCA reconstruction of six representative spectra of the CSD. The FPCA reconstructions of the corrected spectra (solid black curves) are plotted as colored solid lines. The six spectra are selected from three observational epochs: $\sim$ 1 week before maximum, maximum light, and $\sim$ 2 weeks past maximum. Note that a relatively noisy spectrum is presented  with a spectrum of very high S/N for each epoch. The spectra are arbitrarily shifted in the vertical direction for clarity of display.}
\end{figure}

The mean absolute percentage error (MAPE) is used to evaluate the difference between a {\it corrected} spectrum and its FPCA reconstruction.
As shown in Figure~\ref{fig:FPCA-ReconAccuracy}, where the statistics are drawn from the reconstruction of the CSD, the errors are considerably smaller when more principal components are used. A closer look at the trend (see the lower panel of Figure~\ref{fig:FPCA-ReconAccuracy}) shows that the improvement of reconstruction accuracy decelerates as the number of components increases. The decrease of the median MAPE is larger than 1\% from 5 to 10 components but falls to less than 0.01\% from 85 to 90 components.
The median MAPE of the FPCA reconstruction over the CSD can reach $\sim$1.1$\%$ using 90 components. Note that MAPE is a pixel-by-pixel measurement without taking into account the fluctuations due to observational noise. 
Thus, a spectrum with low S/N is more likely to have a larger MAPE. This trend is partially responsible for the broad and skewed distribution of MAPE, shown in the box plot of Figure~\ref{fig:FPCA-ReconAccuracy}.
Figure~\ref{fig:FPCA-ReconExamples} demonstrates the performance of FPCA reconstruction by six representative examples selected from the set of CSD. These examples show excellent agreement across the prominent spectral features of SNe~Ia.

\section{Network Architecture and Sample Weights} \label{sec:lstm}

%% ***** Section: LSTM NN
\begin{figure}[ht!]
    \centering
    \includegraphics[trim=7cm 2cm 7cm 2cm,clip=true,width=8.5cm]{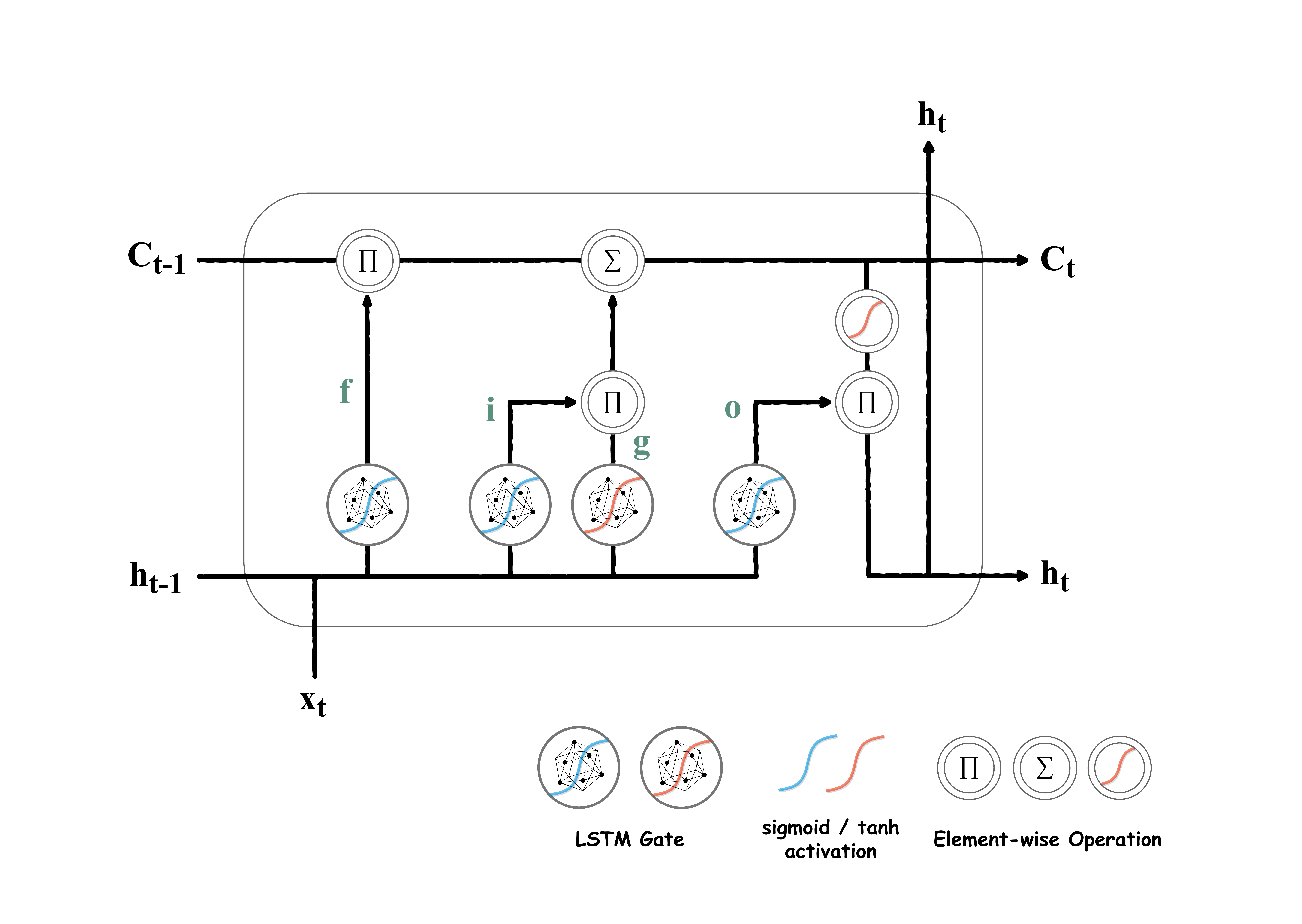}
    \caption{\label{fig:LSTM-RepeatingModule} Repeating module in LSTM that contains four interacting layers. The definitions of the nodes in the module are shown at the bottom of the figure. The detailed calculations in the four LSTM gates are given in Equation~(\ref{eqn:lstm1}). The element-wise operations are described in Equation~(\ref{eqn:lstm2}).
}
\end{figure}

The proposed model is built upon a multilayer LSTM \citep{LSTM} neural network, a widely used subclass of recurrent neural network (RNN). 
Previous works like \citet{Stahl2020deepSIP} and \citet{Chen2020AIAI} adopted CNNs to analyze the spectral data of SNe~Ia, where one-dimensional convolution processes are applied to the wavelength axis. 
In contrast, the LSTM architecture in our work is not convolutional. 
The spectral data have been compressed into isolated FPCA-encoded arrays, and additional dimensions are needed to store the phase information required for training the neural network.
Therefore, the input data cannot maintain a structure with a concept of spatial correlation as in the original spectra, thus becoming incompatible with the CNN approach.

Like all other subtypes of RNN, LSTM has a chain-like structure with a repeating module that allows the algorithm to learn from spectral time sequences.
Figure~\ref{fig:LSTM-RepeatingModule} shows the repeating module in our LSTM architecture. We will always use the subscript ``$t$" to denote the index of a spectral time sequence but use the subscript ``$\text{tar}$" to indicate an attribute of the target spectrum. 
LSTM is characterized by four gates: an input gate ($\textbf{i}$), a forget gate ($\textbf{f}$), an output gate ($\textbf{o}$) and an input modulation gate ($\textbf{g}$). These gates conduct the following operations:

\begin{align}
    \label{eqn:lstm1}
    \begin{split}
    i &= \textbf{sigm}(h_{t-1}U_i + x_tW_i)
    \\
    f &= \textbf{sigm}(h_{t-1}U_f + x_tW_f)
    \\
    o &= \textbf{sigm}(h_{t-1}U_o + x_tW_o)
    \\
    g &= \textbf{tanh}(h_{t-1}U_g + x_tW_g)
    \end{split}
\end{align}
and
\begin{align}
    \label{eqn:lstm2}
    \begin{split}
    c_t = f \circ c_{t-1} + i \circ g   \qquad
    h_t = o \circ \textbf{tanh}(c_t)
    \end{split},
\end{align}
where $\textbf{sigm}$ ($\textbf{tanh}$) is the sigmoid (hyperbolic tangent) activation function and the circle symbol refers to an element-wise product. Here $x_t$ is the input data delivered into the module, and the internal state $c_t$ ($h_t$) is known as the cell (hidden) state at time step $t$.
The weight matrices $\textbf{w} = \{W_i, U_i, W_f, U_f, W_o, U_o, W_g, U_g\}$ are repeatedly used by each time step.

Our study aims to predict the spectrum at a specific target phase $p_{\text{tar}}$ by feeding a spectral sequence with arbitrary time sampling. 
However, for typical discrete-time dynamic systems, such as in text classification \citep{Dai2015} and stock price prediction \citep{Eapen2019}, LSTM was applied to ordered data without time labels or with constant time-sampling rates.
Observations of SNe~Ia are usually irregularly time-sampled. Consequently, the proposed predictive model needs to include the specific phases of the spectra as part of the input parameters, yet LSTM does not contain a channel in its structure to incorporate the spectral phases.
A variant LSTM known as phased LSTM, which is proposed to handle unevenly sampled time series by adding a new time gate \citep{PhasedLSTM} is a likely choice. Unfortunately, the target phase {$p_{\text{tar}}$} of the predicted spectrum is also a variable in our framework, making it infeasible to apply phased LSTM directly.

Our solution to this problem is a straightforward integration of the phase information into the input spectral data.
Recall that each {\it corrected} spectrum has been compressed and encoded into a much shorter array denoted as $[V_{\text{FPCA}}]$. 
We concatenated the phase of each input spectrum and the target phase for spectral prediction at the beginning of the FPCA-encoded array,
namely, the input data of the first LSTM layer at the $t$th time step $x_t = (p_{\text{tar}}, p_t, [V_{\text{FPCA}}]_t)$.
As a result, the LSTM acquires an input layer with 186 neural processing units.

\begin{figure}[ht!]
    \centering
    \includegraphics[trim=2cm 2cm 2cm 1.8cm,clip=true,width=7cm]{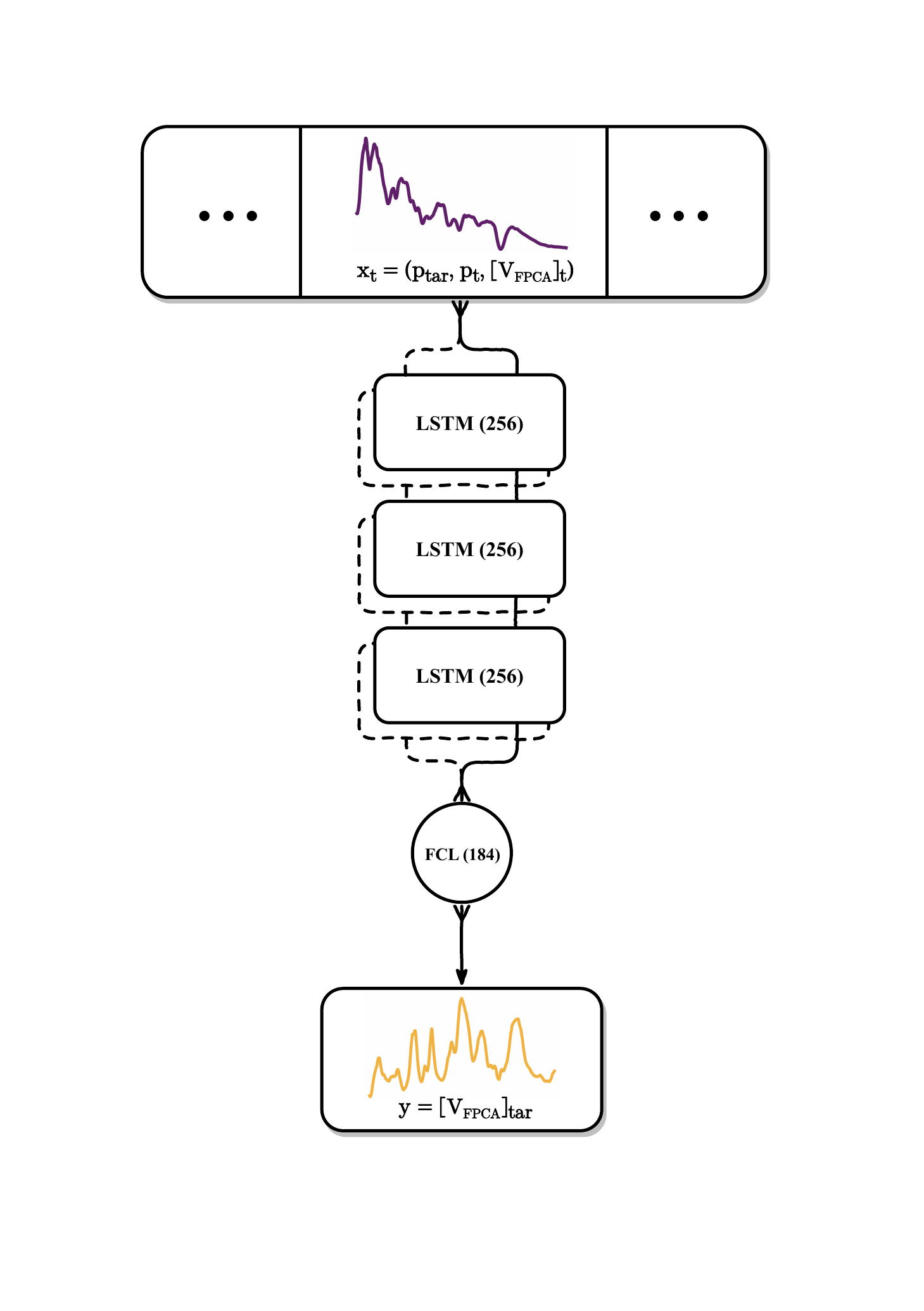}
    \caption{\label{fig:LSTM-Architecture} Architecture of the proposed LSTM model. The input data (shown by the boxes at the top) are one or more one-dimensional arrays describing the target phase for which the neural network will make spectral predictions ($p_{\text{tar}}$), the phase of the input spectral data ($p_t$), and the FPCA-encoded array of the $t$th spectral data ($[V_{\text{FPCA}}]_t$). The three boxes labeled with LSTM refer to the repeating module described in Figure~\ref{fig:LSTM-RepeatingModule}, where each layer in the module comprises 256 neurons. Note that the dashed lines indicate bidirectional operations which make the proposed architecture symmetric to the sequence inversion of the input data.
    A fully-connected layer (FCL) follows the last LSTM module and converts the 256 neural processing units back to the 180 FPCA scores and the four scaling parameters. The last step is a reconstruction of the spectroscopic data for the phase $t_{\text{tar}}$ during the predicting process. Note that the architecture itself can accommodate any variable-length spectral time series as input, but our model used herein currently supports a fixed length of 2.}
\end{figure}

Figure~\ref{fig:LSTM-Architecture} shows the architecture of the {proposed} LSTM implemented using the {\tt\string Python} module $\texttt{keras}$ \citep{chollet2015keras}.
For each SN, the input spectra are a given number of $K$ spectra from a total of $L$ spectra, allowing repetitions but with the phases only in nondecreasing order.
The neural network contains three bidirectional LSTM layers \citep{BiRNN} and a fully-connected output layer. 
The bidirectional LSTM allows the neural network to learn from the time sequence data in reverse order and strengthens the robustness of the neural network.
Note that the last layer is time distributed ($\texttt{keras.TimeDistributed}$), thereby yielding an output for each time step.
The difference between the output of each time step and the FPCA-encoded array of the target spectrum contributes equally to the loss function during the training process.
The Nadam algorithm ($\texttt{keras.Nadam}$) is used to optimize the network, with the loss function being the mean squared error (MSE) weighted by the data quality (see Equation~(\ref{eqn:wspec})-(\ref{eqn:lossfunc}) and discussions below for details).
The last box shown in Figure~\ref{fig:LSTM-Architecture} reconstructs the spectrum at the target phase during the predicting process, the outputs from different time steps were averaged to generate the predicted spectrum via FPCA reconstruction, followed by a final flux normalization.

To alleviate overfitting in the training process, we adopted the commonly used dropout method \citep{Hinton2012,Zaremba2014}. The efficient regularization technique suppresses the co-adaptations amongst the neurons by stochastically dropping rows of the weight matrices.
The dropout can also be activated in the testing step (aka Monte Carlo dropout) to make the network probabilistic \citep{Gal2015,Gal2016}.
With Monte Carlo dropout, the model is thereby no longer deterministic after the training process. A forward pass can generate a different result by feeding the same input, which is now determined by the applied dropout mask as a realization of a Bernoulli process. In this scenario, one can estimate the model uncertainty through the variance of predictions from multiple forward passes.
In our study, we set the kernel dropout rate (namely, for $W$ matrices in Equation~(\ref{eqn:lstm1})) to 0.14 and the recurrent dropout rate (namely, for $U$ matrices in Equation~(\ref{eqn:lstm1})) to 0.16, respectively. The only exception is that we disabled the kernel dropout for the first LSTM layer, and no dropout is performed on the final fully-connected layer.

Conceptually, a dynamic number of time sequence $K$ is allowed for RNNs; this is also our initial motivation to adopt LSTM to handle spectral sequences with arbitrary time coverages.
However, the complexity of {training} the neural network increases drastically with $K$ due to the vast amount of spectral combinations for large $K$. Our current model only supports $K\ = \ 2$, i.e., a spectral pair (with nondecreasing phases) as input. In particular, the structure also allows predictions from a single spectrum through one-time duplication. Moreover, following the selection limit on the phases of the SNe in our dataset, the target phase is restricted to be in the range from -15 to +33 days.

To mitigate the effect of the heterogeneity of the spectral data, each spectrum in the CSD was assigned an approximate statistical weight $w_{\text{spec}}$ based on the S/N of that spectrum {defined in Section~\ref{ssec:snr}} and a formula that is a function of $|\Delta_{B-V}|$ defined in Section~\ref{ssec:colorcal}.
The weight $w_{\text{spec}}$ is expected to be positively correlated with the SNR, which measures approximately the pixel-to-pixel fluctuations due to shot noise, and negatively correlated to $|\Delta_{B-V}|$.
In the practical implementation of the neutral networks, the weight $w_{\text{spec}}$ for each spectrum in the CSD is given as follows:

\begin{align}
    \label{eqn:wspec}
    \begin{split}
    &w_{\text{spec}} = w_{\text{S/N}} \times w_{\text{color}}
    \\
    &w_{\text{S/N}} = 0.7 + 0.3 \times \min\{\text{S/N}/100, 1\}
    \\
    &w_{\text{color}} = 0.7 + 0.3 \times (10/7 - 
    \\
    &\qquad\quad\,\,\,\,\,\min\{\max\{\abs{\Delta_{B-V}}/{\sigma_{\text{GP}}}, 3\}, 10\}/7)
    \end{split}
\end{align}
where $\sigma_{\text{GP}}$ denotes the photometric error of the color $B-V$ which is given by $\sqrt{(\sigma_B^2 + \sigma_V^2)}$, with $\sigma_B$ and $\sigma_V$ being the photometric errors of the $B$ and $V$ bands, respectively.
About 3/4 of the spectra in the CSD had been calibrated by their corresponding photometric colors for which the $\abs{\Delta_{B-V}}/{\sigma_{\text{GP}}}$ values could be computed.
For the rest of the spectra, the ratio $\abs{\Delta_{B-V}}/{\sigma_{\text{GP}}}$ was set to 5 artificially.
Notice that $w_{\text{S/N}}$ and $w_{\text{color}}$ were restricted to values in the range from 0.7 to 1.0 such that their weightings do not differ drastically for data with very different S/N and $\abs{\Delta_{B-V}}$.

Finally, the statistical weight for a training sample containing a total of $K+1$ spectra from the CSD, i.e., $K$ input spectra together with the output spectrum, was constructed by multiplying the $K+1$ spectral weights,

\begin{align}
    \label{eqn:wsamp}
    \begin{split}
    w_{sample} = \alpha * \prod w_{\text{spec}}
    \end{split}
\end{align}
where the factor $\alpha$ was introduced as a penalty factor to account for the heterogeneous instrumental sources of the $K+1$ spectra, which was set to 1 if all of the $K+1$ spectra are observed by the same instrument (i.e., no penalty), 0.6 if two instruments are involved, and 0.4 if more than two instruments are used.
Similar to the weights used for individual spectra, the penalty factors used herein avoid making extreme sample weights. The loss function of the neural networks is computed as

\begin{align}
    \label{eqn:lossfunc}
    \begin{split}
    \mathcal{L} = w_{sample} \times \sum_{i,t}(Y_{t}^{i} - [V_{\text{FPCA}}]_{\text{tar}}^{i})^2,
    \end{split}
\end{align}
where $Y_{t}$ is the neural network output at time step $t$, and $[V_{\text{FPCA}}]_{\text{tar}}$ is the FPCA-encoded array of the target spectrum. The superscript $i$ denotes the dimension index.

\section{LSTM applied to the construction spectral templates} \label{sec:lstm-spectemp}

In this section, we apply the proposed neural networks to construct the spectral template for each of the 361 SNe~Ia in our dataset. Here a spectral template is defined as a spectral time sequence from -15 to +33 days with a constant cadence of 3hr. The goal of the neural networks is to enable accurate predictions of the time evolution of spectral features based on spectral observations of limited spectral time coverage. In this section, the accuracy of the template construction using LSTM is evaluated using a test spectral set. Its performance is also compared with other published spectral templates of SNe~Ia.

\subsection{Training the neural networks for spectral template constructions} \label{ssec:train4temp}

We split the CSD into a training set of 90\% (2782) spectra and a testing set of 10\% (309) spectra, where the testing set was randomly drawn from the CSD with the following restrictions. 
\begin{itemize}
    \item The S/N of the spectral data is higher than 15.
    \item The phase difference between the selected spectrum and the nearest spectral observation of the same SN in the CSD is larger than two-thirds of a day. This avoids the selection of multiple spectra of the same SN taken at almost the same night.
    \item The testing set does not contain all of the observed spectra of any SN in CSD as is required by the algorithm of spectral template construction shown in Algorithm~\ref{algo:ctemp}.
\end{itemize}

The remaining spectra of the CSD after the selection of the test dataset were employed to train an LSTM model for the spectral template constructions using the framework shown in Figure~\ref{fig:LSTM-Architecture}.
Due to the simplicity of the proposed LSTM architecture (Figure~\ref{fig:LSTM-Architecture}) and the limited sample size, no validation set is created to fine-tune the hyperparameters of the LSTM neural networks.

In the algorithm, a training sample consists of $K+1$ {\it training spectra} of a particular SN with their corresponding phases.
The first $K$ spectra form a time sequence with phases in nondecreasing order as neural network input, and the last spectrum is the target spectrum for the neural network prediction. Repetitions in the $K+1$ spectra are allowed.
The training samples are generated by exhausting all possible permutations with repetition of the $K+1$ spectra over the training set.

\begin{algorithm} \label{algo:ctemp}
    \SetAlgoLined
    \SetKwInOut{Input}{input}
    \SetKwInOut{Output}{output}
    \Input{The trained LSTM model}
    \Input{The spectral training set: $\textbf{X}$}
    define the set of ordered spectral pairs for one SN $I_0 \leftarrow $ all $(\textbf{x}_l, \textbf{x}_m)$ with $p_{\textbf{x}_l} \leq p_{\textbf{x}_m}$, where $\textbf{x}_l,\textbf{x}_m \in \textbf{X}$\;
    \If{the size of set $I_0 \leq 128$}{
        let $I = I_0$
    }
    \If{the size of set $I_0 > 128$}{
        \For{$(\textbf{x}_l, \textbf{x}_m)_r \in I_0$}{
            initialize $\beta \leftarrow 1$\;
            initialize $\gamma \leftarrow 1$\;
            \If{$\textbf{x}_l$ and $\textbf{x}_m$ were taken by different instruments}{
                $\beta \leftarrow 0.3$\;
            }
            \If{$p_{\textbf{x}_m} - p_{\textbf{x}_l} \leq 3/2$}{
                $\gamma \leftarrow 0.2$\;
            }
            assign selection probability ${\phi}_r = \beta * \gamma * w_{\text{spec}}\big|_{\textbf{x}_l} * w_{\text{spec}}\big|_{\textbf{x}_m}$\;
        }
        randomly select $I \subseteq I_0$ of size 128 with selection weights ${\phi}_r$\;
    }

    \For{$(\textbf{x}_l, \textbf{x}_m)_i \in I$}{
        \For{$k \in [1,2, ..., 24]$}{
            feed $(\textbf{x}_l, \textbf{x}_m)_i$ to LSTM model to predict spectral time sequence $y_{i,k}(\lambda, p)$, where $p \in [-15, -14.875, ..., +29.875, +33]$
        }
        compute predictive mean: $\mu_i(\lambda, p) = \texttt{MEAN} (y_{i,1}, y_{i,2}, ..., y_{i,24})$\; 
        compute predictive uncertainty: $\sigma_i(\lambda, p) = \texttt{STD} (y_{i,1}, y_{i,2}, ..., y_{i,24})$\;
    }
    compute the final spectral template: $T(\lambda, p) = \sum_{i} \sigma_i^{-2}(\lambda, p) * \mu_i(\lambda, p)$\;
    renormalize the final spectral template: $T(\lambda, p) \leftarrow T(\lambda, p) / \sum_{\lambda} T(\lambda, p)$
    \caption{Spectral Template Construction}
\end{algorithm}

\subsection{Construct the spectral templates} \label{ssec:constemp}

The construction of the spectral templates follows the procedures given in Algorithm~\ref{algo:ctemp}. The LSTM model obtained over the {\it training spectra} was used as the input model. A spectral template for each SN in the CSD was constructed by feeding the {\it training spectra} of the SN into the trained LSTM model.
Throughout this study, $K$ is set to 2, allowing spectral sequences at given target phases to be constructed using two spectra.
It also allows for the special cases that the two spectra are identical; i.e., the predictions can be triggered by only one spectrum.
The restriction on $K$ makes it infeasible to apply the trained LSTM model directly when an SN has more than two {\it training spectra}. Nevertheless, Algorithm~\ref{algo:ctemp} provides a general strategy to construct spectral templates for all SNe in the CSD.

For SNe with large numbers of observed spectra such as SN~2011fe which has a total of 65 {\it training spectra}, the total number of possible combinations (the size of $I_0$) is as large as 2145, which makes it very time-consuming to exhaustively calculate all possible input spectral combinations. 
To construct the input spectral pairs $I$, a subset of possible combinations is sufficient, as the information from such well-observed SNe is also likely redundant.
A weighted selection scheme is adopted by assigning a selection probability for each possible spectral pair (see line~15 of Algorithm~\ref{algo:ctemp}). 
A spectral pair with higher spectral weights $w_{\text{spec}}$ has a higher probability of being selected in predicting the spectral sequence. We also introduced two penalty factors, $\beta$, and $\gamma$, which preferably select spectral pairs taken by the same instrument and disfavor the spectral pairs observed almost at the same night, respectively.

%% ***** Section: SpecTemplate Construction [LSTM]
\begin{figure*}[ht!]
    \centering
    \gridline{
        \fig{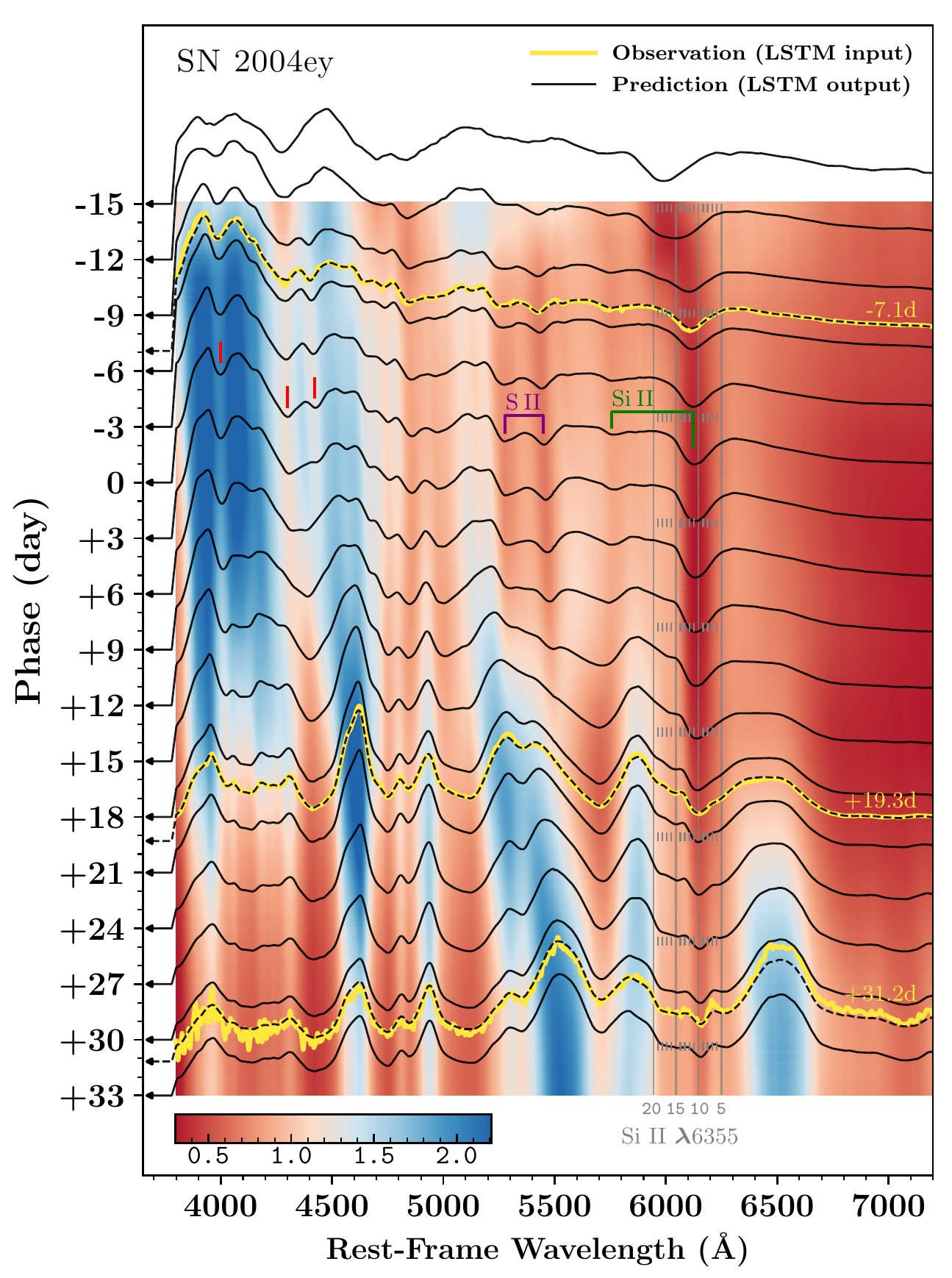}{0.48\textwidth}{(a) SN~2004ey (NV)}
        \fig{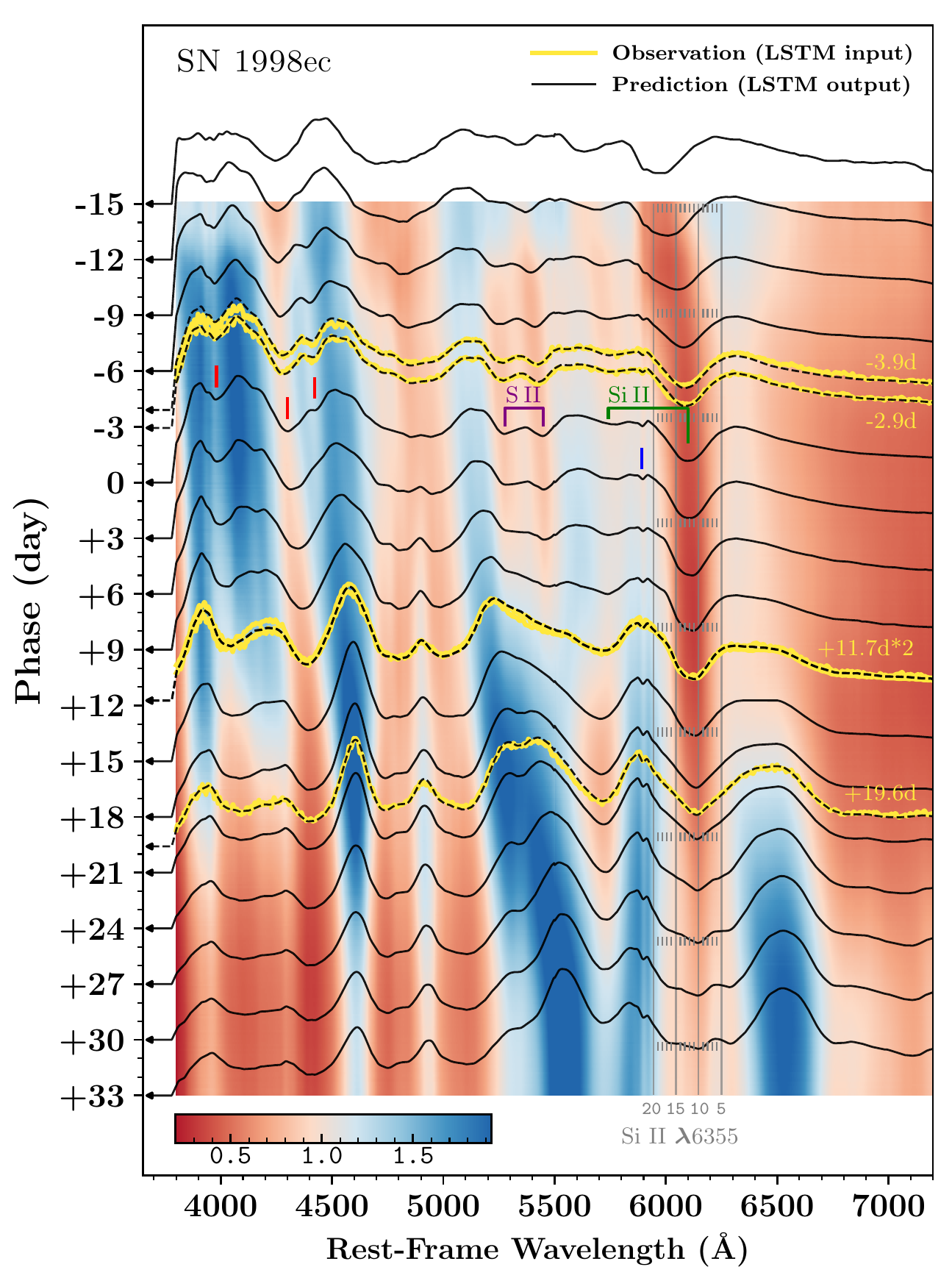}{0.48\textwidth}{(b) SN~1998ec (HV)}
        }
    \caption{\label{fig:SpecTemp-Schematic-04ey98ec} Predicted spectral templates of SN~2004ey (\textit{left}) and SN~1998ec (\textit{right}) using LSTM neural networks. The {\it training spectra} (yellow curves) of SN~2004ey (SN~1998ec) are fed into the LSTM model to construct a spectral template (the heat map). To improve the readability of the two-dimensional image, a selected spectral sequence (black curves) that constitutes certain discrete slices of the template spectral surface is overplotted to demonstrate the spectral features of certain phases as shown by the arrows on the left. Some typical absorption features of SNe~Ia are marked by short vertical lines, including the Si~II lines (at $6355$ and $5972$\AA, green lines), the $\mathsf{W}$-shape S~II lines (at $5400$ and $5600 $\AA, purple lines), Si~II $\lambda 4130$\AA\ (left red line), Fe II $\lambda 4404$\AA\ and Mg II $\lambda 4481$\AA\ (middle red line) and Si~III $\lambda 4560$\AA\ (right red line). Moreover, in the case of SN~1998ec, the persistent interstellar absorption Na~I~D lines (blue line) is also reproduced by the spectral template. Note that there are two spectra of SN~1998ec observed at the same epoch (+11.7 days) by two different instruments. The vertical gray lines indicate the positions of the blueshifted Si~II $\lambda 6355$\AA\ at different velocities (the velocity unit is 1000 km $\text{s}^{-1}$).}
\end{figure*}

\begin{figure*}[ht!]
    \centering
    \gridline{\fig{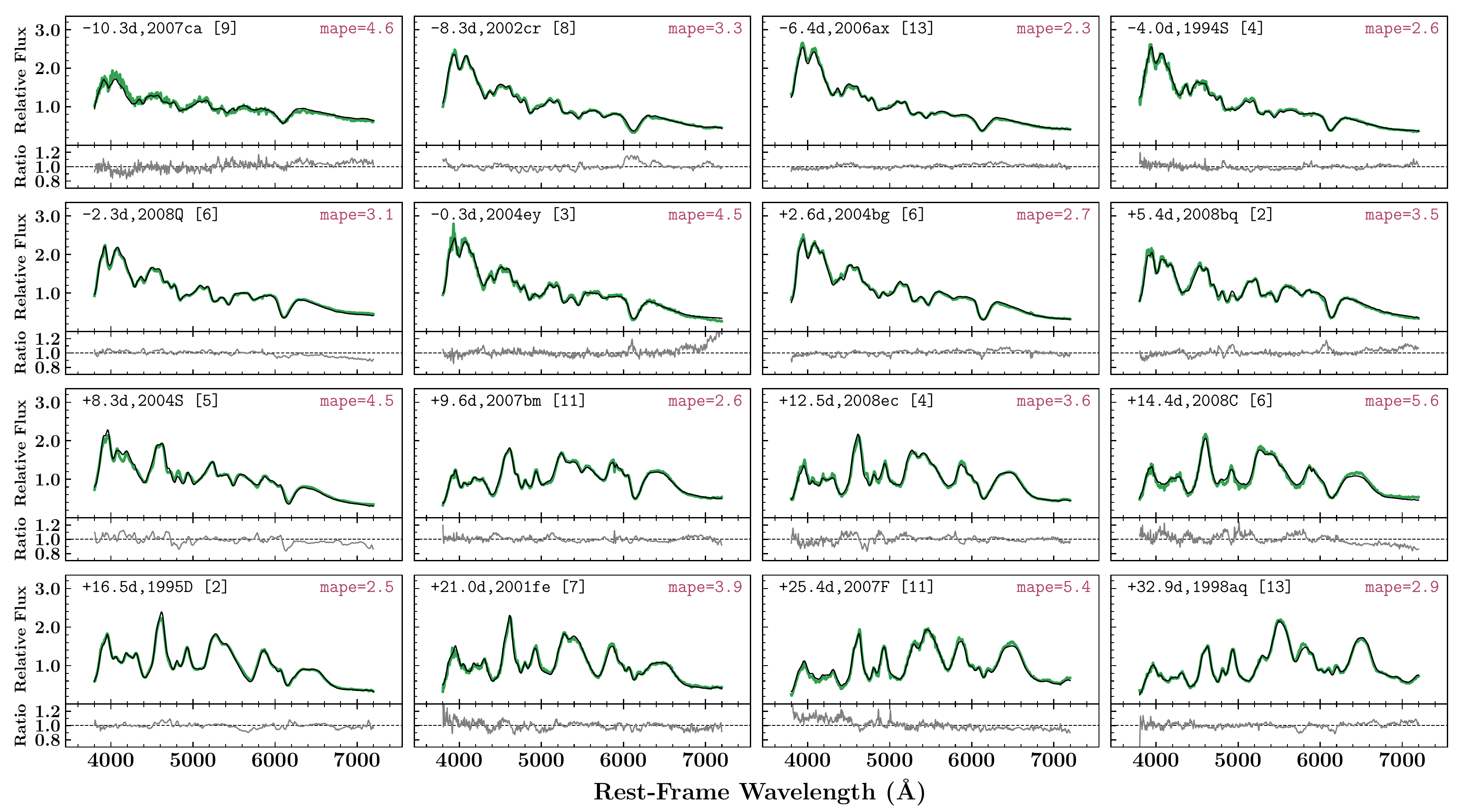}{0.86\textwidth}{(a) NV}}
    \gridline{\fig{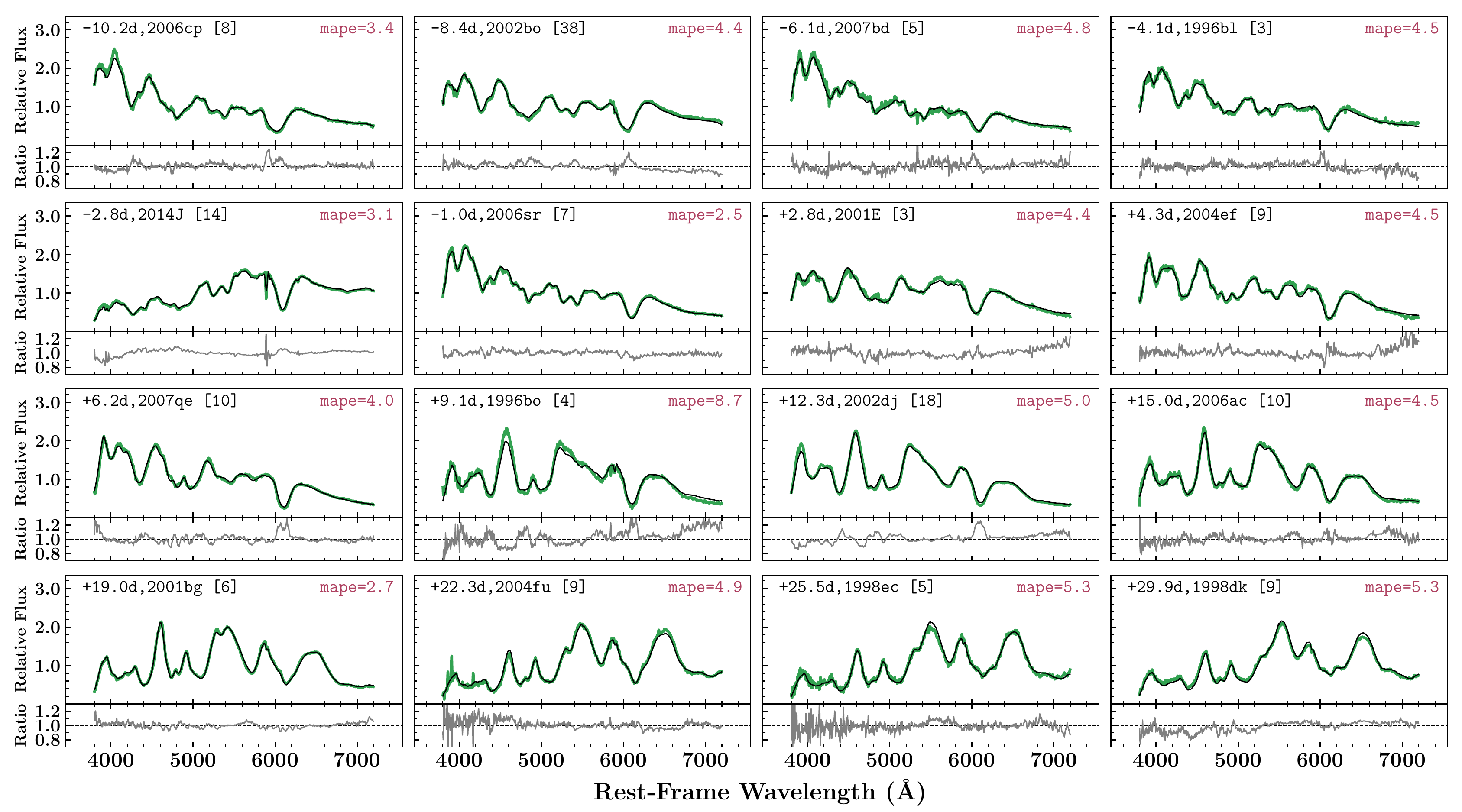}{0.86\textwidth}{(b) HV}}
    \caption{\label{fig:SpecTemp-Validate-Example-NVHV} Comparisons of the LSTM-predicted template spectra (black curves) with the observed spectra in the test set (green curves) for 16 NV objects and 16 HV objects. The flux ratios of the predictions to observations are shown in the lower attached frame for each panel. The number at the top left of each panel shows the phase from $B$-band maximum. The number in the square brackets in the top left corner of each panel indicates the number of {\it training spectra} of the SN. The MAPEs between the predicted template spectra and the observed {\it testing spectra} are shown in the top right corner of each panel.}
\end{figure*}

\begin{figure*}[ht!]
    \centering
    \gridline{\fig{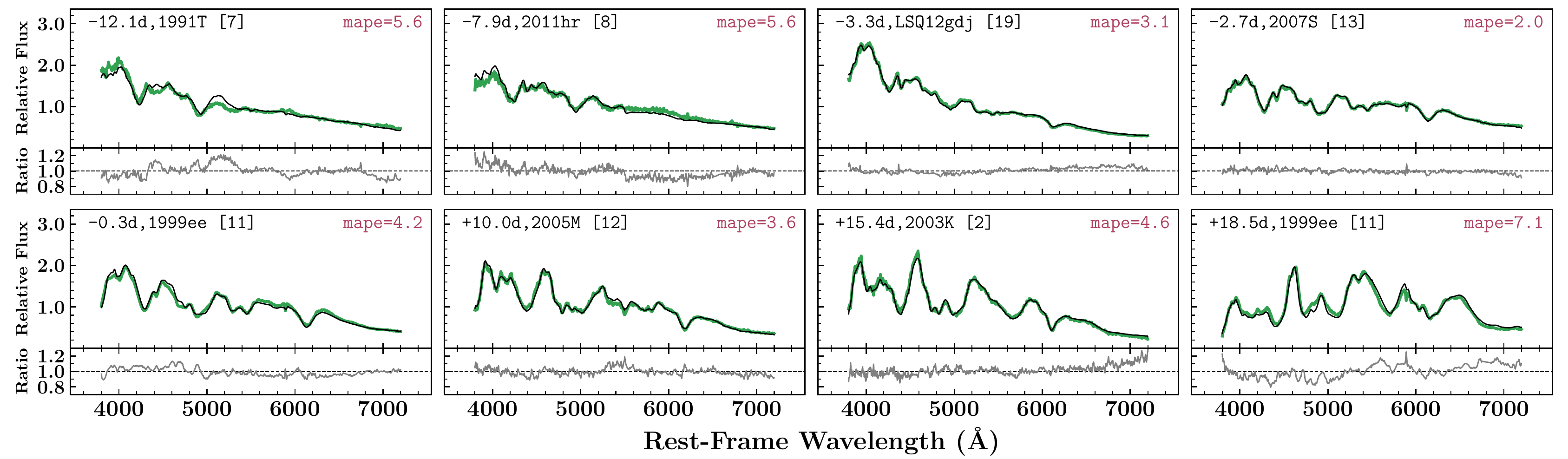}{0.85\textwidth}{(a) Ia-91T}}
    \gridline{\fig{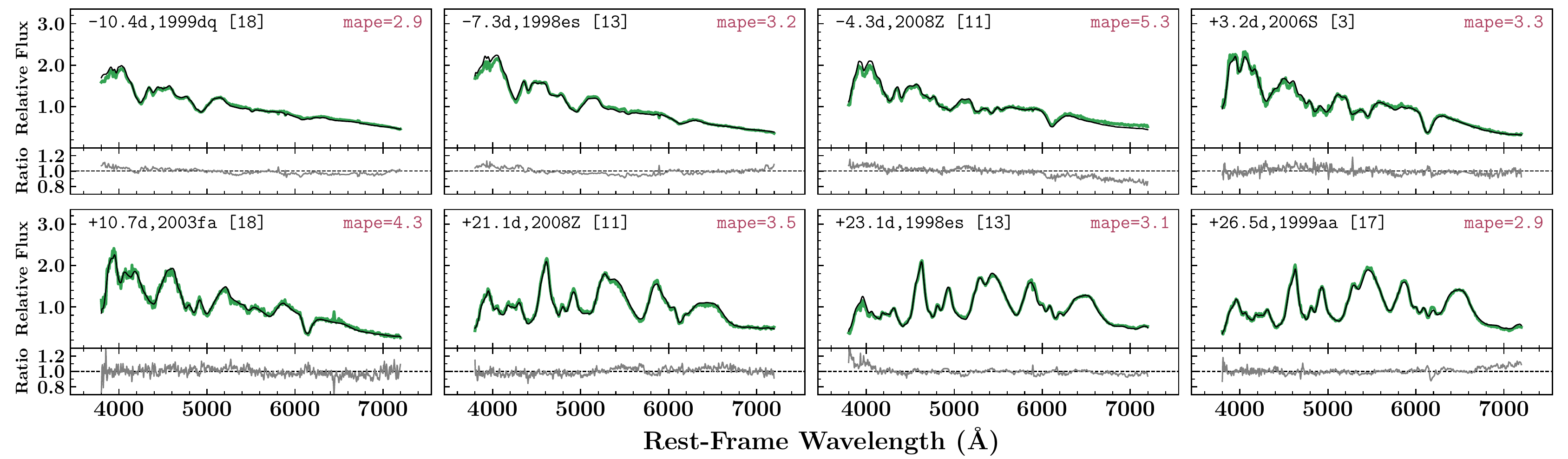}{0.85\textwidth}{(b) Ia-99aa}}
    \gridline{\fig{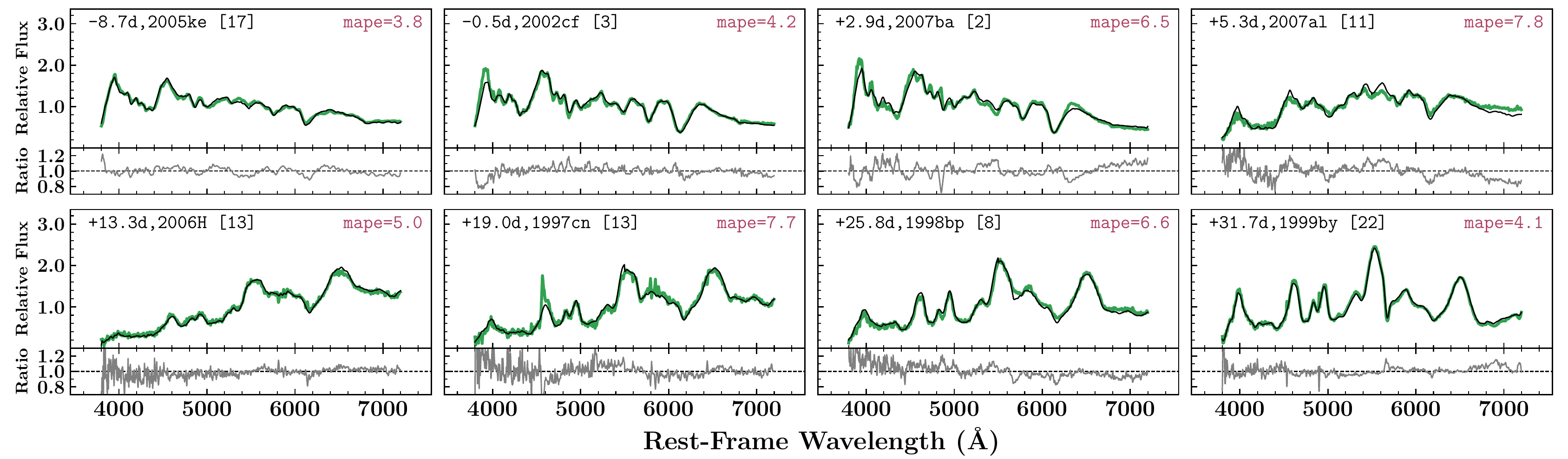}{0.85\textwidth}{(c) Ia-91bg}}
    \gridline{\fig{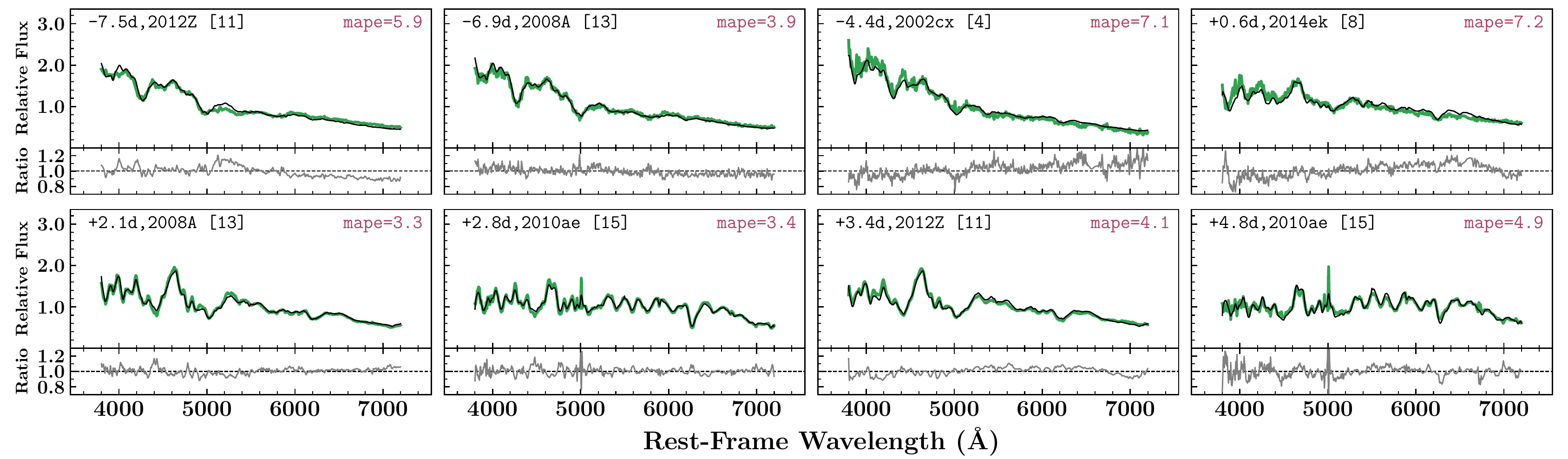}{0.85\textwidth}{(d) Iax}}
    \caption{\label{fig:SpecTemp-Validate-Example-pec} Comparison of the predicted template spectra (black curves) with the observed {\it testing spectra} (green curves) for (a) Ia-91T, (b) Ia-99aa, (c) Ia-91bg, and (d) Iax. The panel format is the same as in Figure~\ref{fig:SpecTemp-Validate-Example-NVHV}.}
\end{figure*}

\begin{figure*}[ht!]
    \centering
    \includegraphics[trim=0cm 0cm 0cm 0cm,clip=true,width=12.5cm]{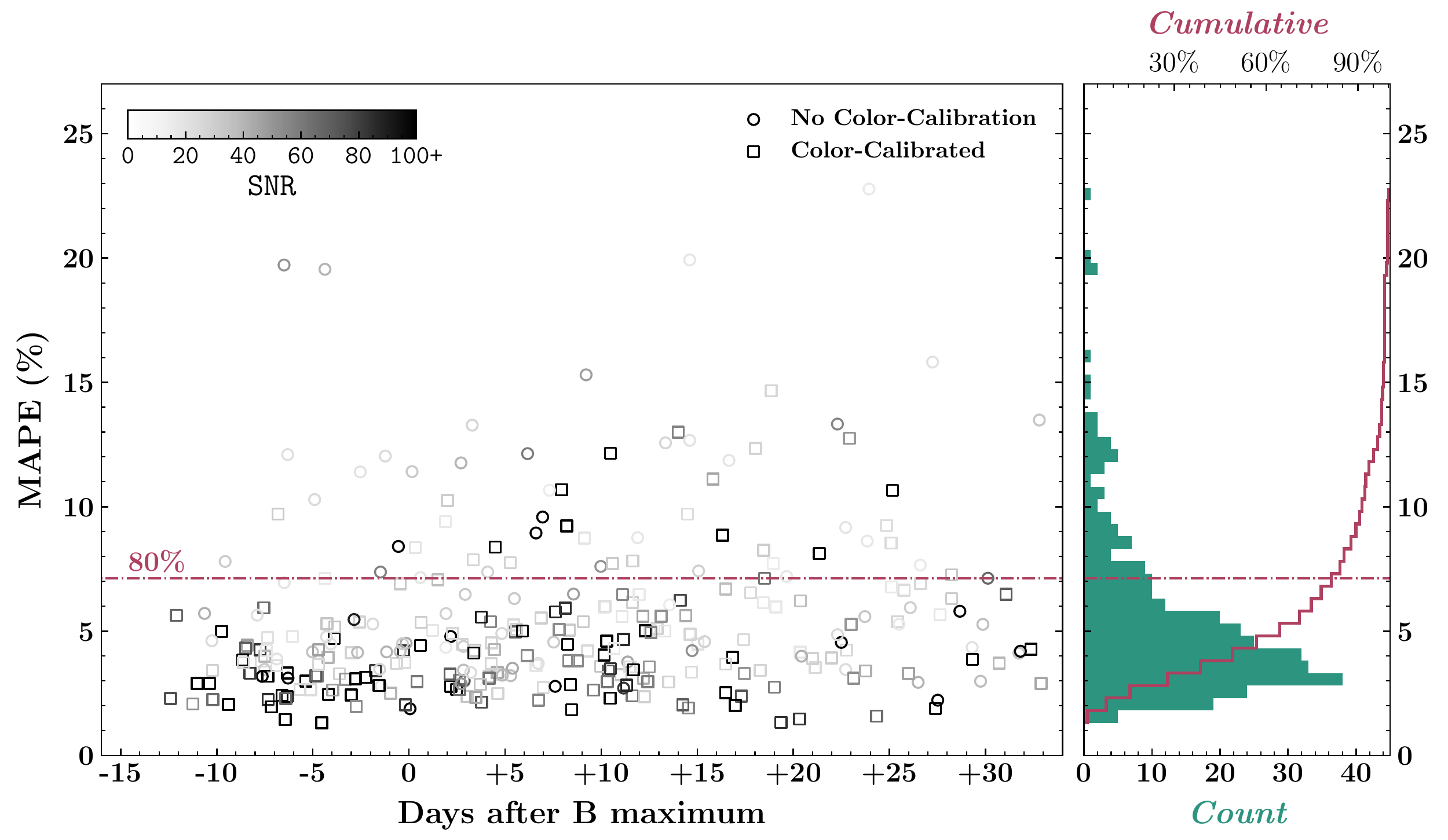}
    \caption{\label{fig:SpecTemp-Validate-TargetPhase} The MAPEs between the observed spectra in the test set and the predicted template spectra as a function of phase. We use squares (circles) to indicate the {\it testing spectra} with (without) available color calibration. The gray-scale colors of the data points represent the S/N of the {testing \it spectra}. A histogram of total MAPEs with a cumulative curve (red line) is attached to the right side. The horizontal dashed line shows the 80th percentile of the total MAPE distribution.}
\end{figure*}

\begin{figure*}[ht!]
    \centering
    \includegraphics[trim=0cm 0cm 0cm 0cm,clip=true,width=11cm]{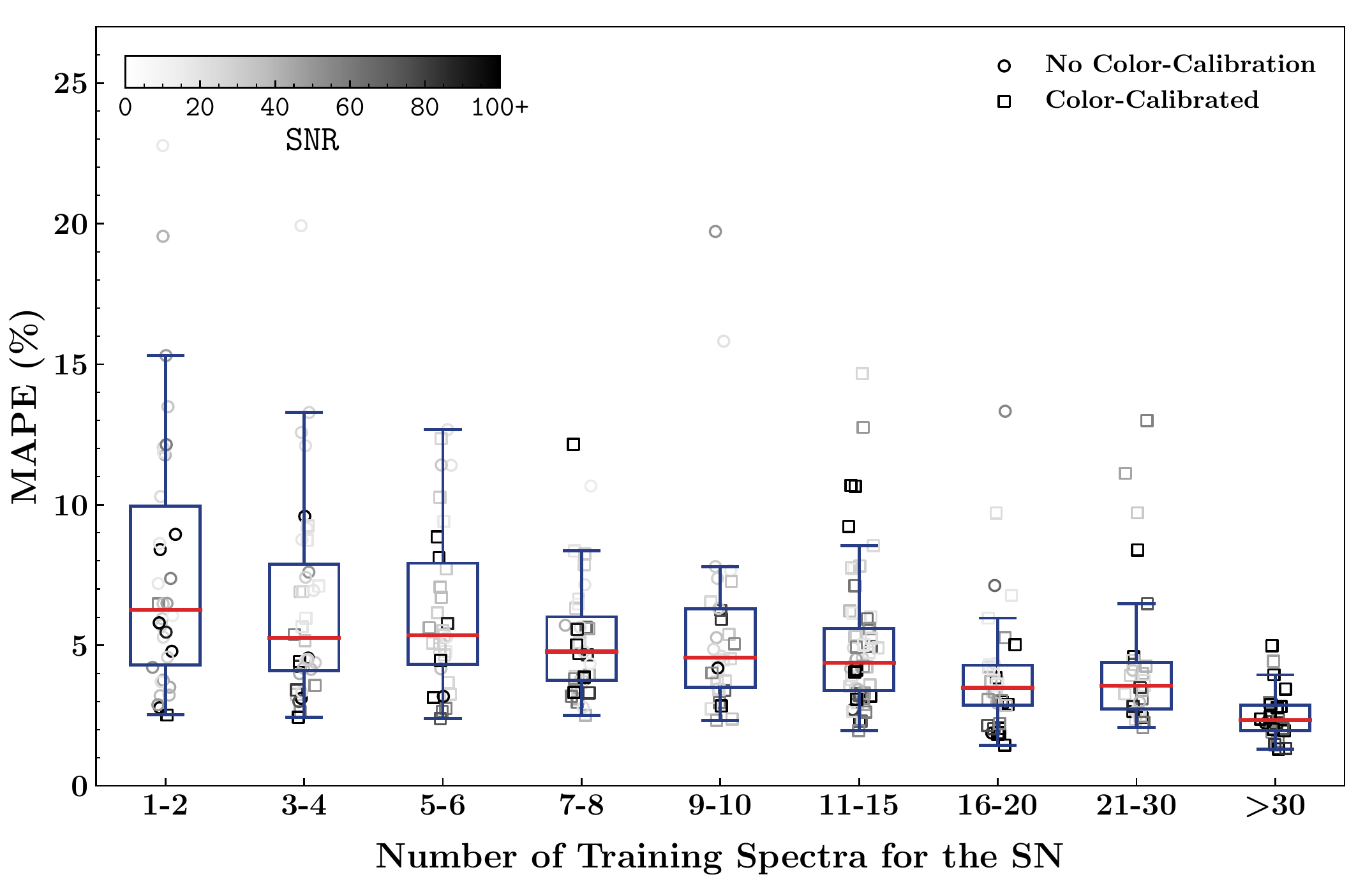}
    \caption{\label{fig:SpecTemp-Validate-NumGSpec} The MAPEs between the observed spectra in the test set and the predicted template spectra versus the number of the {\it training spectra} for the SN. The symbols are the same as in Figure~\ref{fig:SpecTemp-Validate-TargetPhase}. The data points are arbitrarily shifted with small displacements in the horizontal direction for clarity of display. A box plot is overplotted onto each group of data points (see the definition of a box plot in Figure~\ref{fig:FPCA-ReconAccuracy}).}
\end{figure*}

Each spectral pair in this selection was fed to the neural networks for the spectral sequence prediction that covers spectral phases between -15 to +33 days (see line~21 of Algorithm~\ref{algo:ctemp}).
The activation of dropout can make the predictions probabilistic (see Section~\ref{sec:lstm}) and allows for estimates of the uncertainties of the predictions.
For every input spectral pair, the prediction process is repeated 24 times with a new stochastic dropout mask applied each time.
The final target spectrum prediction is given by the average of these predictions weighted by their predictive uncertainties (see line~26 of Algorithm~\ref{algo:ctemp}).
At last, each spectrum of the spectral template was renormalized by dividing its average flux (see line~27 of Algorithm~\ref{algo:ctemp}).

As examples, Figure~\ref{fig:SpecTemp-Schematic-04ey98ec} shows the predicted spectral templates of two SNe with no more than five {\it training spectra}. 
Of the two SNe, SN~2004ey is a normal-velocity (NV) \citep{2015ApJS..220...20Z} and SN~1998ec is a high-velocity (HV) object \citep{2012AJ....143..126B}. 
Figure~\ref{fig:SpecTemp-Schematic-04ey98ec} offers a glimpse of SN~Ia diversities in their spectral evolution. Around the $B$-band maximum, for instance, SN~1998ec has a distinctly broader (relative to SN~2004ey) and stronger Si $\lambda$6355, Fe II $\lambda$4404, and Mg II $\lambda$4481 lines, in line with typical broad-line (BL) SNe~Ia \citep{2013Sci...340..170W, 2011ApJ...732...30P} in Branch classes \citep{2006PASP..118..560B}. 
As the objects evolve from -10 to +5 days, SN~1998ec shows a steeper velocity gradient across the Si~II features than SN~2004ey. SN~1998ec was classified as a high velocity gradient (HVG) object \citep{2011ApJ...732...30P}. The rapid line profile evolution is well captured by the neural network predictions. 
Moreover, the predicted spectral templates reconstruct the smooth spectral evolution without any unexpected discontinuities over the phase dimension. The LSTM neural networks also demonstrate excellent consistency at the boundaries of the blue and the red sections at 5500 \AA.

\subsection{Test the spectral template constructions} \label{ssec:test4ctemp}

The test set was used to evaluate the performance of the neural networks.
The observed spectra in the testing set were compared with the spectral template the constructed from the neural network for the same SN at the same phase.
To precisely match the phases of the {\it testing spectra} (hereafter testing phases) to those of the spectral template, 
the spectral template used for comparison was constructed at the phases that match exactly the phases of observations of an SN instead of the fixed time grid, as shown in line~21 of Algorithm~\ref{algo:ctemp}. Examples of the spectral comparisons are shown in Figures~\ref{fig:SpecTemp-Validate-Example-NVHV}, and \ref{fig:SpecTemp-Validate-Example-pec} for normal SNe (NV and HV) and some peculiar SNe, respectively.

Figure~\ref{fig:SpecTemp-Validate-TargetPhase} demonstrates the results of the spectral comparison by measuring the MAPE between the {\it testing spectra} and the corresponding neural network constructions.
We found 80\% of the {\it testing spectra} show an MAPE smaller than 7.1\%, and the overall median MAPE is 4.4\%. The MAPE error does not show any significant dependence on the phases of the SNe, which means the performance of the neural network is not statistically biased at any particular phases.
We noticed that the MAPEs over the {\it testing spectra} without normalizing to the observed photometric colors are systematically larger than MAPE over those spectra with photometric color calibrations; 35 of 94 unnormalized {\it testing spectra} have MAPE larger than 7.1\%; By contrast, the ratio is 27 in 215 for the color-calibrated {\it testing spectra}.
We surmise that the MAPEs for {\it testing spectra} without color calibrations are more likely to be dominated by the inaccuracy of the observational colors of the spectra. 
Alternatively, we note that the median S/Ns over the test spectra without and with color calibrations are 31.9 and 46.7, respectively. The uncalibrated {\it testing spectra} generally have lower S/Ns, which could also contribute to their larger MAPEs.

The reliability of the prediction may also depend on the number of available spectra in the training set. Figure~\ref{fig:SpecTemp-Validate-NumGSpec} shows the MAPEs between the observed spectra in the test set and the neural network-predicted spectra for SNe with different numbers of spectra in the training set. The figure confirms the general trend that the errors decrease as the number of spectra used in the training set increases.

%% ***** Section: SpecTemplate Construction [Comp]
\begin{figure*}[ht!]
    \centering
    \includegraphics[trim=0cm 0cm 0cm 0cm,clip=true,width=15.8cm]{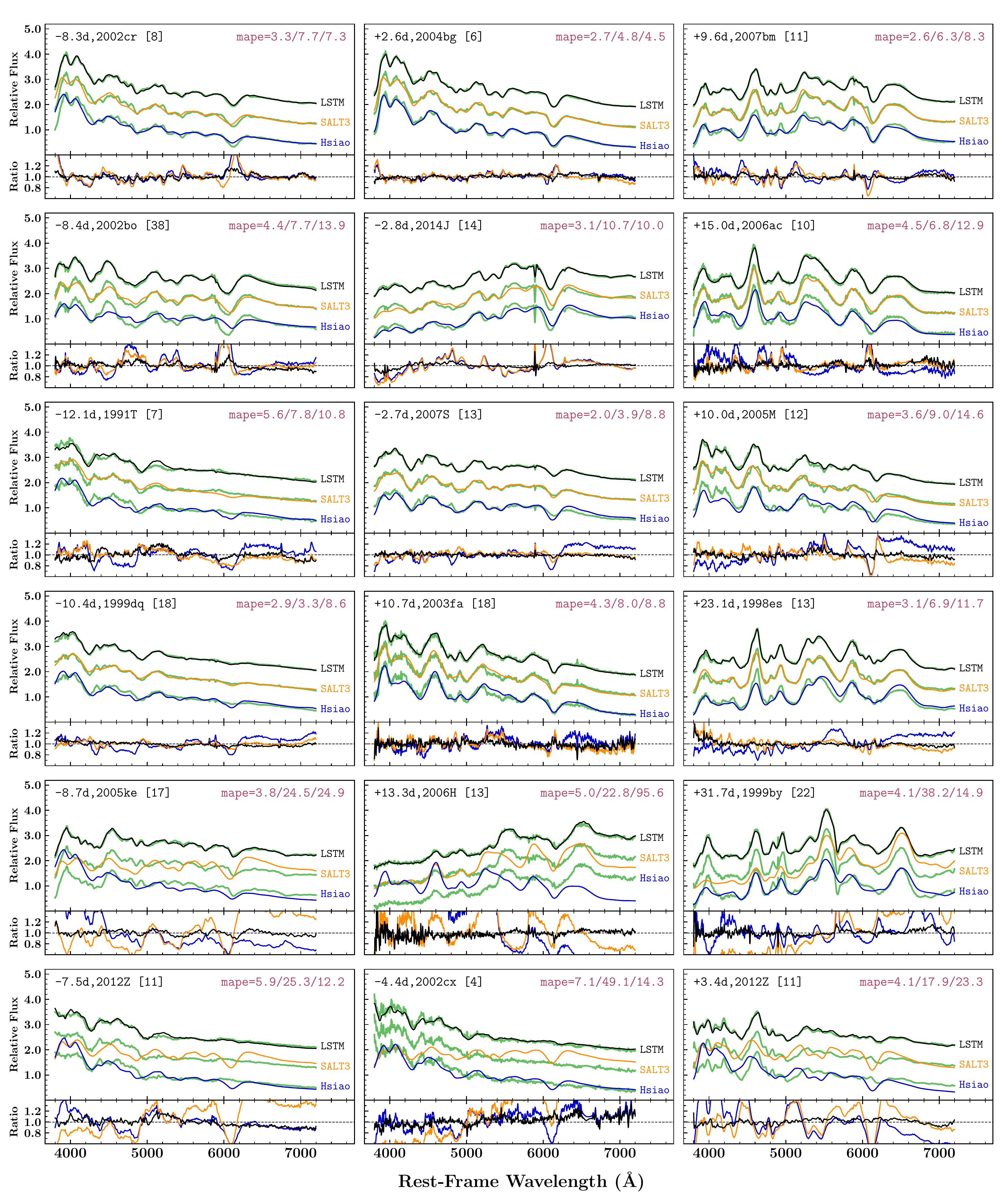}
    \caption{\label{fig:SpecTemp-BaselineValidate-Example} Comparisons of the template spectra generated by different methods (black curves for LSTM; yellow curves for SALT3; blue curves for the Hsiao template) with the observed spectra in the test set (green curves). Each row shows three cases at different phases for a specific subtype, from top to bottom, NV, HV, Ia-91T, Ia-99aa, Ia-91bg, and Iax. The panel format is the same as in Figure~\ref{fig:SpecTemp-Validate-Example-NVHV}, but additional comparisons for the baseline models are also presented. In each panel, the template spectra have been arbitrarily shifted in the vertical direction for display clarity, and the same shifts are applied to the observed spectra (green curves). The MAPEs between the template spectra and the observed {\it testing spectra} are given in the top right corner of each panel; from left to right, the numbers are for the LSTM, the SALT3, and the Hsiao template.}
\end{figure*}

\subsection{Comparsions with Other Models} \label{ssec:baseline4ctemp}

To showcase the fidelity improvement of our method, we compared the spectral templates constructed by LSTM with those generated by two  empirical models of SNe~Ia, i.e., the template of \cite{Hsiao2007Kcorrection} and the SALT3 model \citep{SALT3}.

The spectral template built by \citep[][hereafter the Hsiao template;]{Hsiao2007Kcorrection} is a phase-dependent SED model based on a compilation of $\sim$100 SNe~Ia with $\sim$600 spectra. This uniform template $H(p,\lambda)$, as a function of rest-frame wavelength $\lambda$ and phase $p$, was obtained by averaging the observed spectra at different epochs after correcting their spectral colors to align to a typical normal SN~Ia with a ``stretch" value of 1 \citep{Hsiao2007Kcorrection,Knop2003}.
SALT2 \citep{SALT2} typifies the SN~Ia spectral templates that are portrayed by a few free parameters. The spectral flux of the SALT2 model is given by
\begin{align}
    \label{eqn:salt2}
    \begin{split}
    F(p,\lambda) =\,\,& x_0[M_0(p,\lambda) + x_1M_1(p,\lambda)]
    \\
    &\cdot\exp[c\cdot CL(\lambda)]
    \end{split}
\end{align}
where $M_0$ and $M_1$ as principal components are flux surfaces derived from a training spectral sample, and $CL(\lambda)$ represents the average color correction law. 
A SALT2 template is determined by three parameters: $x_0$ (the overall flux normalization), $x_1$ (linked to the light-curve stretch) and $c$ (a color-law coefficient).
Beyond giving an average spectral evolution of SNe~Ia, SALT2 also models the variations from SN stretch and includes a modulation term encoding the time-invariant color component.
SALT3 \citep{SALT3} is an improved version of SALT2 using the same framework but has better uncertainty estimation and better disentanglement of color and SN stretch. 
The SALT3 baseline model used in our work is the one presented in \cite{SALT3}, which was trained on a dataset of 1207 spectra from 1083 SNe. Throughout the paper, we used the {\tt\string Python} implementation of the Hsiao template and SALT3 model in the {\tt\string sncosmo} library \citep{SncosmoSoftware} for the baseline comparisons.

In this section, we applied the two \cite{Hsiao2007Kcorrection} and SALT3 models in the CSD. The reddenings by interstellar dust are needed for these models. For SALT3, photometric light curves are required. Only 118 SNe out of the 361 SNe in the CSD have photometric coverages and published host galaxy reddenings that are appropriate for such comparisons. 
The Milky Way (MW) foreground extinction of each SN is derived using the \cite{SFD2011} reddening map. For 108 of the 118 SNe, we directly adopted the values of host extinction provided by the Kaepora database \citep{2019MNRAS.486.5785S} which lists $A_V$ derived from MLCS2k2 \citep{Jha_2007} assuming $R_V=2.5$. The host extinctions for the remaining 10 SNe are not available in the Kaepora database and are adopted from the following publications: SN~2011iv \citep{SN2011iv_Extinction}, LSQ~12gdj \citep{LSQ12gdj_Extinction}, SN~2010ae \citep{SN2010ae_Extinction}, SN~2014J \citep{SN2014J_Extinction}, SN~2012cu \citep{SN2012cu_Extinction}, SN~2012Z \citep{SN2012Z_Extinction}, SN~2010jn \citep{SN2010jn_Extinction}, SN~2014ek \citep{SN2014ek_Extinction}, ASASSN-14lp \citep{ASASSN-14lp_Extinction} and SN~2011hr \citep{SN2011hr_Extinction}.

A model by the Hsiao template is uniquely determined by the assumed dust extinctions from the MW and the host galaxy. We generated the model Hsiao templates for the 118 SNe after corrections for both the MW and the host reddening, assuming the Cardelli, Clayton $\&$ Mathis (CCM) extinction law \citep{1989ApJ...345..245C}.
Unlike the Hsiao template, the SALT3 model requires the observed spectra as the input data to fit the parameterized spectral time sequence. The SALT3 template of each SN is derived by fitting the LSTM {\it training spectra} corrected by their photometric $B-V$ color.

The spectra in the CSD are all flux normalized by their average flux (see Section~\ref{ssec:colorcal}), whereas the SALT3 model requires the spectra being fitted to preserve the time evolution of the flux. Therefore, the flux levels of the input spectra are renormalized to their corresponding $V$-band magnitudes before SALT3 fitting. 
The SALT3 model allows for fitting spectra with given spectral uncertainties. Here the uncertainty spectra are deduced from the approximate error spectra described in Section~\ref{ssec:snr} by taking into account the flux scaling factors.

Figure~\ref{fig:SpecTemp-BaselineValidate-Example} shows examples of the template spectra at testing epochs generated by the LSTM, the \cite{Hsiao2007Kcorrection}, and SALT3 models. The SNe of distinct spectral properties at different phases are shown. In all cases, the LSTM method generally offers a considerable improvement over the other two models in reconstructing the spectral features. 
The Hsiao templates are unable to capture the diversity of the spectral profiles of most of the SNe. The SALT3 model only shows moderate improvement over the Hsiao template in most cases. The advantages of LSTM become even more obvious for the peculiar SNe such as Ia-91T, Ia-91bg, and Iax. For these peculiar events, not only are the spectral features poorly fit, but also the continuum levels are missed by the Hsiao and SALT3 templates. These models also have severe difficulties matching the spectral features of HV SNe.

\begin{figure}[ht!]
    \centering
    \includegraphics[trim=0cm 0cm 0cm 0cm,clip=true,width=6.5cm]{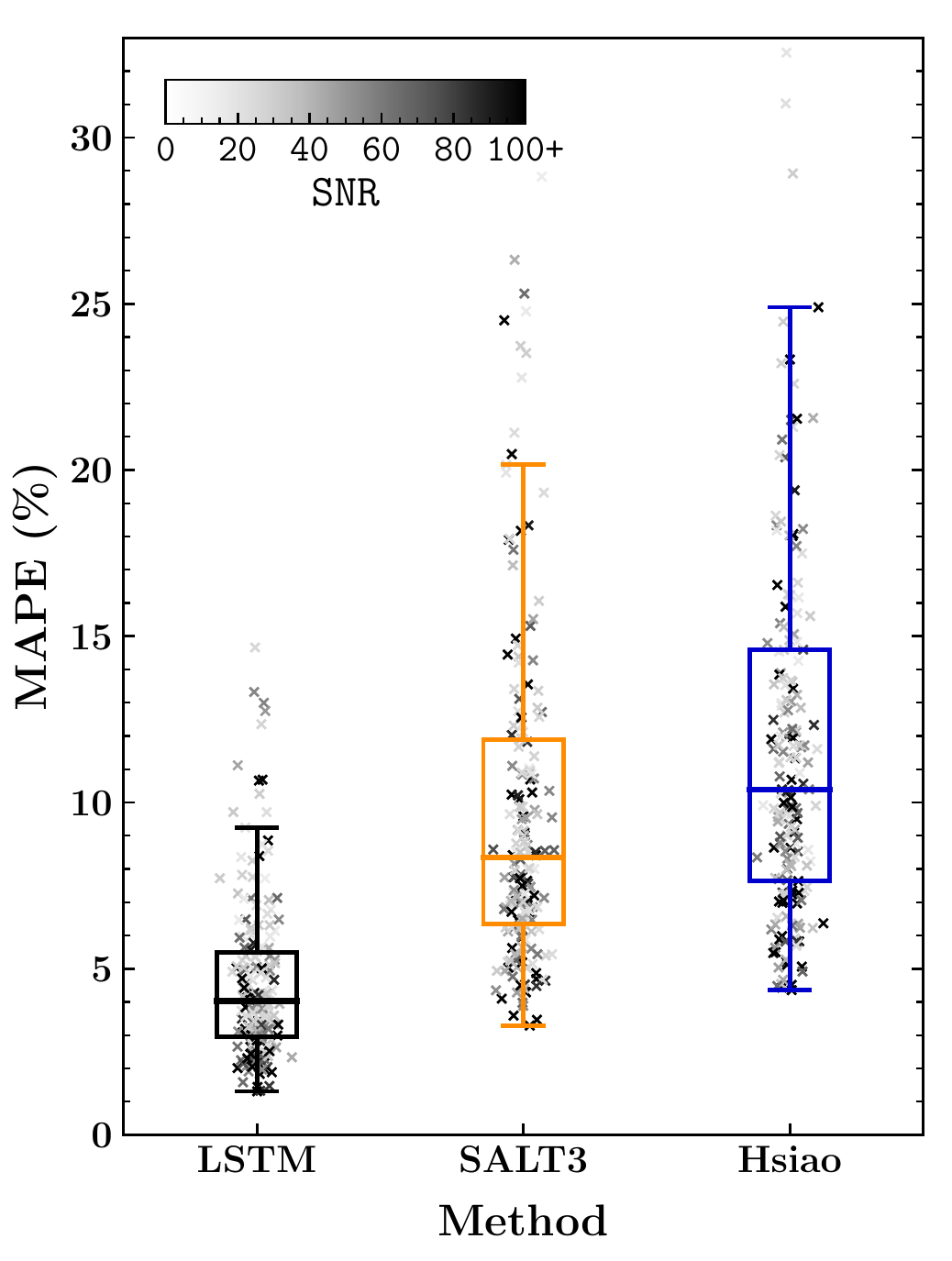}
    \caption{\label{fig:SpecTemp-BaselineValidate-Stat} The MAPEs between the testing spectra from the 118 SNe and the template spectra versus the template generation methods. The data points are arbitrarily shifted with small displacements in the horizontal direction for clarity of display. A box plot is overplotted onto each group of data points (see the definition of a box plot in Figure~\ref{fig:FPCA-ReconAccuracy}).}
\end{figure}

Figure~\ref{fig:SpecTemp-BaselineValidate-Stat} shows the MAPEs between the {\it testing spectra} from the 118 SNe and their corresponding template spectra versus the template generation methods. This statistics also confirms that the LSTM templates with a median MAPE error of 4.0\% outperform the two other models. The median MAPEs are 8.3\% and 10.4\% for the SALT3 and Hsiao templates, respectively.

%% ***** Section: Spec Prediction [Normal]
\begin{figure*}[ht!]
    \centering
    \gridline{
        \fig{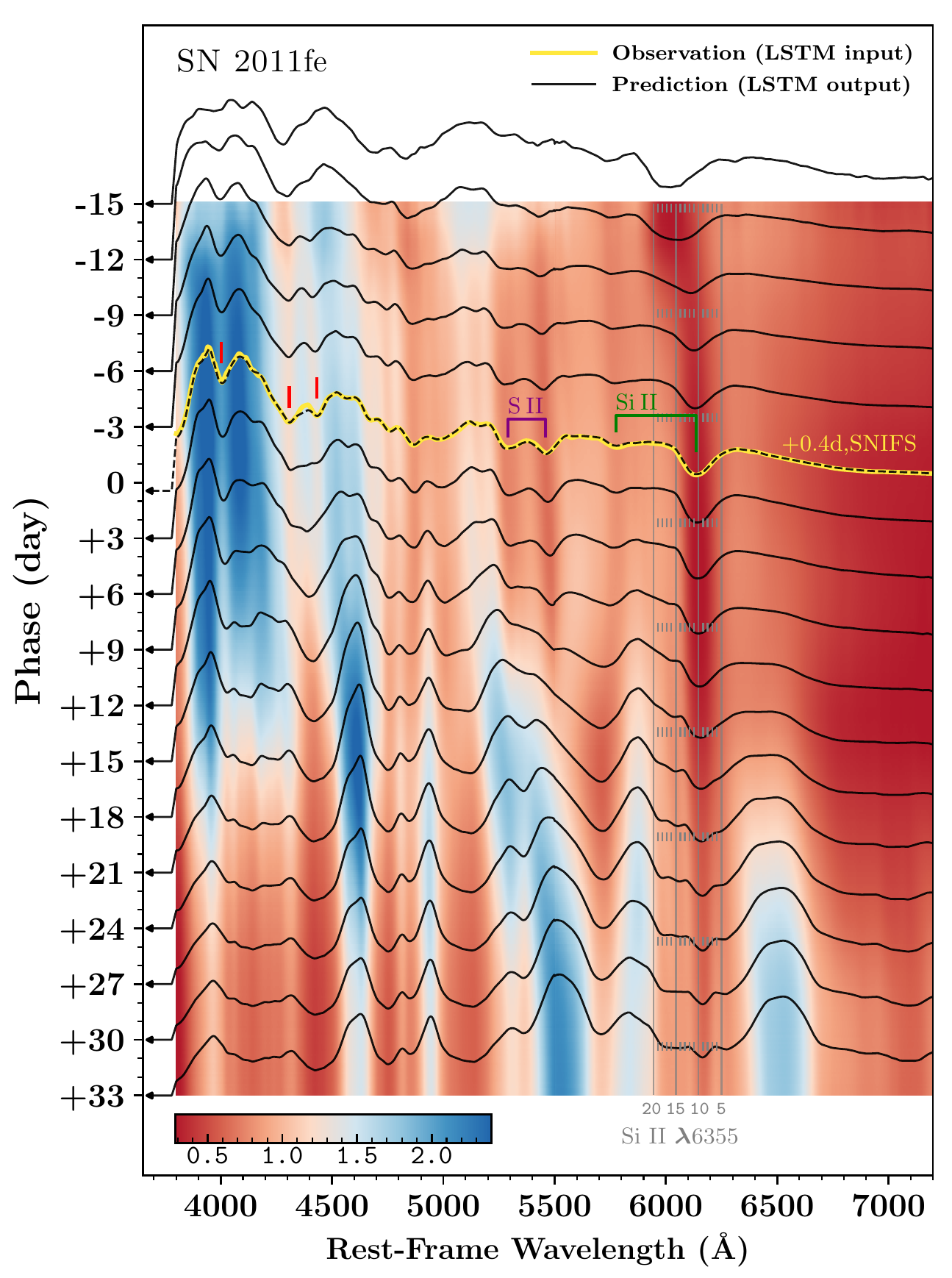}{0.48\textwidth}{(a) SN~2011fe (NV)}
        \fig{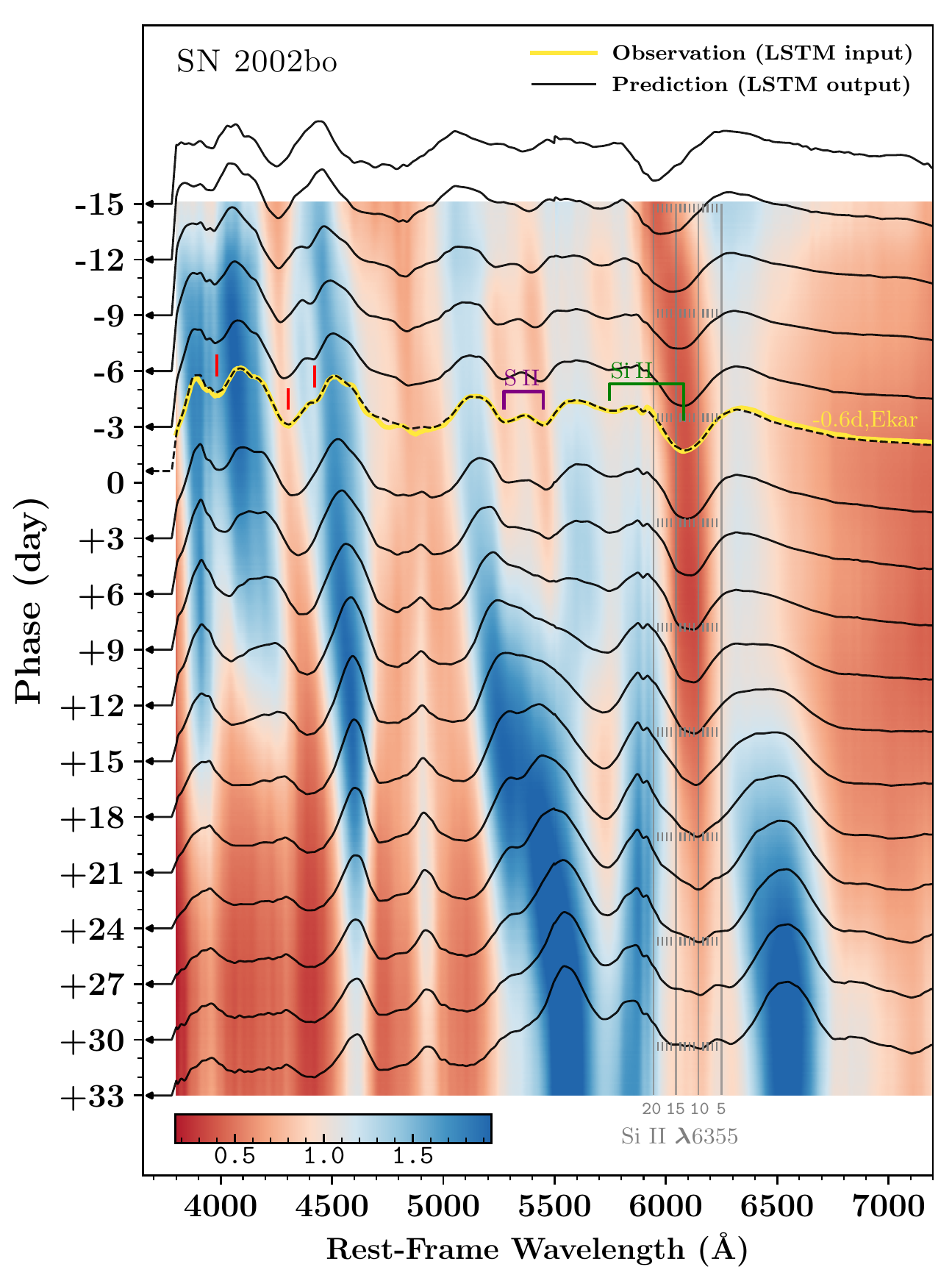}{0.48\textwidth}{(b) SN~2002bo (HV)}
        }
    \caption{\label{fig:OneSpec-Schematic-11fe02bo} Predictive mean sequences of SN~2011fe (\textit{left}) and SN~2002bo (\textit{right}) derived from one single observed spectrum at maximum light using LSTM neural networks. For each SN, the spectrum (yellow curve) observed at maximum light is fed into the LSTM model (trained on the data excluding this SN), then the predictive mean sequence (heat map) is obtained by averaging all results from multiple forward passes. The overplotted discrete spectra (black curves) are drawn from the full sequence. The panel format is the same as in Figure~\ref{fig:SpecTemp-Schematic-04ey98ec}.}
\end{figure*}

\begin{figure*}[ht!]
    \centering
    \gridline{\fig{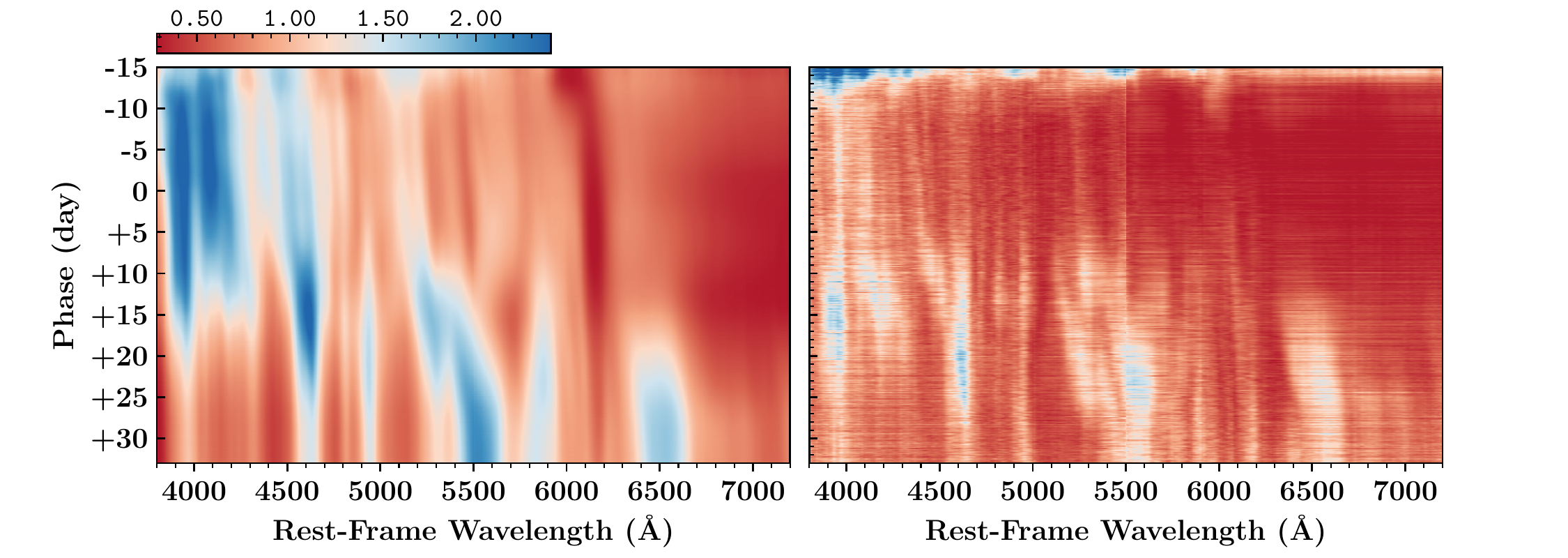}{0.95\textwidth}{(a) SN~2011fe (NV)}}
    \gridline{\fig{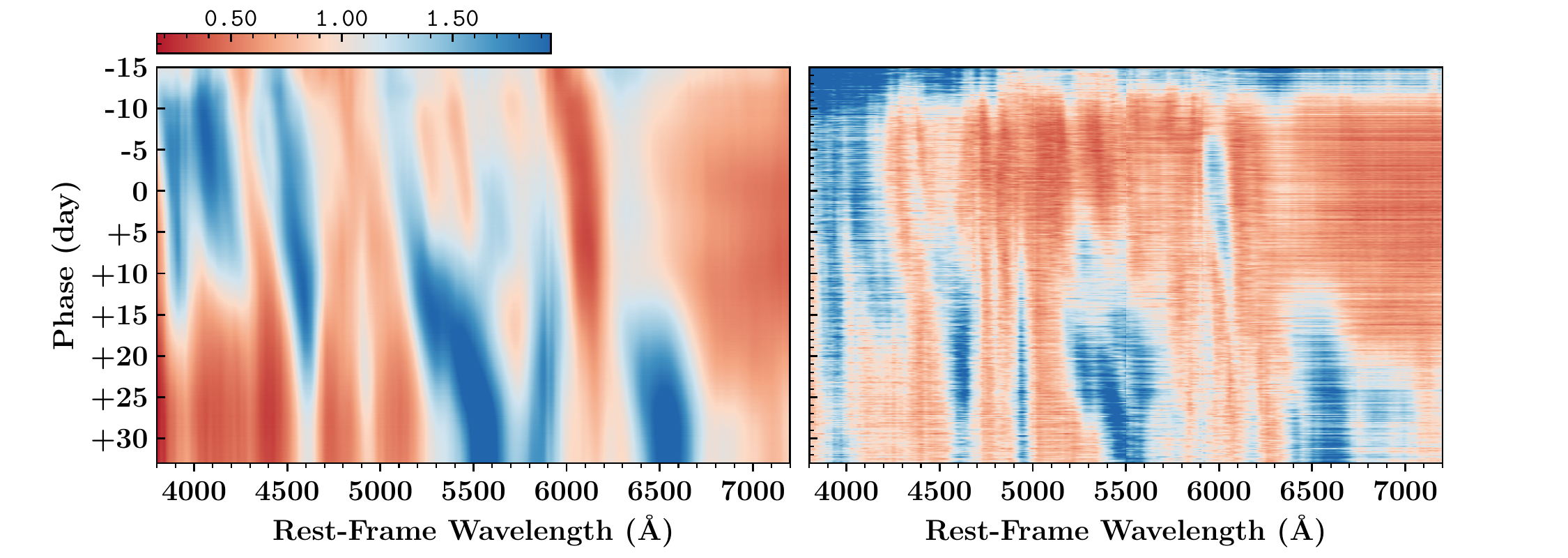}{0.95\textwidth}{(b) SN~2002bo (HV)}}
    \caption{\label{fig:OneSpec-Uncertainty-11fe02bo} Comparison of the predictive mean sequences with their corresponding uncertainty sequences of SN~2011fe (\textit{top}) and SN~2002bo (\textit{bottom}) derived from one single observed spectrum at maximum light. The left heat maps are identical to those shown in Figure~\ref{fig:OneSpec-Schematic-11fe02bo} with adjusted sizes. The right panels show the 1-$\sigma$ uncertainties amplified by a factor of 50.}
\end{figure*}

\begin{figure*}[ht!]
    \centering
    \gridline{
        \fig{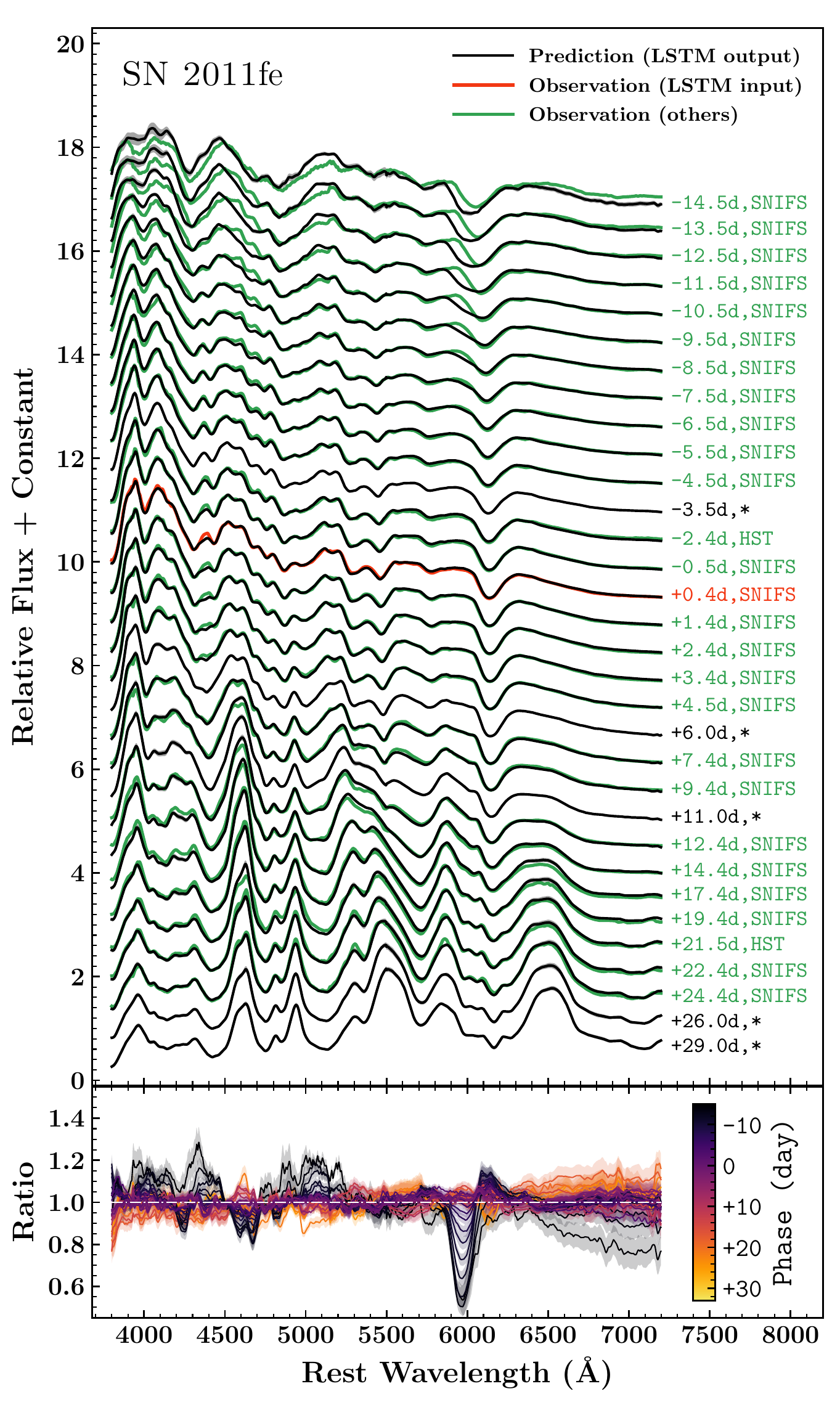}{0.48\textwidth}{(a) SN~2011fe (NV)}
        \fig{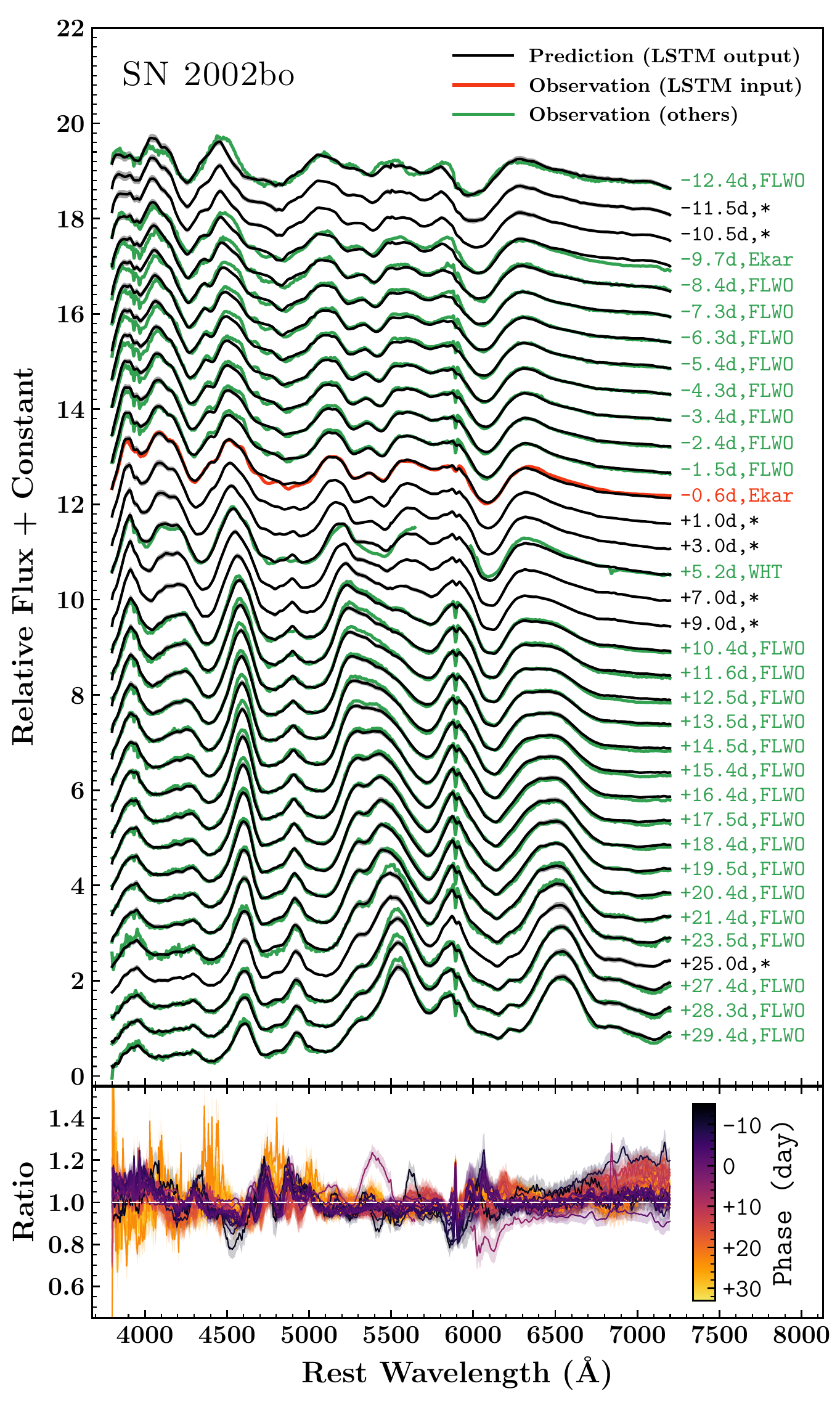}{0.48\textwidth}{(b) SN~2002bo (HV)}
        }
    \caption{\label{fig:OneSpec-Validate-NVHV-direct} Spectral sequence predicted from one spectrum at maximum light for SN~2011fe (\textit{left}) and SN~2002bo (\textit{right}) using LSTM neural networks. The solid black lines show the predictive mean spectra, and the gray shaded areas indicate 2$\sigma$ standard deviations, representing 95$\%$ confidence. The corresponding observations (photometric color-corrected) are plotted as green curves, except for the input spectrum around the peak highlighted in red. All of these spectra have been arbitrarily shifted in the vertical direction for clarity of display. They are labeled with phase and instrumental source, where the asterisk is a placeholder for the cases without corresponding spectroscopic observations. The flux ratios of predictions to observations for the spectral sequence are shown in the lower panel, where the shades with lighter colors represent 95$\%$ confidence caused by the predictive uncertainty. In the case of SN~2002bo, the spectrum contributed by WHT at +5.2 days is a {\it homogenized} spectrum from the extended dataset and notice that the segment from 5700 to 6000 \AA\ of this WHT spectrum is not available.}
\end{figure*}

\begin{figure*}[ht!]
    \centering
    \gridline{\fig{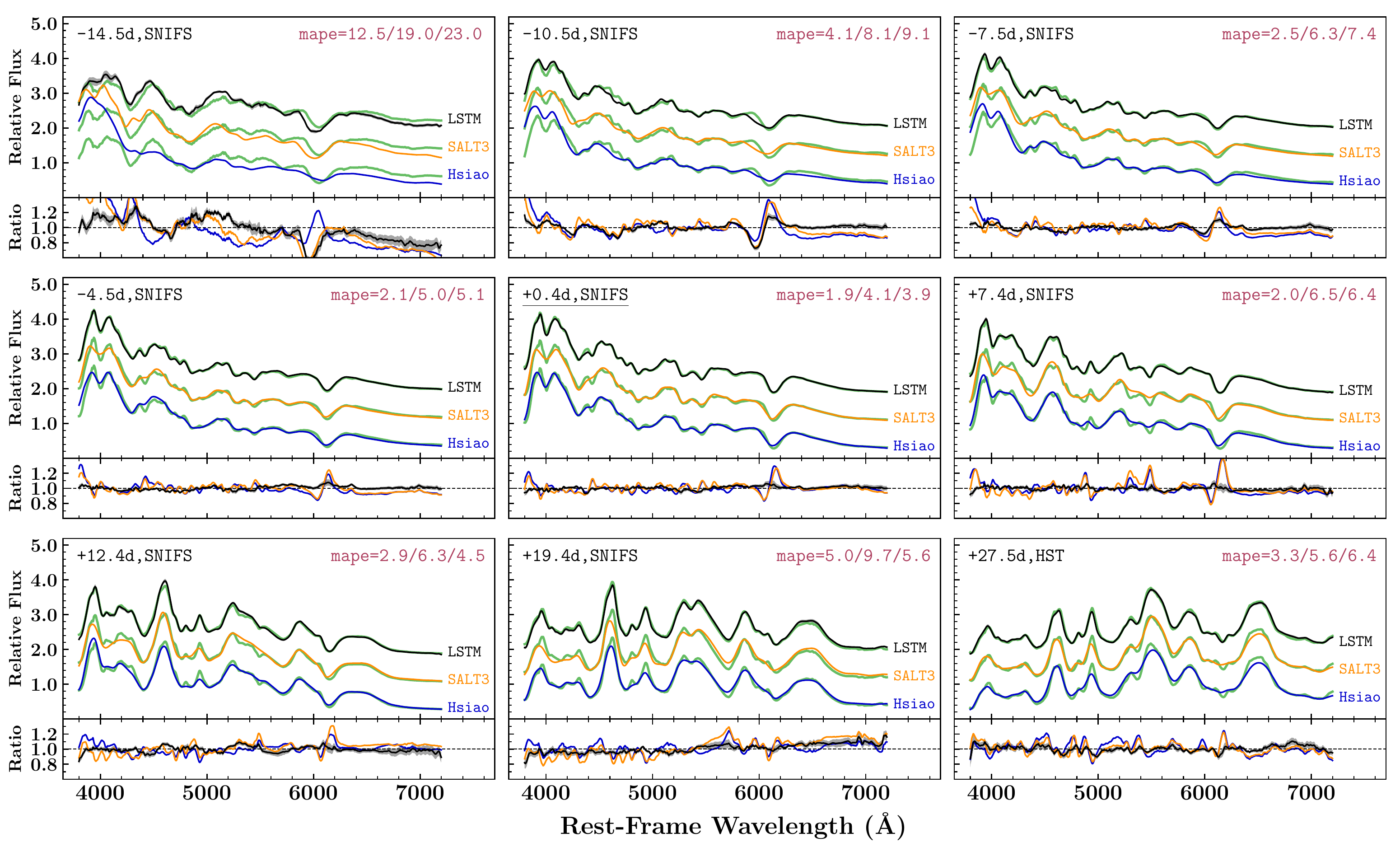}{0.83\textwidth}{(a) SN~2011fe (NV)}}
    \gridline{\fig{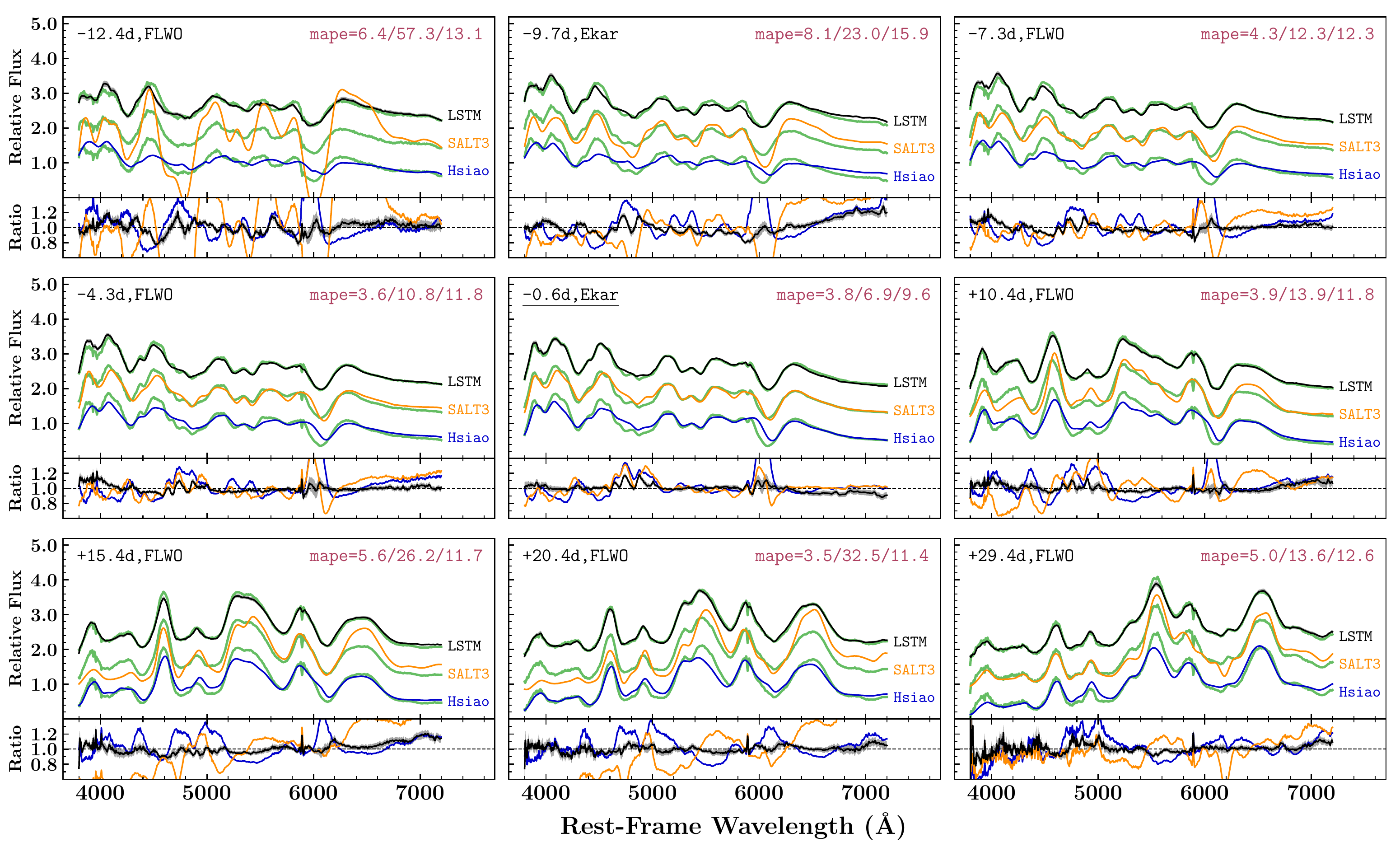}{0.83\textwidth}{(b) SN~2002bo (HV)}}
    \caption{\label{fig:OneSpec-BaselineValidate-NVHV-direct} Comparison of the predicted spectra generated by different methods (black curves for LSTM; yellow curves for SALT3; blue curves for the Hsiao template) with the observed spectra (green curves) for SN~2011fe (\textit{top}) and SN~2002bo (\textit{bottom}). For each SN, the predicted spectra from LSTM and SALT3 are obtained by fitting on the single spectrum at maximum light, while the Hsiao template is the one already constructed in Section~\ref{ssec:baseline4ctemp}. Each panel shows a comparison at a different epoch, and the panel format is the same as in Figure~\ref{fig:SpecTemp-BaselineValidate-Example}, but the SN name is replaced by the instrument source in the top left corner of each panel. The underlined text in the central panel of each SN highlights the epoch of the spectra employed in the fits.}
\end{figure*}

\begin{figure*}[ht!]
    \centering
    \gridline{
        \fig{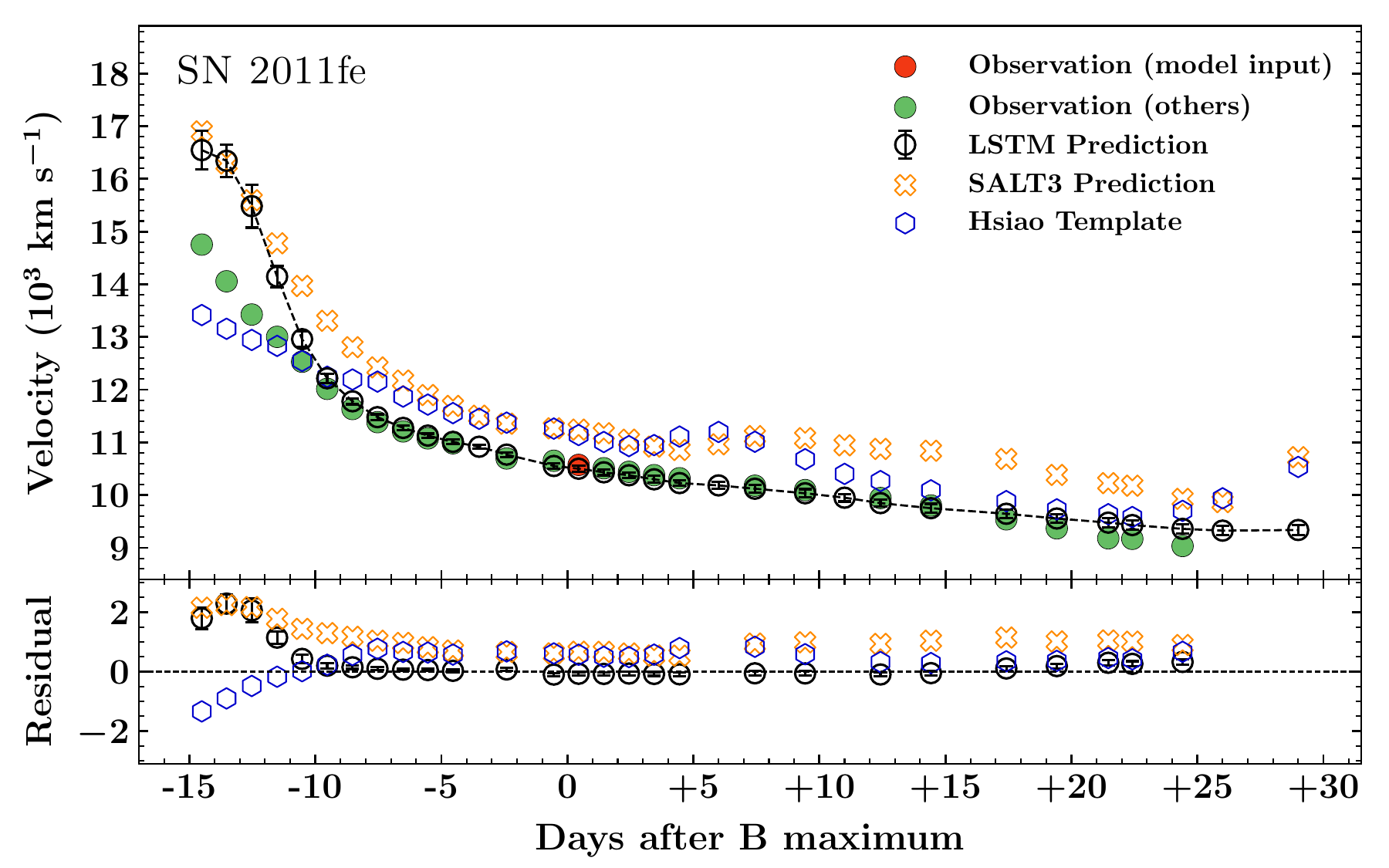}{0.48\textwidth}{(a) SN~2011fe (NV)}
        \fig{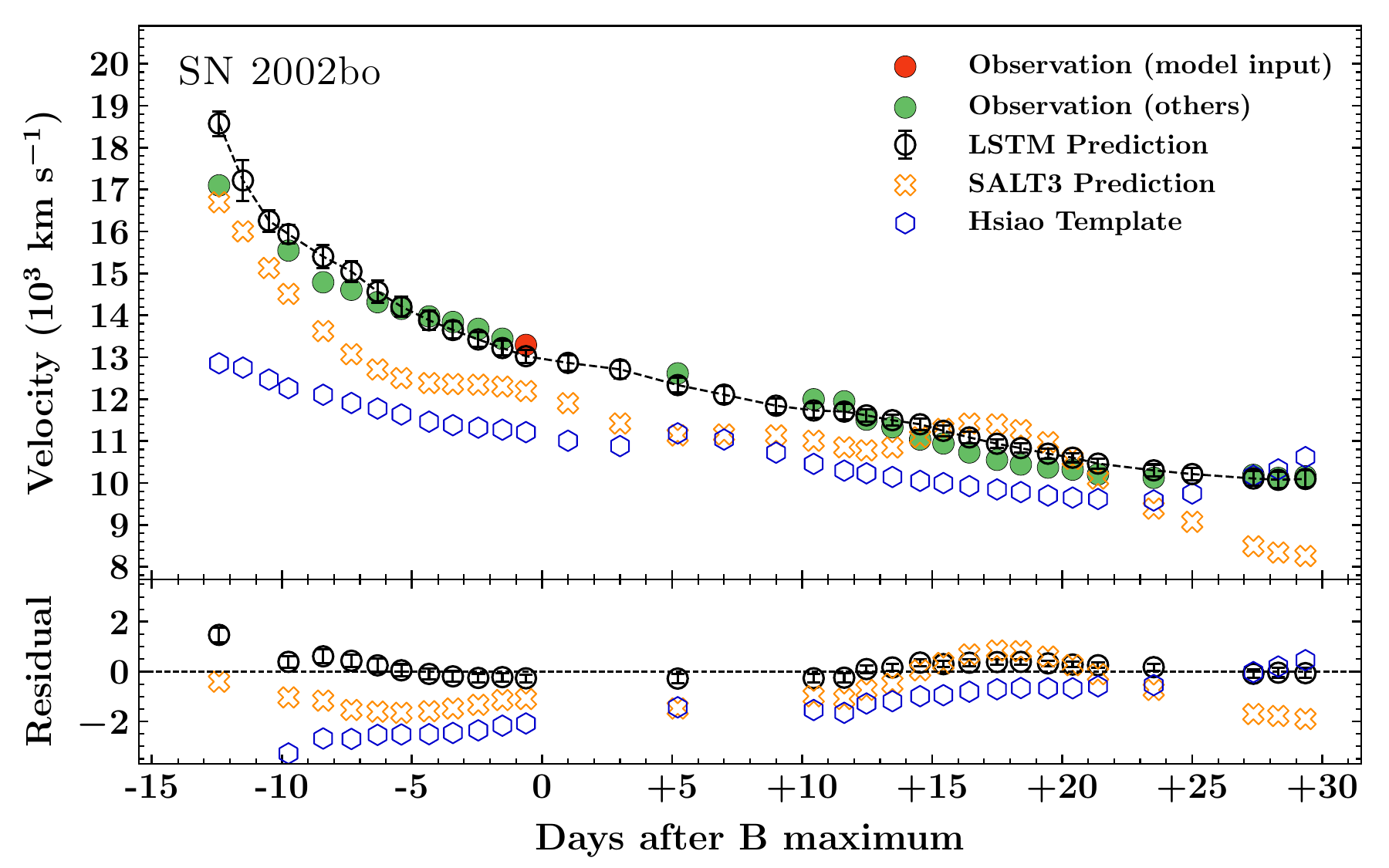}{0.48\textwidth}{(b) SN~2002bo (HV)}
        }
    \caption{\label{fig:OneSpec-Validate-11fe02bo-velocity} The Si~II $\lambda 6355$\AA\ velocity measured on the spectral sequence predicted from a single spectrum at maximum light using LSTM neural networks (black) and the corresponding observed spectra (green) for SN~2011fe (\textit{left}) and SN~2002bo (\textit{right}). The LSTM-predicted spectra and the observed spectra are the same as those shown in Figure~\ref{fig:OneSpec-Validate-NVHV-direct}, and the measurements from the input observed spectra at maximum light have been highlighted in red. Here the error bars are 1$\sigma$ uncertainties of the measurements over the LSTM-predicted spectra from different forward passes. For comparison, the velocities measured on the SALT3-predicted spectra (yellow) and Hsiao template spectra (blue) at the epochs of LSTM predictions are overplotted. As the broken WHT spectrum of SN~2002bo has well covered the silicon feature, we also present its Si $\lambda 6355$ velocity measurement in the figure.}
\end{figure*}

\begin{figure*}[ht!]
    \centering
    \gridline{
        \fig{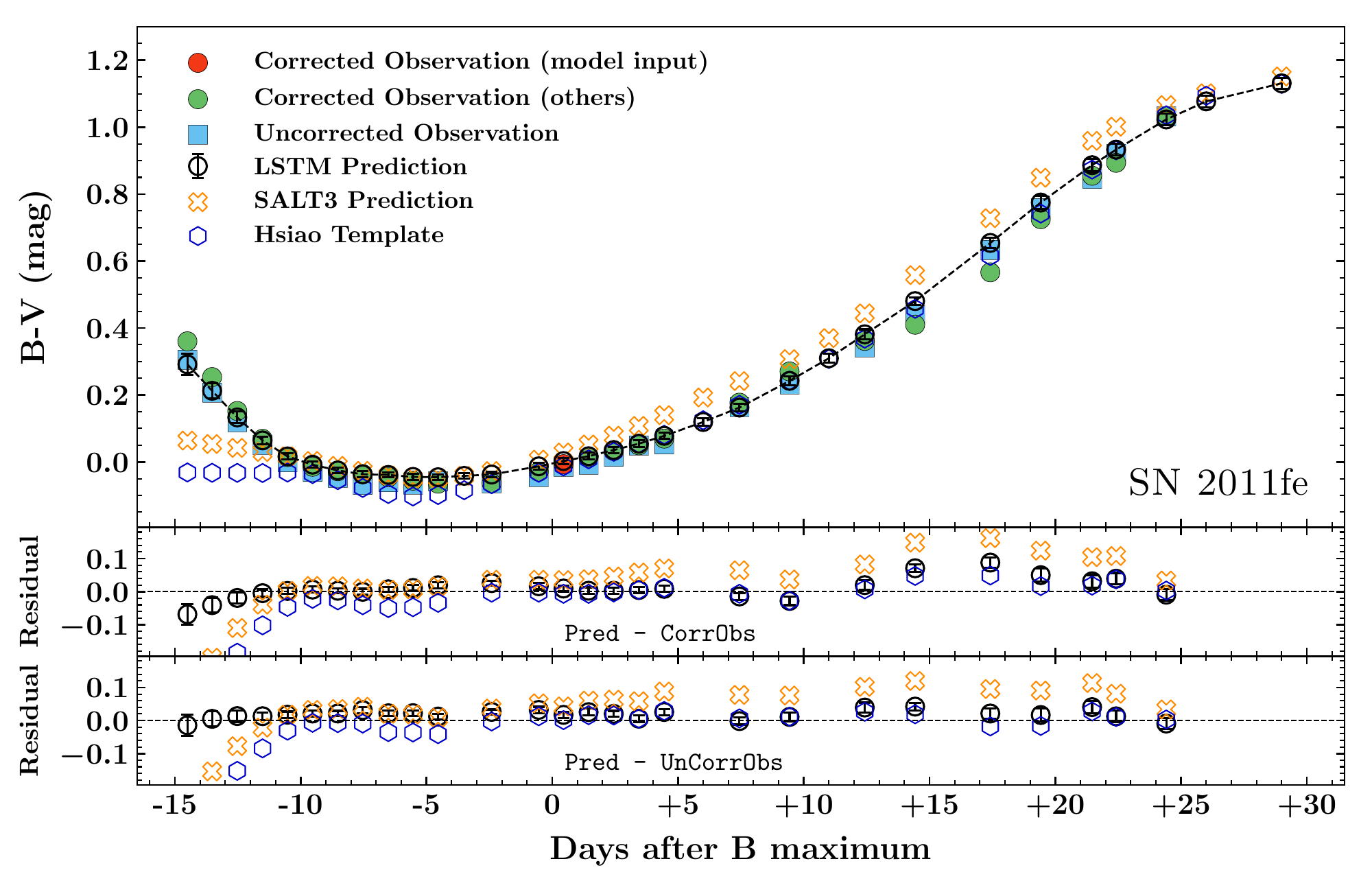}{0.48\textwidth}{(a) SN~2011fe (NV)}
        \fig{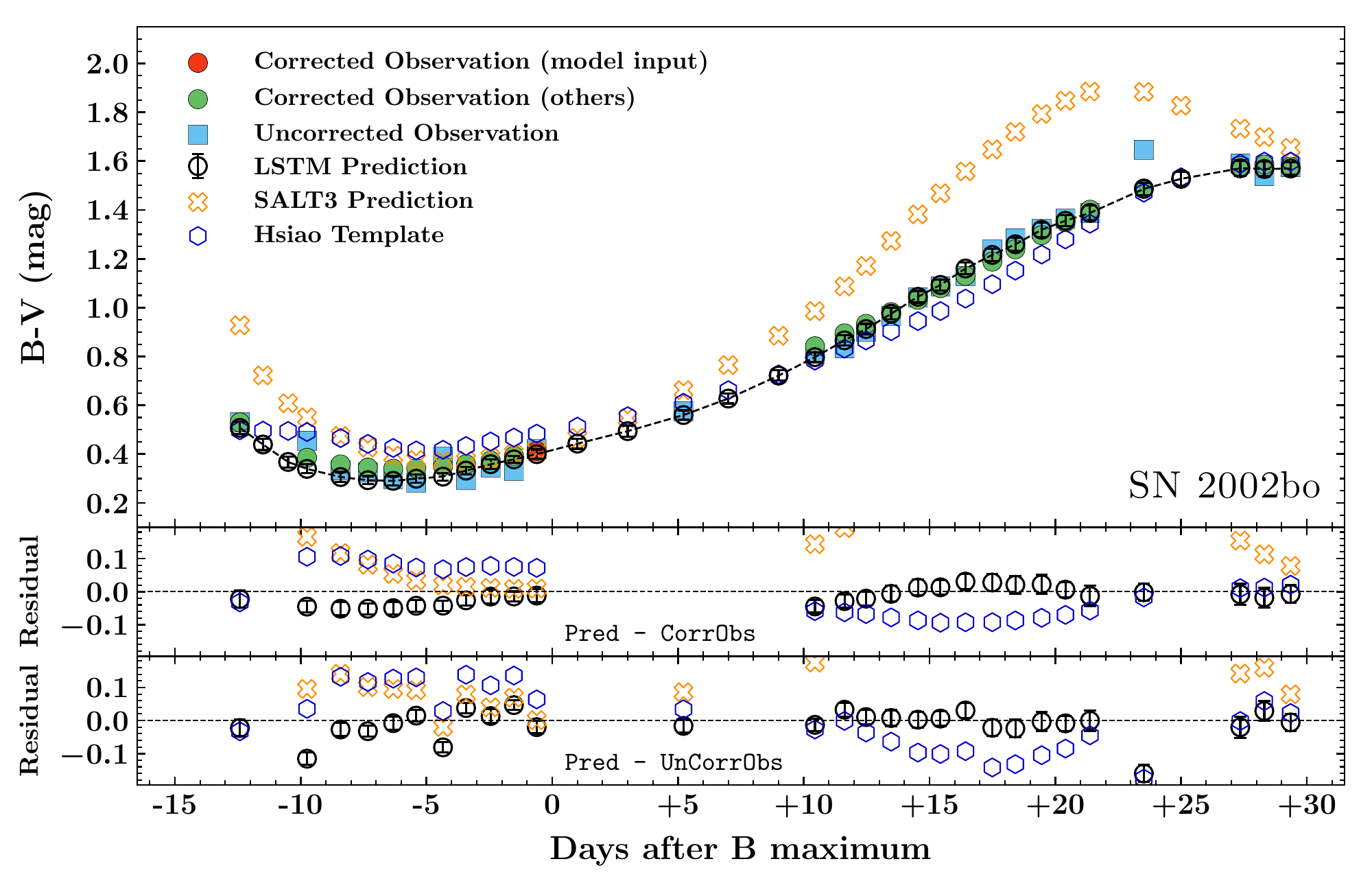}{0.48\textwidth}{(b) SN~2002bo (HV)}
        }
    \caption{\label{fig:OneSpec-Validate-11fe02bo-color} Synthetic $B-V$ color measured from the spectral sequence predicted from a single spectrum at maximum light using LSTM neural networks (black circles) and the corresponding observational data (green circles and blue squares) for (a) SN~2011fe and (b) SN~2002bo. 
    The LSTM-predicted spectra and the observed spectra are those already shown in Figure~\ref{fig:OneSpec-Validate-NVHV-direct}, and the measurements from the input observed spectra at maximum light have been highlighted in red. The error bars are 1$\sigma$ uncertainties of the measurements over the LSTM-predicted spectra from different forward passes. 
    To analyze the effect of photometric color calibration of the spectra, the measurements on both the {\it homogenized} (blue squares, labeled as uncorrected) and {\it corrected} (green circles) spectra are shown in the figure. 
    For comparison, the synthetic $B-V$ colors measured on the SALT3-predicted spectra (yellow crosses) and Hsiao template spectra (blue hexagons) at the epochs of LSTM predictions are also plotted. Note that the spectral $B-V$ color is not measured on the broken WHT spectrum of SN~2002bo. 
    Moreover, the last two {\it corrected} spectra of SN~2011fe and the first {\it corrected} spectrum of SN~2002bo did not go through any color calibration due to the deficiency of photometric coverage. Hence, the measurements from the {\it homogenized} and {\it corrected} spectra are consistent for them.}
\end{figure*}

\section{LSTM applied to the Analyses of SN~Ia Spectra} \label{sec:lstm-12spec}

Given the deficiency of spectroscopic resources, it is challenging for future transient surveys to acquire multiepoch spectroscopy. It is interesting to see how well a neural network-based algorithm can predict the spectral sequence of an SN based only on one or two spectra. Such a prediction is also a direct assessment of the critical information contained in any individual spectrum that can be employed to derive the intrinsic properties of an SN Ia. 

A more detailed quantitative study of SN properties based on the neural networks constructed here will be presented in an upcoming paper. Here we only show examples of the neural network predictions to demonstrate its potential in SN Ia spectral studies.

\begin{algorithm} \label{algo:makeseq-12spec}
    \SetAlgoLined
    \SetKwInOut{Input}{input}
    \SetKwInOut{Output}{output}
    \Input{The trained LSTM model}
    \Input{A spectral pair ($\textbf{x}_{l}$, $\textbf{x}_{m}$) with $p_{\textbf{x}_{l}} \leq p_{\textbf{x}_{m}}$ ($\textbf{x}_{l} = \textbf{x}_{m}$ is allowed)}
    \For{$k \in [1,2, ..., 64]$}{
            feed $(\textbf{x}_{l}, \textbf{x}_{m})$ to LSTM model to predict spectral time sequence $y_{k}(\lambda, p)$, where $p \in [-15, -14.875, ..., +29.875, +33]$
        }
    compute predictive mean: $\mu(\lambda, p) = \texttt{MEAN} (y_{1}, y_{2}, ..., y_{64})$\; 
    compute predictive uncertainty: $\sigma(\lambda, p) = \texttt{STD} (y_{1}, y_{2}, ..., y_{64})$
    \caption{Spectral Sequence Construction from One or Two Spectra}
\end{algorithm}

\subsection{The SNe with Normal and High Velocity} \label{ssec:onesepc-NVHV}

Two representative examples of NV and HV SNe~Ia are SN~2011fe \citep{2015ApJS..220...20Z} and SN~2002bo \citep{2012AJ....143..126B}.
Both SNe are well observed with extensive multi-phase spectroscopic coverage. In this section, we introduce a fictitious scenario: an supernova similar to SN~2011fe or SN~2002bo is newly discovered with only one available spectrum around its maximum light. We are interested in inferring the entire spectral sequence of them using the neural network trained on the data of all of the SNe~Ia excluding these well-observed SNe.

Two separate LSTM models were constructed specifically for these two SNe. The training set of these two different models are different in that for each SN the training set contains all of the spectra in the CSD except the SN being modeled. The training samples for both cases are also generated by exhausting all possible permutations as in Section~\ref{ssec:train4temp}.

The spectral sequence is constructed with the procedures described in Section~\ref{ssec:constemp} but following Algorithm~\ref{algo:makeseq-12spec}. This process is a simplified version of Algorithm~\ref{algo:ctemp}.
In Algorithm~\ref{algo:ctemp}, there is more than one element in the set $I$ (see line 19 in Algorithm~\ref{algo:ctemp}). However, here we only have one or two spectra; thus, the input spectral pair is uniquely determined, and the combination process (see line 26 and 27 in Algorithm~\ref{algo:ctemp}) is no longer necessary.
Unlike Section~\ref{ssec:constemp}, the process bears a much lower computational cost, so we adjust the repeating times from 24 to 64 (see line~20 in Algorithm~\ref{algo:ctemp} and line~1 in Algorithm~\ref{algo:makeseq-12spec}).

Figure~\ref{fig:OneSpec-Schematic-11fe02bo} shows the resulting mean spectral sequences of SN~2011fe and SN~2002bo derived from the neural networks using a single spectrum at maximum light following Algorithm~\ref{algo:makeseq-12spec}.
The Si~II $\lambda$6355\AA\ at the maximum of SN~2002bo (HV object) is obviously broader and stronger than that of SN~2011fe (NV object).
Figure~\ref{fig:OneSpec-Uncertainty-11fe02bo} presents the comparisons of the predictive mean sequences shown in Figure~\ref{fig:OneSpec-Schematic-11fe02bo} with their corresponding  uncertainties.
The uncertainties seem to have footprints broadly consistent with the mean templates, making them reminiscent of the observational noise dominated by photon shot noise. The uncertainties at the earliest phases are relatively large due to the scarcity of spectroscopic data of young SNe~Ia in the training dataset. Overall, the uncertainties are slightly lower for SN~2011fe than for SN~2002bo.

Figure~\ref{fig:OneSpec-Validate-NVHV-direct} shows the spectral comparisons for SN~2011fe and SN~2002bo. The predictions of NV object SN~2011fe at phases $>$ -8 days are in excellent agreement with observations. For earlier phases, an obvious discrepancy across the Si~II $\lambda$6355 feature emerges as the model shows a broader absorption. Meanwhile, the model performs poorly at the blue end with $\lambda < 5200$\AA, where the spectra are dominated by absorption features of Si, Fe, and Mg. Nevertheless, the prominent S~II lines at around $4800$\AA\ seem to be well predicted. For HV object SN~2002bo, the model can properly predict the board and strong Si~II $\lambda$6355 \AA\ line and the prominent S~II lines. However, the performance is less satisfactory between $4500$ and $5000 $\AA.

%% ***** Section: Spec Prediction [Peculiar]
\begin{figure*}[ht!]
    \centering
    \gridline{
        \fig{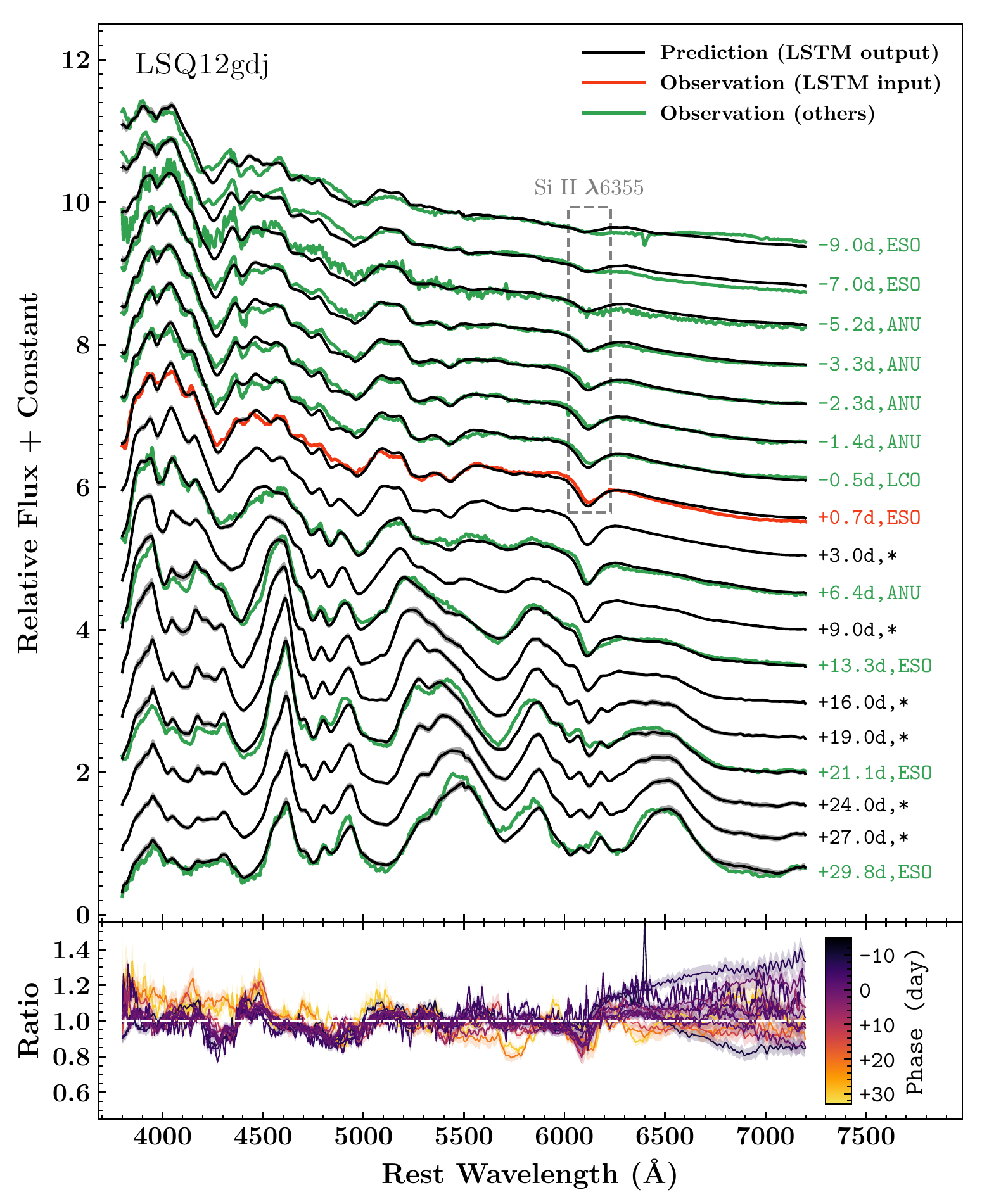}{0.43\textwidth}{(a) LSQ~12gdj (Ia-91T)}
        \fig{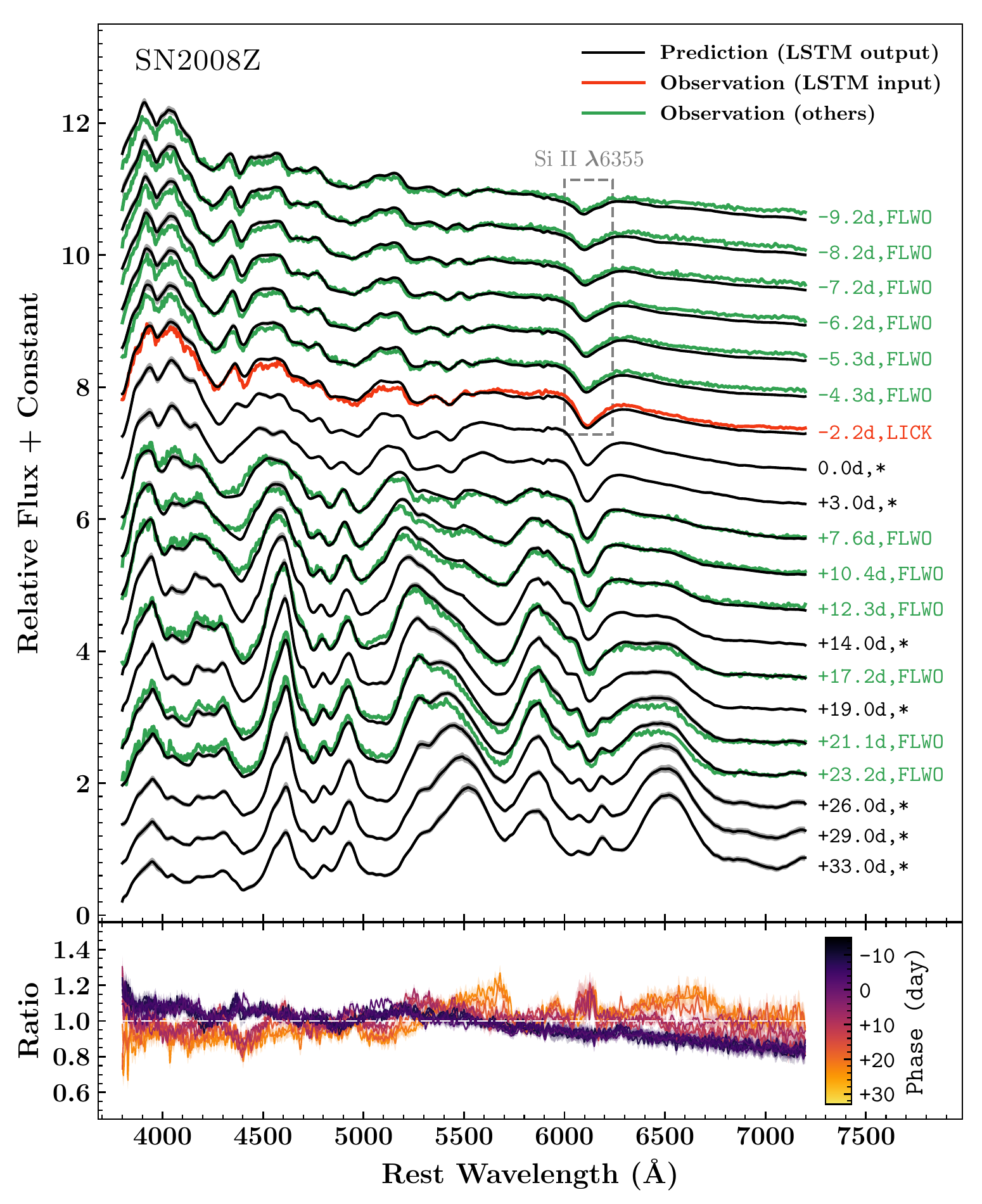}{0.43\textwidth}{(b) SN~2008Z (Ia-99aa)}
        }
    \gridline{
        \fig{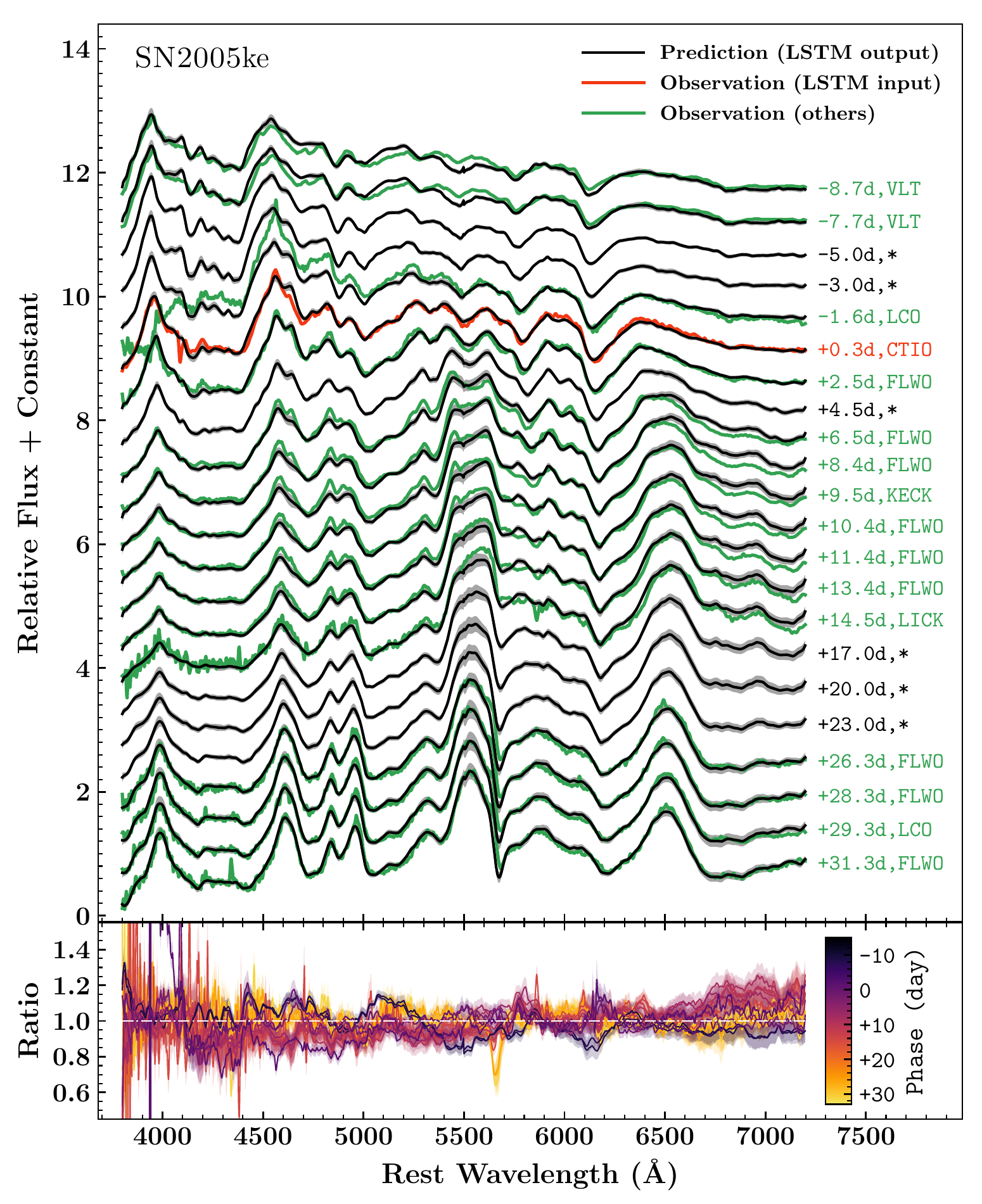}{0.43\textwidth}{(c) SN~2005ke (Ia-91bg)}
        \fig{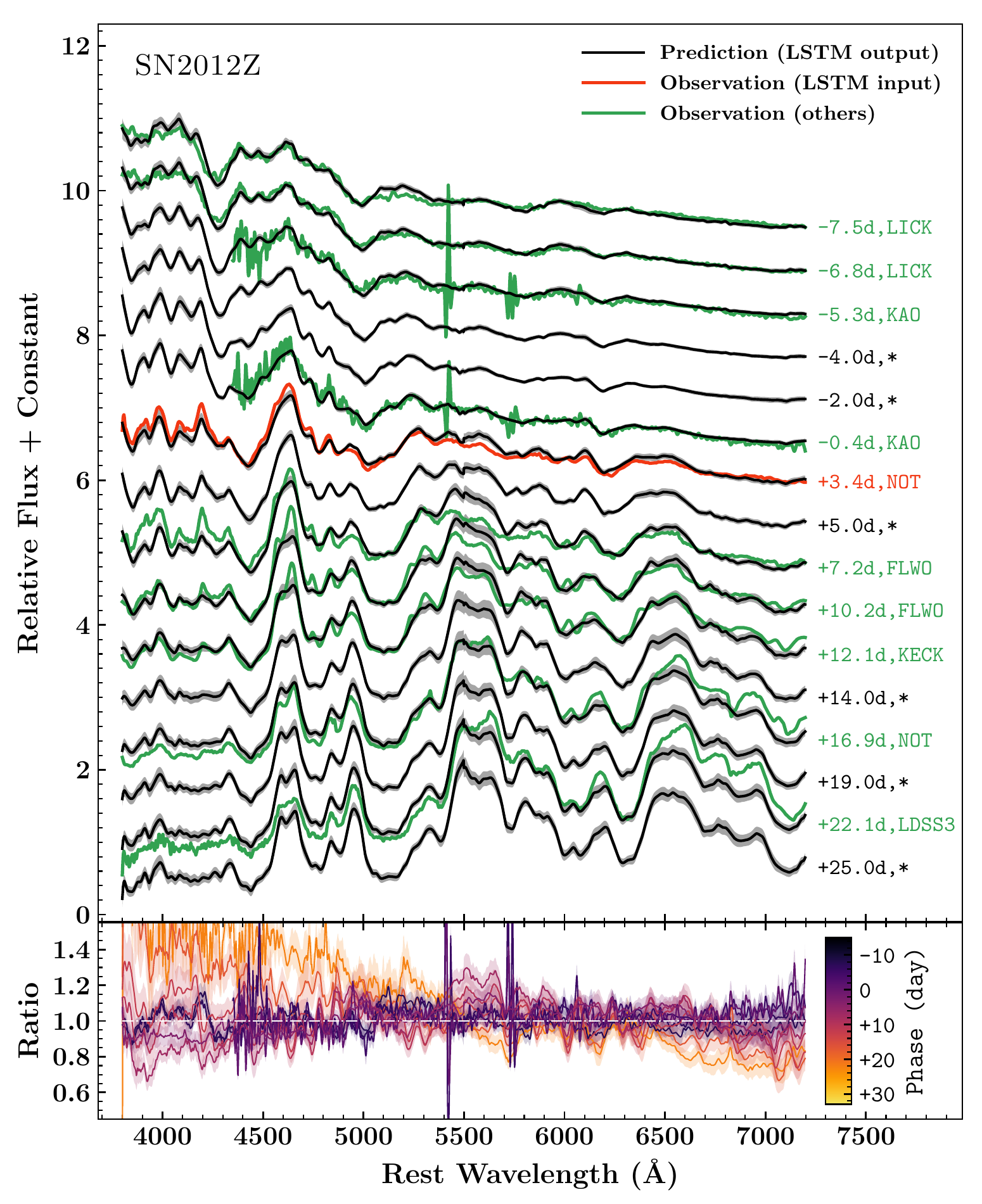}{0.43\textwidth}{(d) SN~2012Z (Iax)}
        }
    \caption{\label{fig:OneSpec-Validate-pec-direct} Same as in Figure~\ref{fig:OneSpec-Validate-NVHV-direct}, but for the spectral sequence predicted from one spectrum at maximum light using LSTM neural networks for (a) Ia-91T object LSQ~12gdj (b) Ia-99aa object SN~2008Z (c) Ia-91bg object SN2005ke, and (d) Iax object SN~2012Z. The absorption features at Si~II $\lambda 6355$\AA\ are highlighted by a gray dashed box for Ia-91T object LSQ~12gdj and Ia-99aa object SN~2008Z.}
\end{figure*}

\begin{figure*}[ht!]
    \centering
    \gridline{\fig{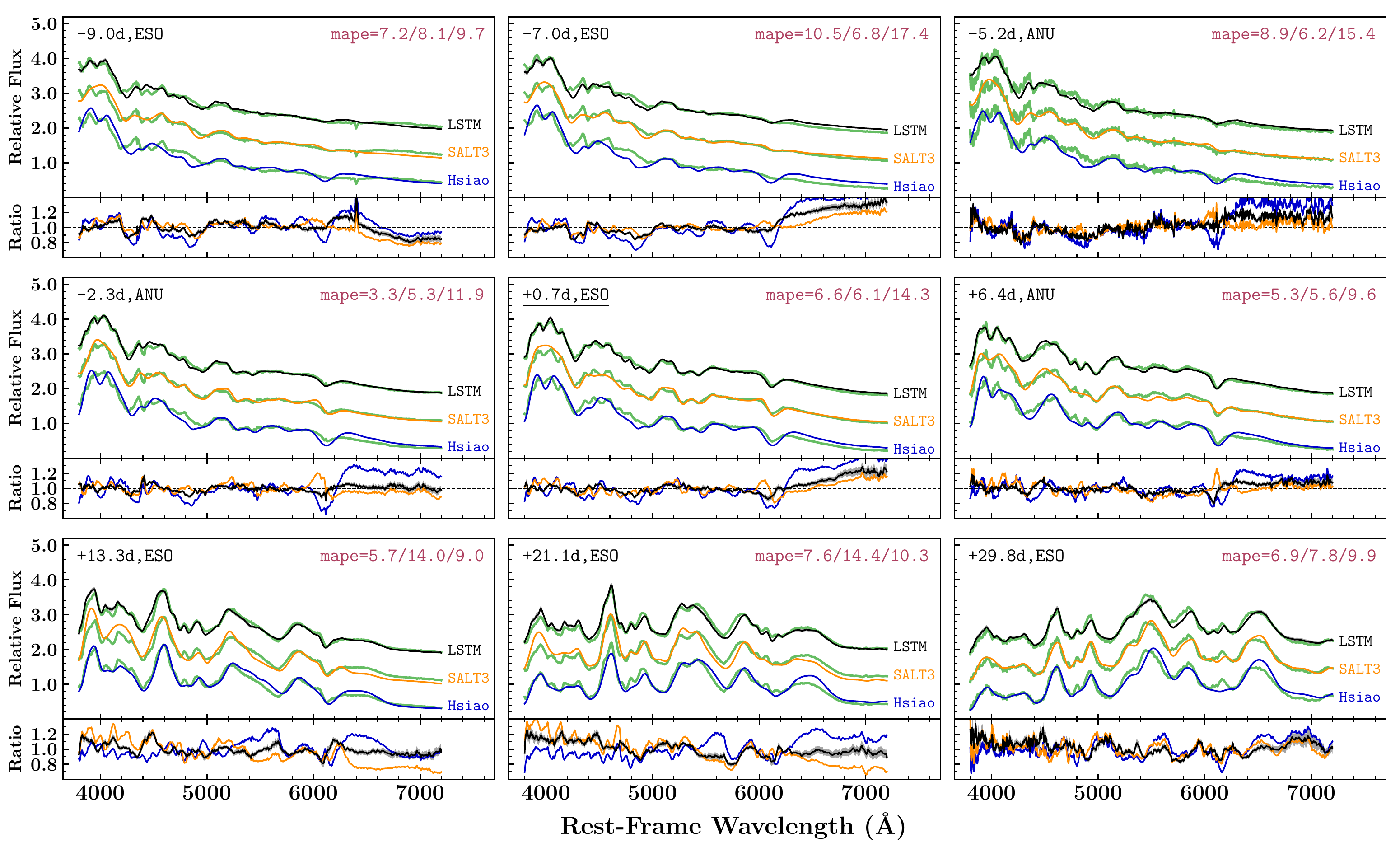}{0.86\textwidth}{(a) LSQ12gdj (Ia-91T)}}
    \gridline{\fig{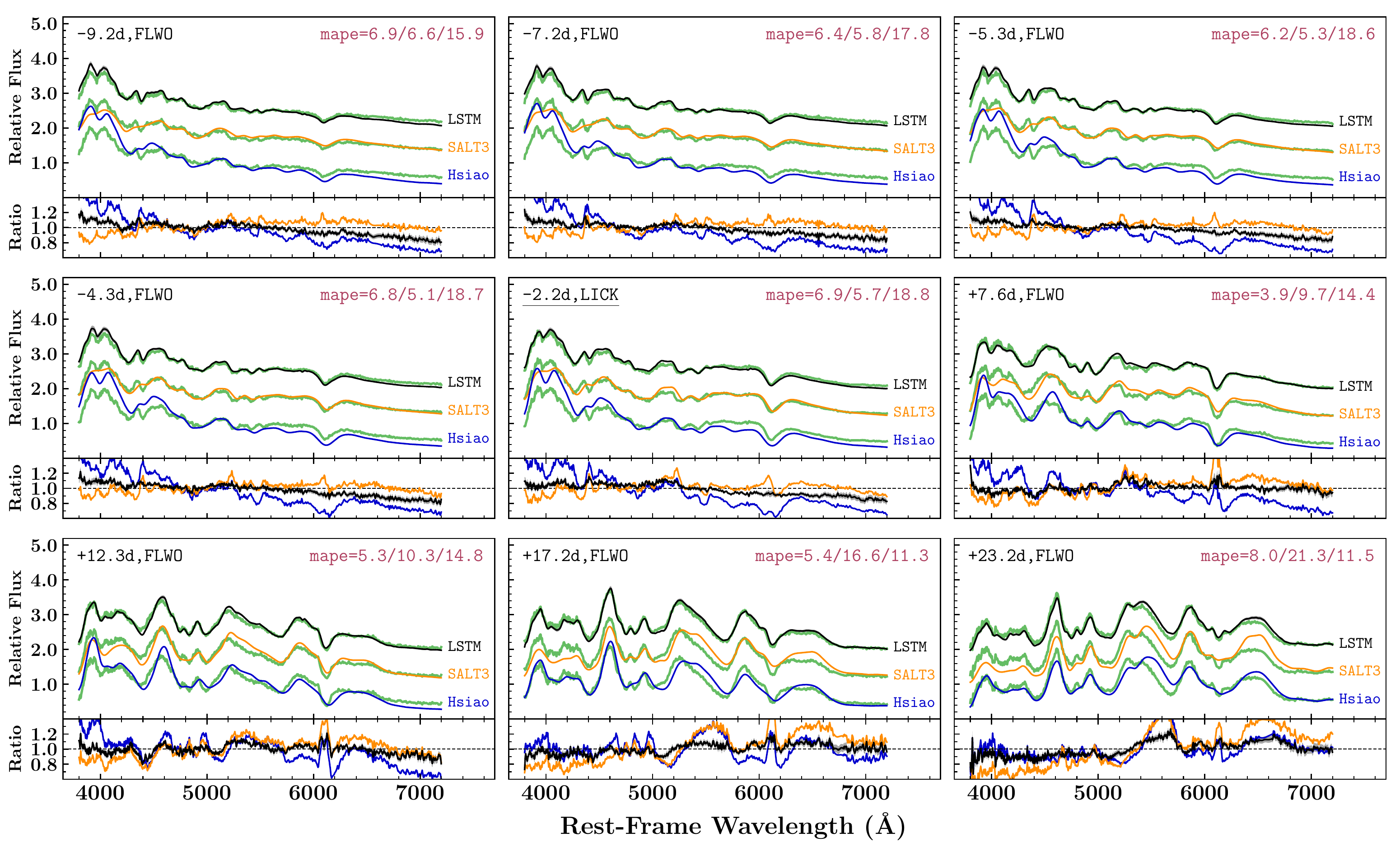}{0.86\textwidth}{(b) SN~2008Z (Ia-99aa)}}
    \caption{\label{fig:OneSpec-BaselineValidate-91T99aa-direct} Comparison of the predicted spectra generated by different methods (black curves for LSTM; yellow curves for SALT3; blue curves for the Hsiao template) with the observed spectra (green curves) for (a) Ia-91T object LSQ~12gdj and (b) Ia-99aa object SN~2008Z. The predicted spectra from LSTM (black) and SALT3 (yellow) were obtained by fitting on the single spectrum at maximum light. The Hsiao template (blue) for LSQ~12gdj is constructed as described in Section~\ref{ssec:baseline4ctemp}. The panel format is the same as in Figure~\ref{fig:OneSpec-BaselineValidate-NVHV-direct}.}
\end{figure*}

\begin{figure*}[ht!]
    \centering
    \gridline{\fig{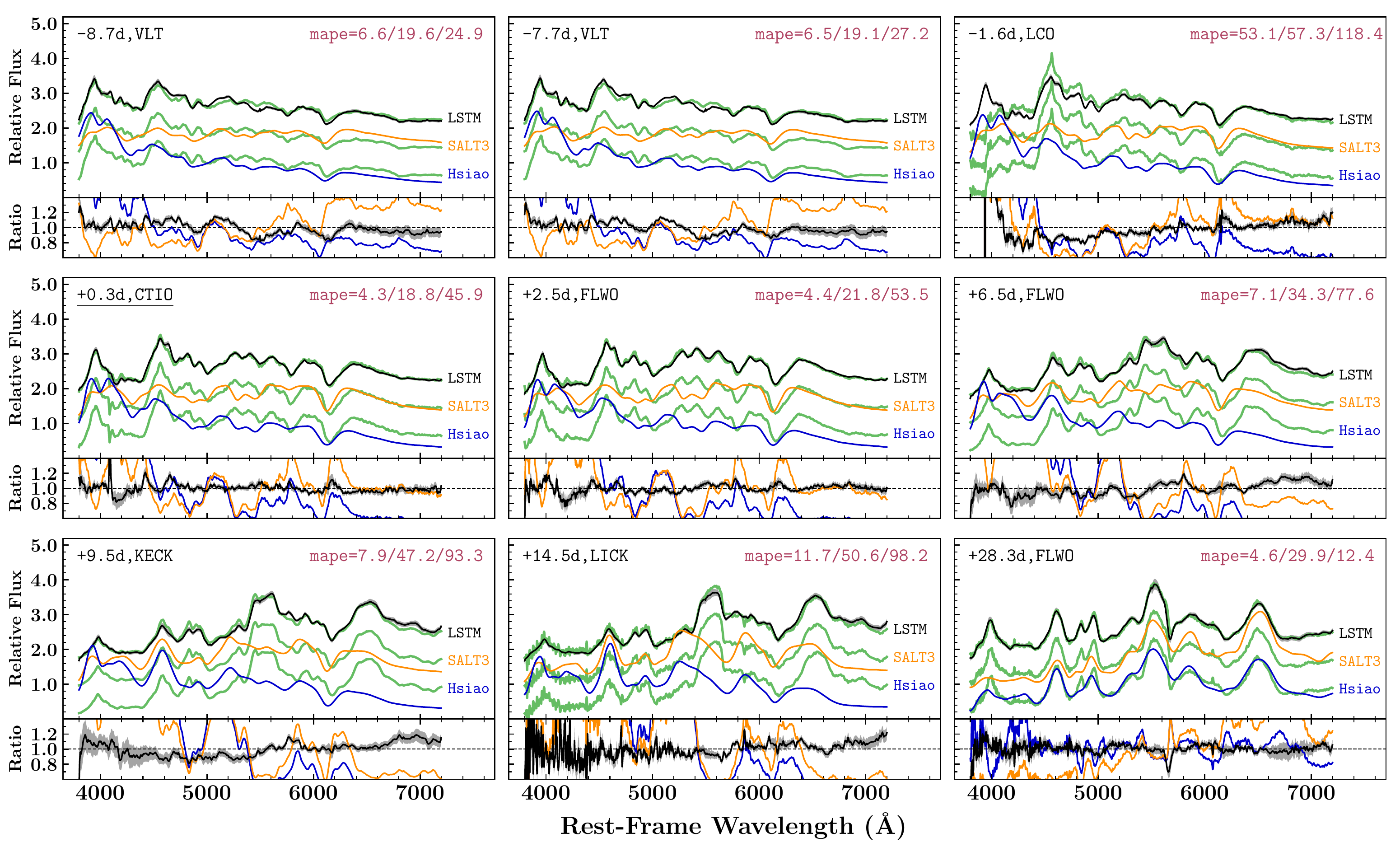}{0.86\textwidth}{(a) SN~2005ke (Ia-91bg)}}
    \gridline{\fig{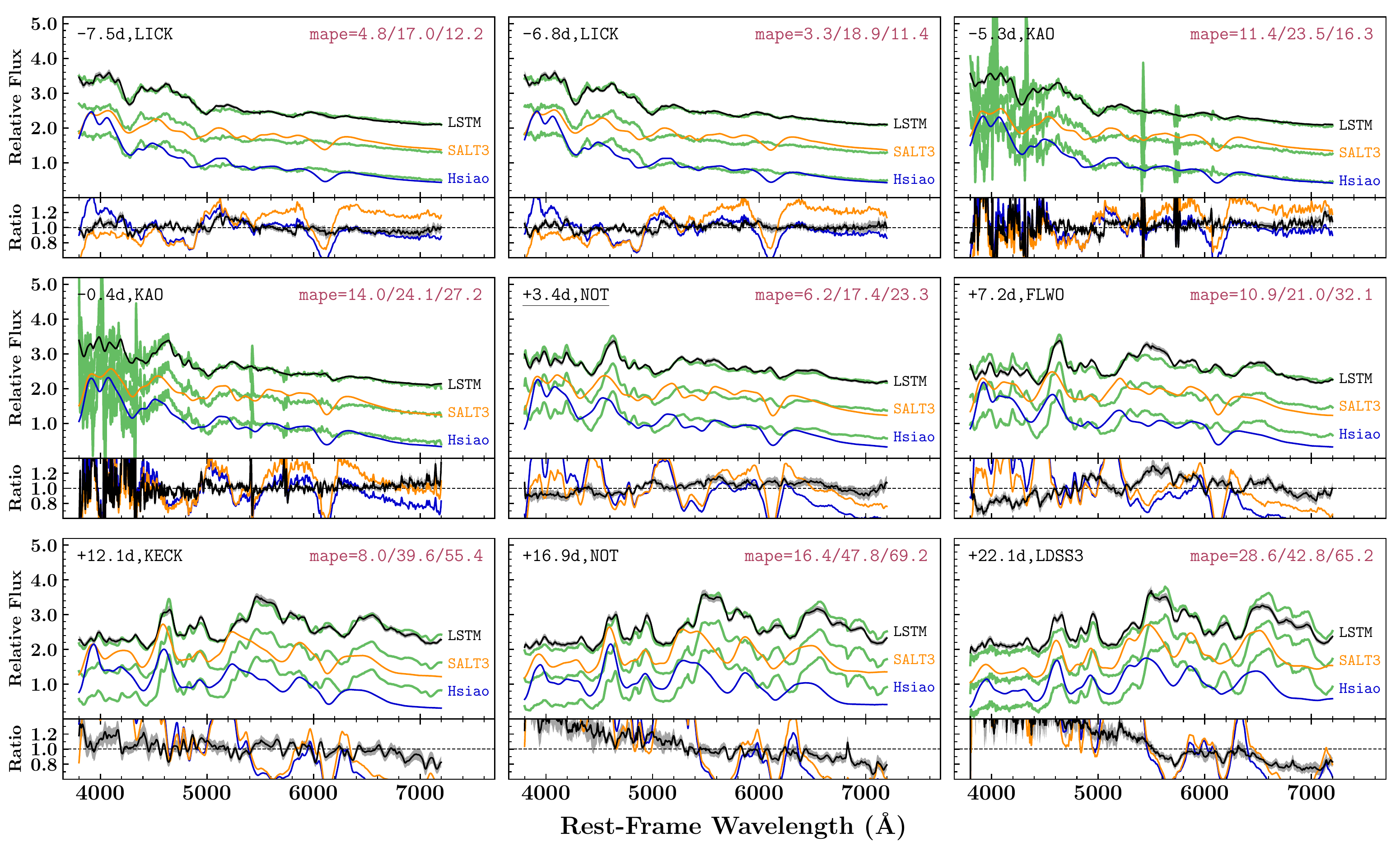}{0.86\textwidth}{(b) SN~2012Z (Iax)}}
    \caption{\label{fig:OneSpec-BaselineValidate-91bgIax-direct} Same as Figure~\ref{fig:OneSpec-BaselineValidate-91T99aa-direct}, but for (a) Ia-91bg object SN~2005ke and (b) Iax object SN~2012Z.}
\end{figure*}

We also generated the spectral models for SN~2011fe and SN~2002bo using the \cite{Hsiao2007Kcorrection} and SALT3 models for comparison. The Hsiao templates of both SNe are generated as in Section~\ref{ssec:baseline4ctemp}. 
For the SALT3 model, we followed the same procedures as described in Section~\ref{ssec:baseline4ctemp} to make the spectral predictions, except that the spectral data are fitted only for the spectrum at the maximum light. These models yield less accurate predictions for both SNe, as shown in Figure~\ref{fig:OneSpec-BaselineValidate-NVHV-direct}. 
Although the same prior knowledge (the single spectrum at maximum light) is used in the fits for the LSTM neural networks and the SALT3 model, the LSTM neural networks show significantly better performance.

The diversity among normal SNe~Ia can be examined by their photospheric velocities. The expansion of the photosphere may evolve quite differently for different SNe. It is interesting to investigate how the neural networks can learn and capture the spectral evolution of different spectral types of SNe~Ia. 
Figure~\ref{fig:OneSpec-Validate-11fe02bo-velocity} shows the velocity of the absorption dip of Si~II $\lambda$6355 \AA\ for SN~2011fe and SN~2002bo. The errors of the neural network predictions are generally less than $200$ km $\text{s}^{-1}$ from day -10 to 30 for SN~2011fe. The errors for SN~2002bo are slightly larger but typically less than $500$ km $\text{s}^{-1}$ from day -10 to about day 30. 
For both SNe, the overall trend of line velocity evolution from about a week before maximum to 4 weeks past maximum is well captured by the neural networks using only a single spectrum around optical maximum as input.
In contrast, the line velocity evolutions measured on the predicted spectra from the other two models have much stronger bias, as shown in Figure~\ref{fig:OneSpec-Validate-11fe02bo-velocity}. The Hsiao template is insensitive to the spectral diversity of SNe~Ia. 
Though SALT3 allows for higher flexibility to model spectral features than the Hsiao template, it still fails to capture the evolution of spectral velocities of any individual SN~Ia.

Accurately predicting the underlying continuum component of the spectra is also a goal of the neural networks we have constructed. Figure~\ref{fig:OneSpec-Validate-11fe02bo-color} shows the comparisons of the integrated spectral $B-V$ colors measured on predictions and observations for SN~2011fe and SN~2002bo. 
The $B$ and $V$ magnitudes were directly integrated over the spectra with standard (Bessell) $B$ and $V$ transmission curves in the rest frame. The missing data are padded by zero before calculating the $B$-band magnitudes.
Recall that no extinction corrections were applied on the spectra of the training set. The $B-V$ colors are not the intrinsic color but a metric to quantify the colors of the predicted spectral sequence. 
The effect of photometric color calibration during preprocessing (see Section~\ref{ssec:colorcal}) is also assessed. The spectral $B-V$ colors of the {\it homogenized} spectra are also shown in Figure~\ref{fig:OneSpec-Validate-11fe02bo-color}.

In general, we find that the $B-V$ color residuals for SN~2011fe and SN~2002bo are typically less than 0.03 mag in both cases, and the overall trend of the color evolution is very well reproduced by LSTM neural networks.
Yet both the \cite{Hsiao2007Kcorrection} and SALT3 models show less satisfactory performance on predicting the color evolution, especially for the HV object SN~2002bo.
The color evolution of the Hsiao template is determined by the uniform average trend of its own training set. It does not accommodate any intrinsic color diversity.
Its performance for modeling the color evolution appears to be better than that of the SALT3. This indicates that using a single spectrum around maximum, the SALT3 model, although it offers more flexibility, is not well constrained to derive a full spectral sequence of SNe~Ia, especially the HV SNe~Ia.

One may notice in Figure~\ref{fig:OneSpec-Validate-11fe02bo-color} that the predicted $B-V$ colors from LSTM neural networks are slightly more consistent with those of the {\it homogenized} spectra which are not calibrated to the observed photometric colors, than with the {\it corrected} spectra. 
This could be an indication that a small portion of original spectral data, e.g., those from SNIFS, may have already achieved excellent flux calibration, and the color calibration we adopted is based on photometric observations from heterogeneous sources, which may, in fact, be less accurate. The oversimplified color calibration process is essential in constructing a uniform dataset but may also inherit the errors of the input photometric data.
The SN~2002bo data shown in Figure~\ref{fig:OneSpec-Validate-11fe02bo-color} illustrate a more common situation: original spectra can exhibit wrong colors and appear as outliers on the $B-V$ evolution curve.
Overall, Figure~\ref{fig:OneSpec-Validate-11fe02bo-color} demonstrates that the color evolution of an SN~Ia can be reliably predicted using only one spectrum taken around optical maximum using LSTM neural networks.

The success in predicting the color evolution of an SN Ia based on spectral data of one or two epochs is a remarkable achievement that may lead to new observational strategies for future SN cosmology.

%% ***** Section: Spec Prediction [Two-Spec Input]

\begin{figure*}[ht!]
    \centering
    \gridline{
        \fig{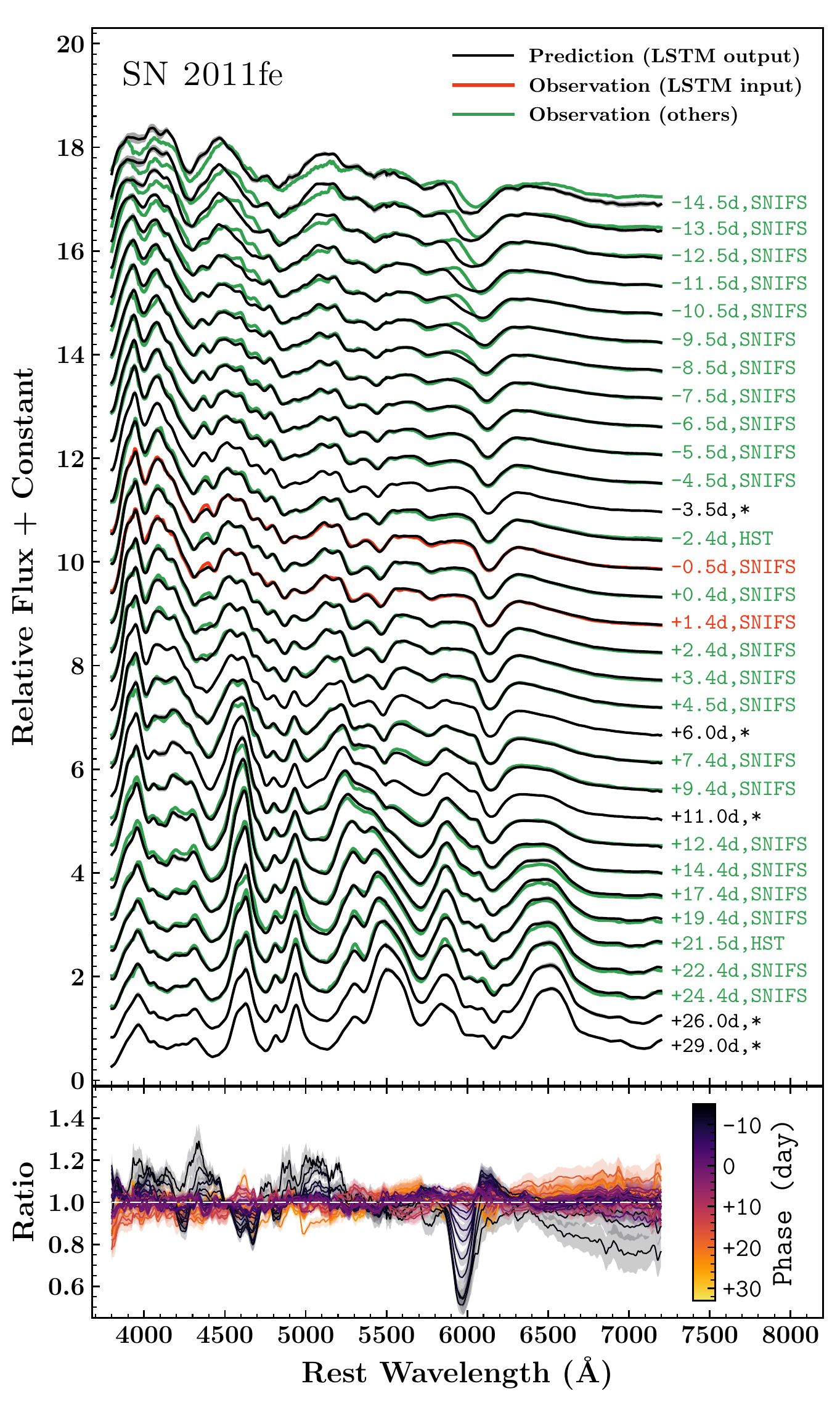}{0.235\textwidth}{(a) SN~2011fe, $\Delta{P}=2\text{d}$}
        \fig{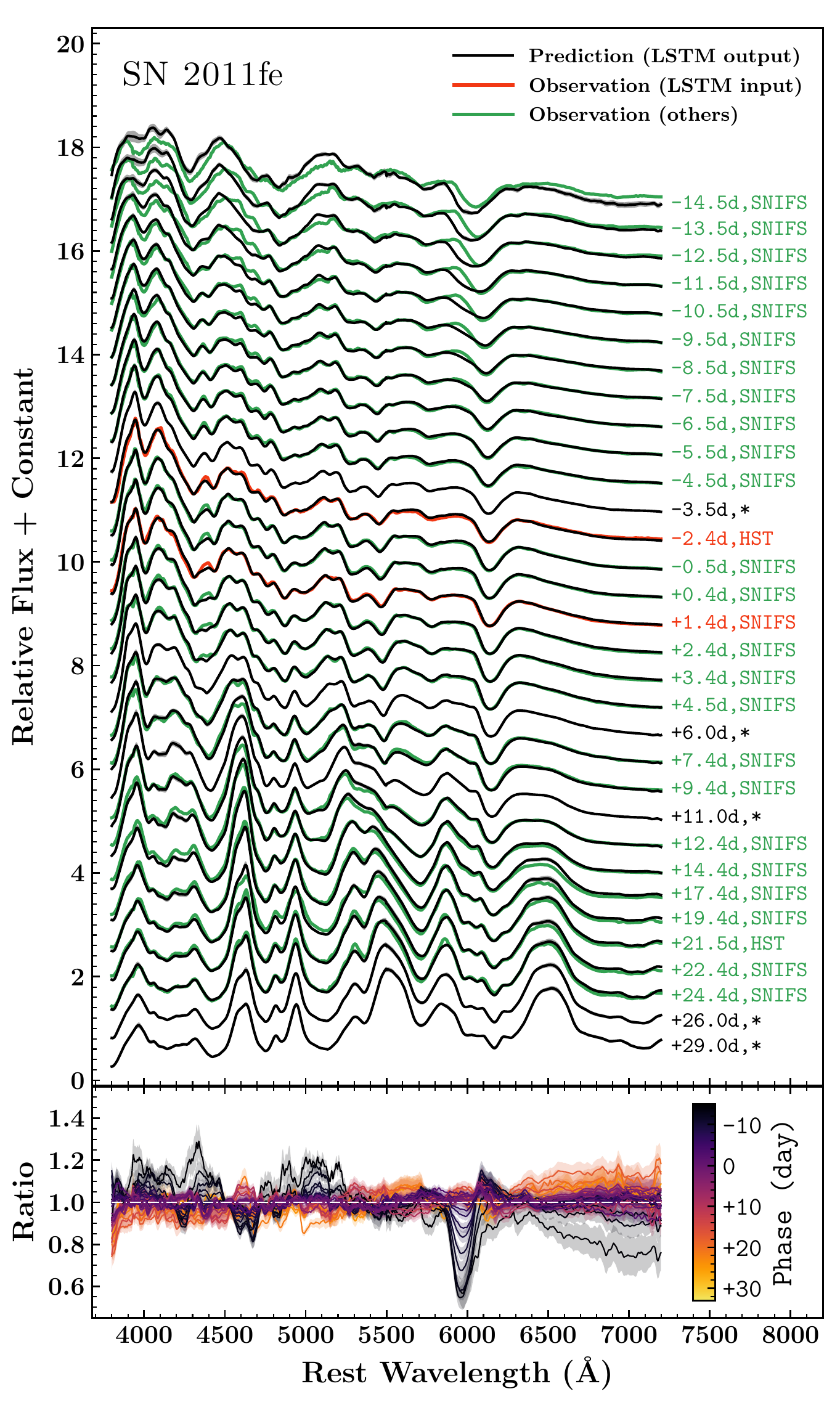}{0.235\textwidth}{(b) SN~2011fe, $\Delta{P}=4\text{d}$}
        \fig{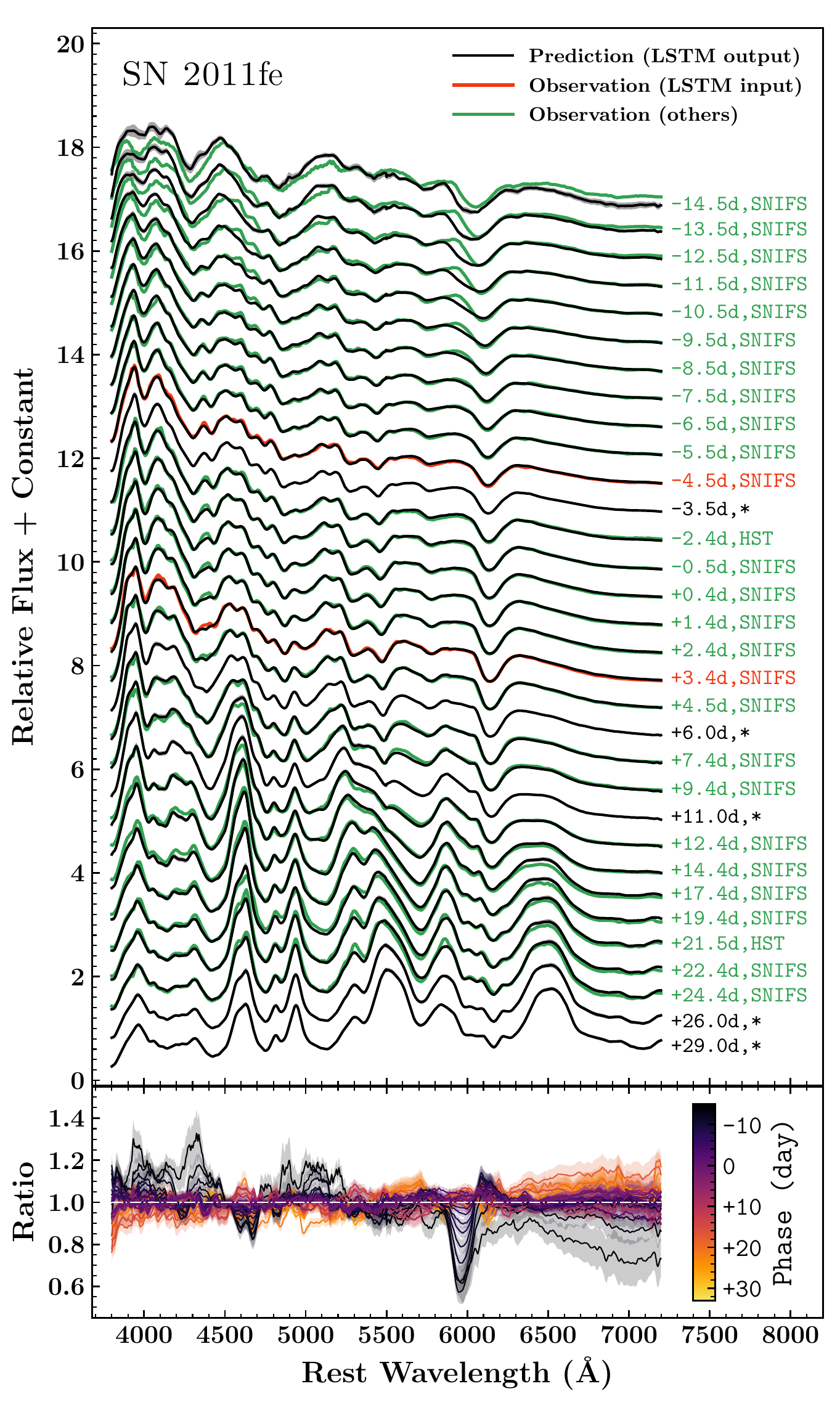}{0.235\textwidth}{(c) SN~2011fe, $\Delta{P}=8\text{d}$}
        \fig{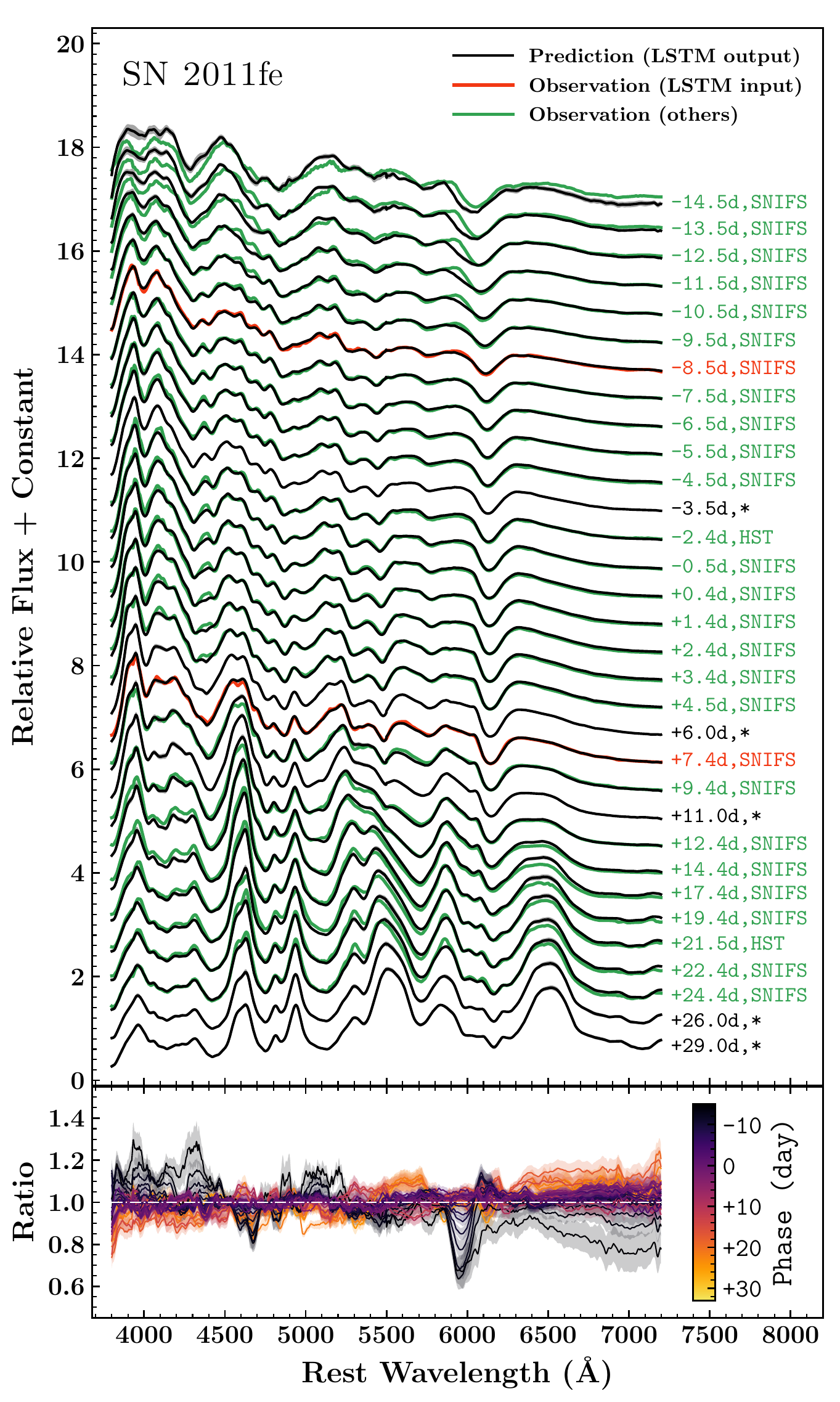}{0.235\textwidth}{(d) SN~2011fe, $\Delta{P}=16\text{d}$}
        }
    \gridline{
        \fig{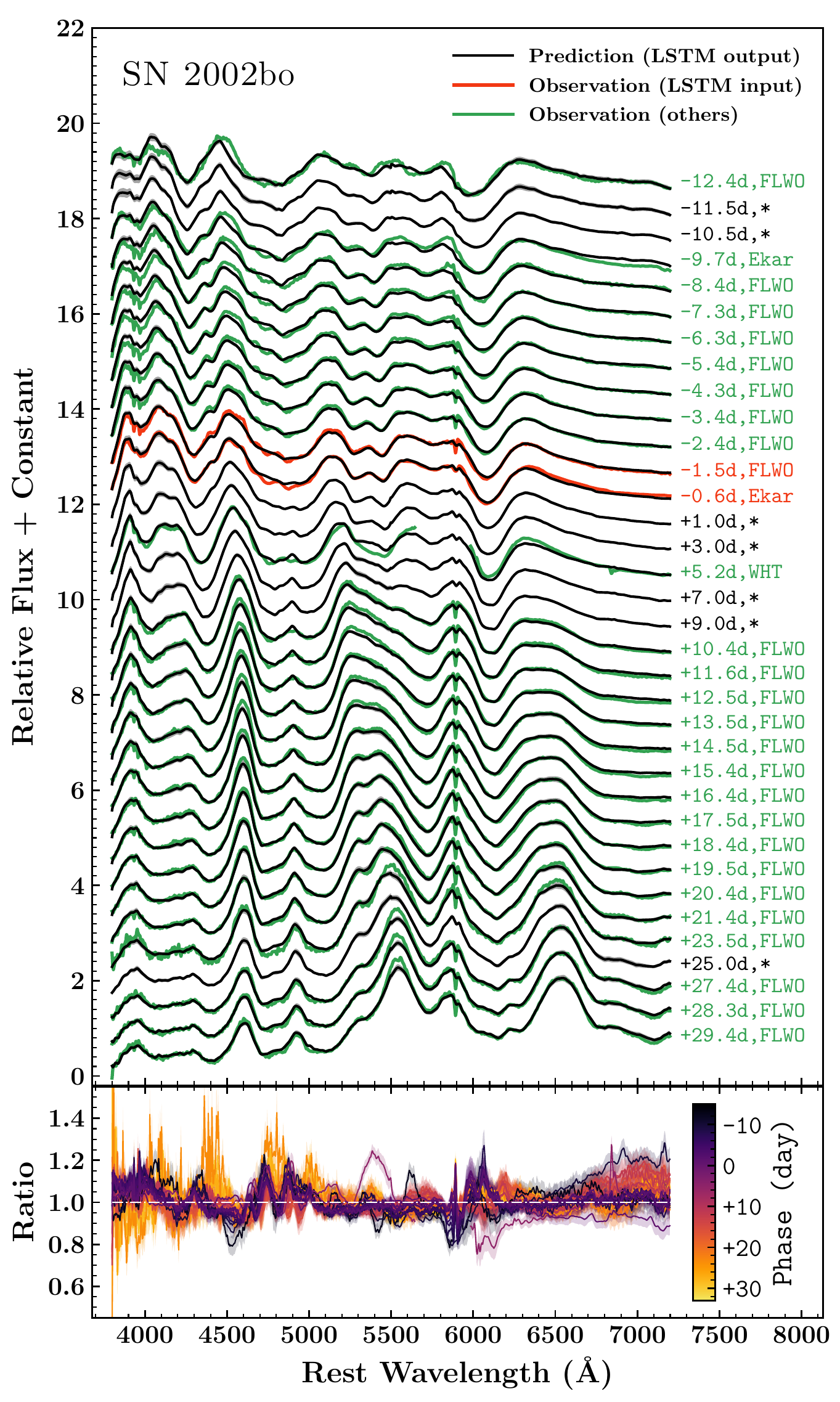}{0.235\textwidth}{(e) SN~2002bo, $\Delta{P}=2\text{d}$}
        \fig{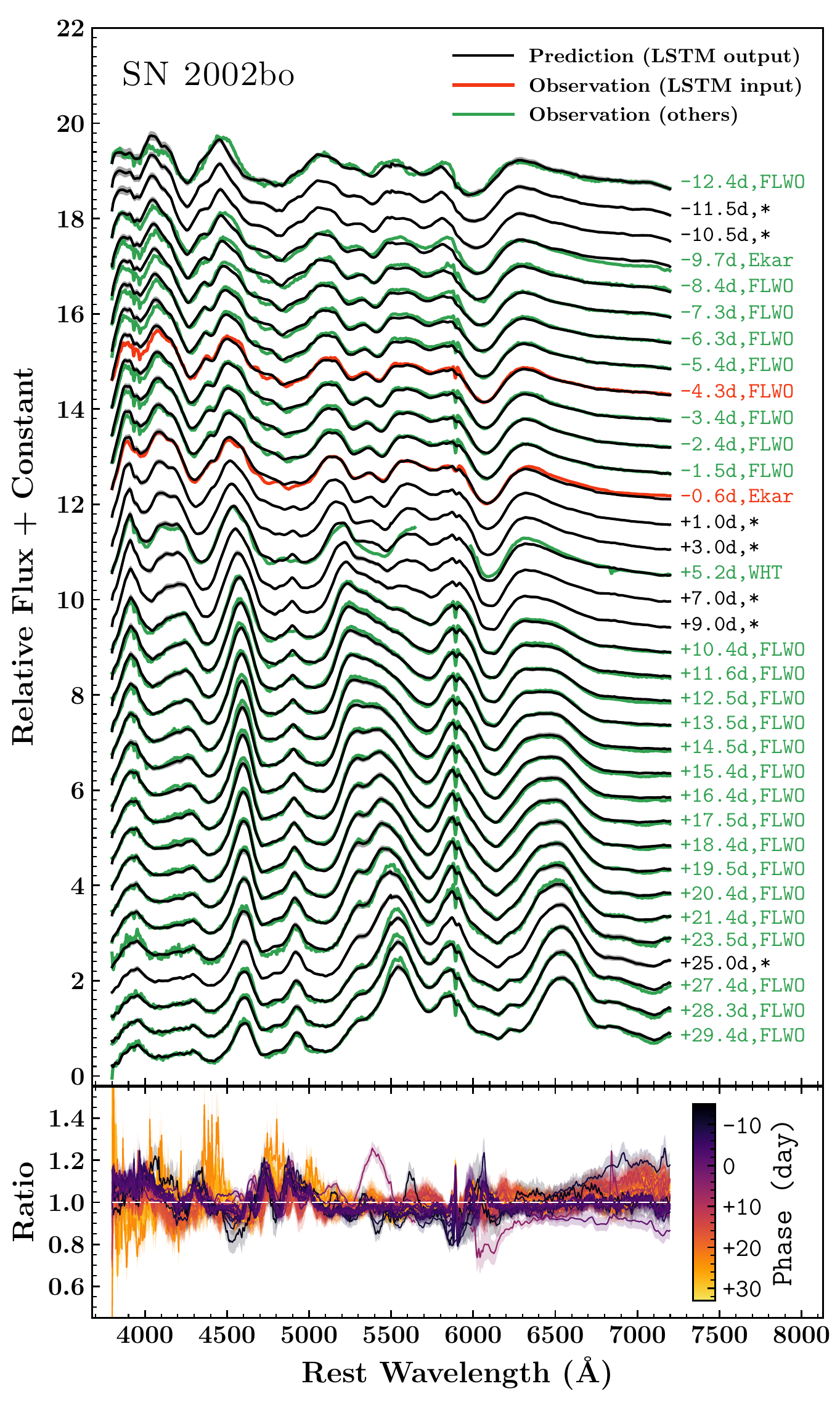}{0.235\textwidth}{(f) SN~2002bo, $\Delta{P}=5\text{d}$}
        \fig{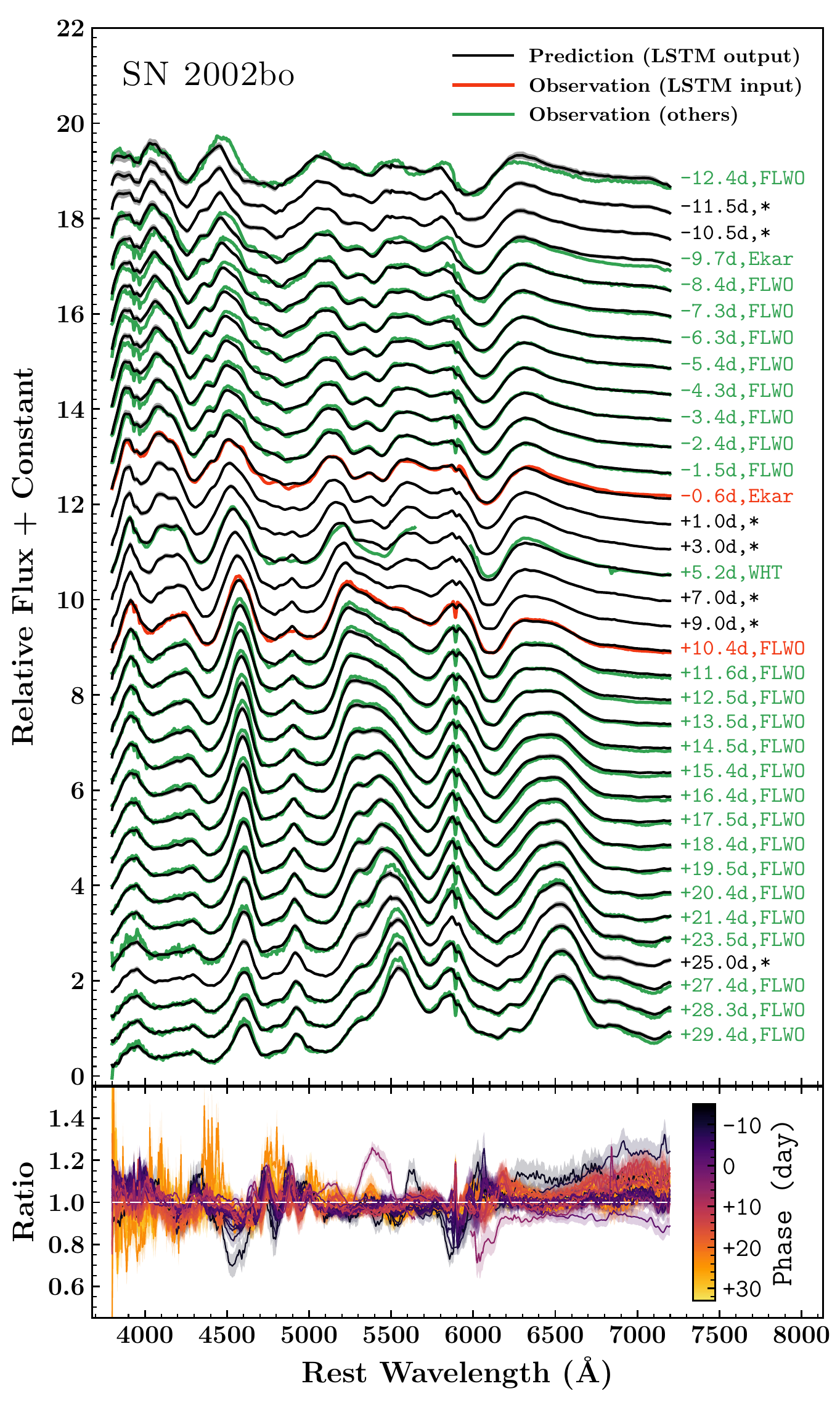}{0.235\textwidth}{(g) SN~2002bo, $\Delta{P}=11\text{d}$}
        \fig{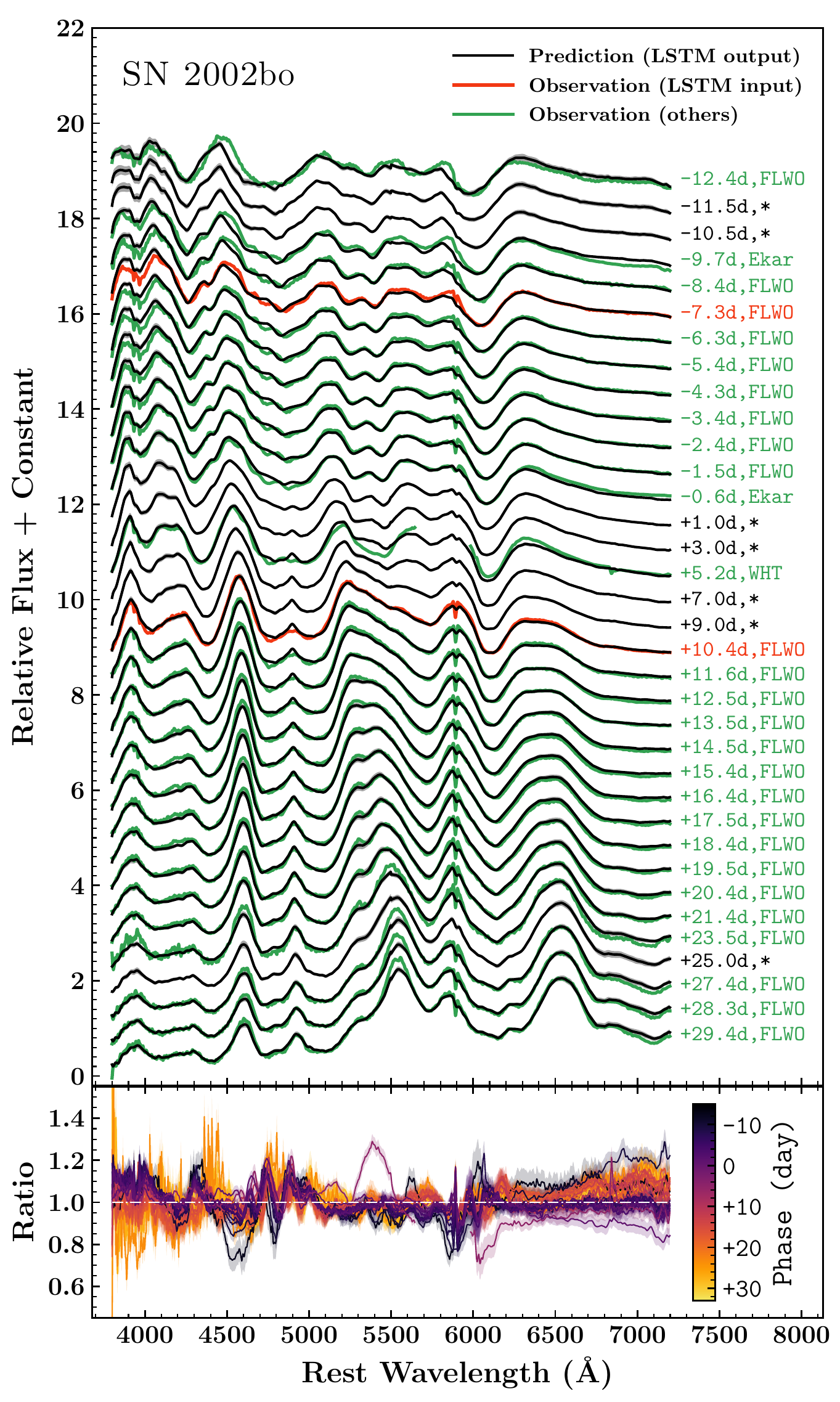}{0.235\textwidth}{(h) SN~2002bo, $\Delta{P}=18\text{d}$}
        }
    \caption{\label{fig:TwoSpec-Validate-NVHV-direct} Spectral sequence predicted from two spectra with phase difference $\Delta{P}$ using LSTM neural networks for SN~2011fe (\textit{top row}) and SN~2002bo (\textit{bottom row}). The panel format of each panel is the same as in Figure~\ref{fig:OneSpec-Validate-NVHV-direct}.}
\end{figure*}

\begin{figure*}[ht!]
    \centering
    \gridline{
        \fig{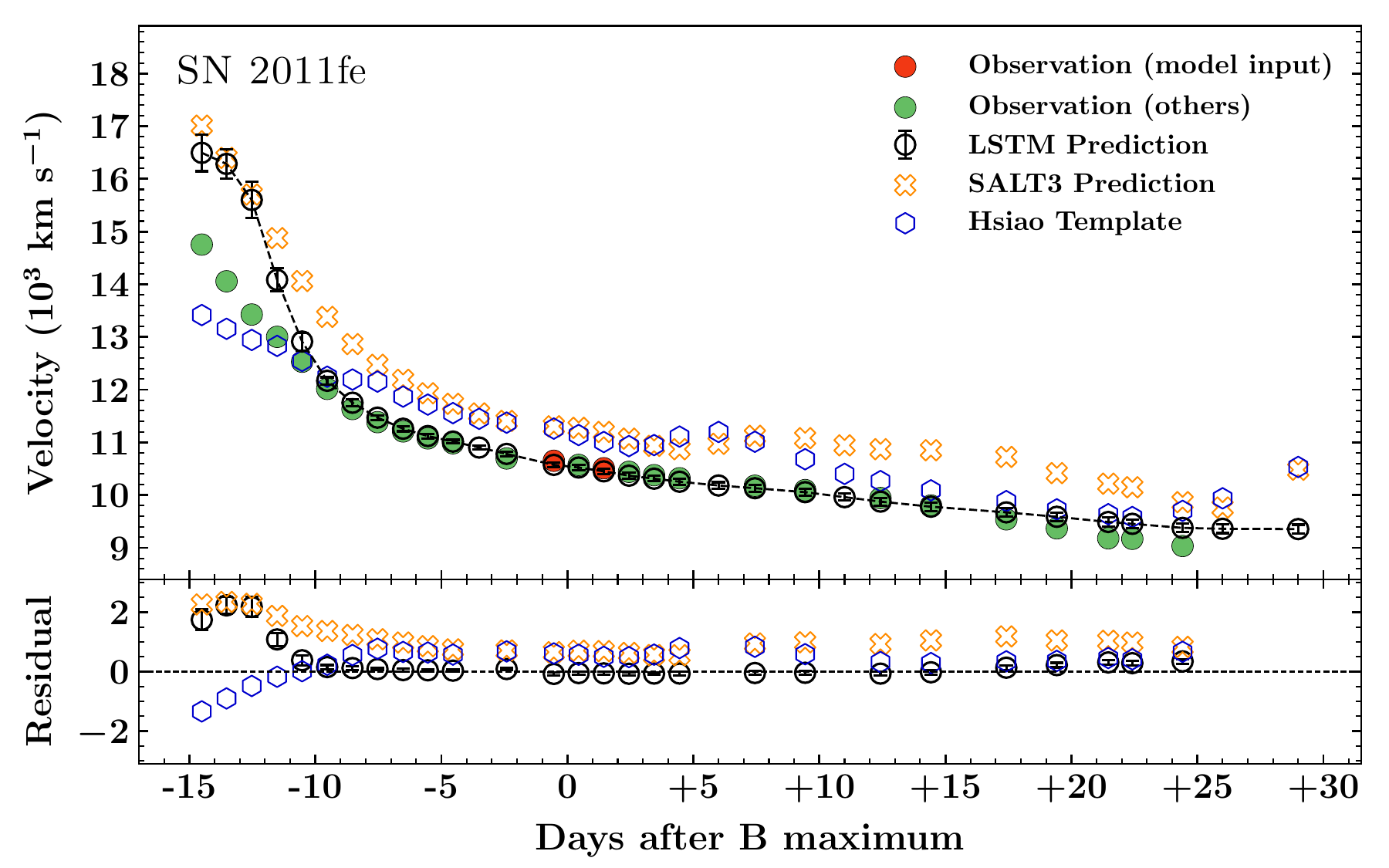}{0.37\textwidth}{(a) SN~2011fe, $\Delta{P}=2\text{d}$}
        \fig{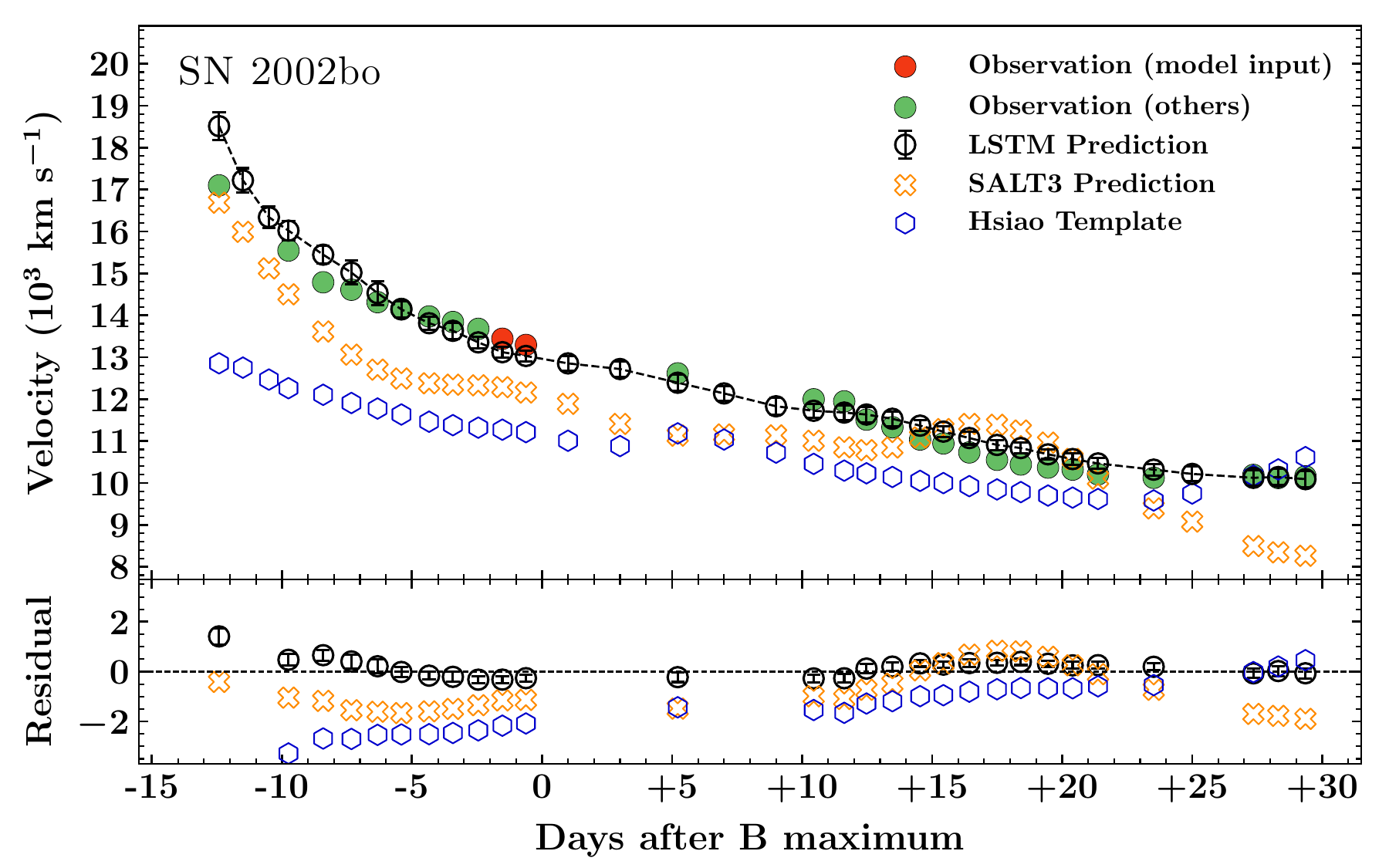}{0.37\textwidth}{(e) SN~2002bo, $\Delta{P}=2\text{d}$}
        }
    \gridline{
        \fig{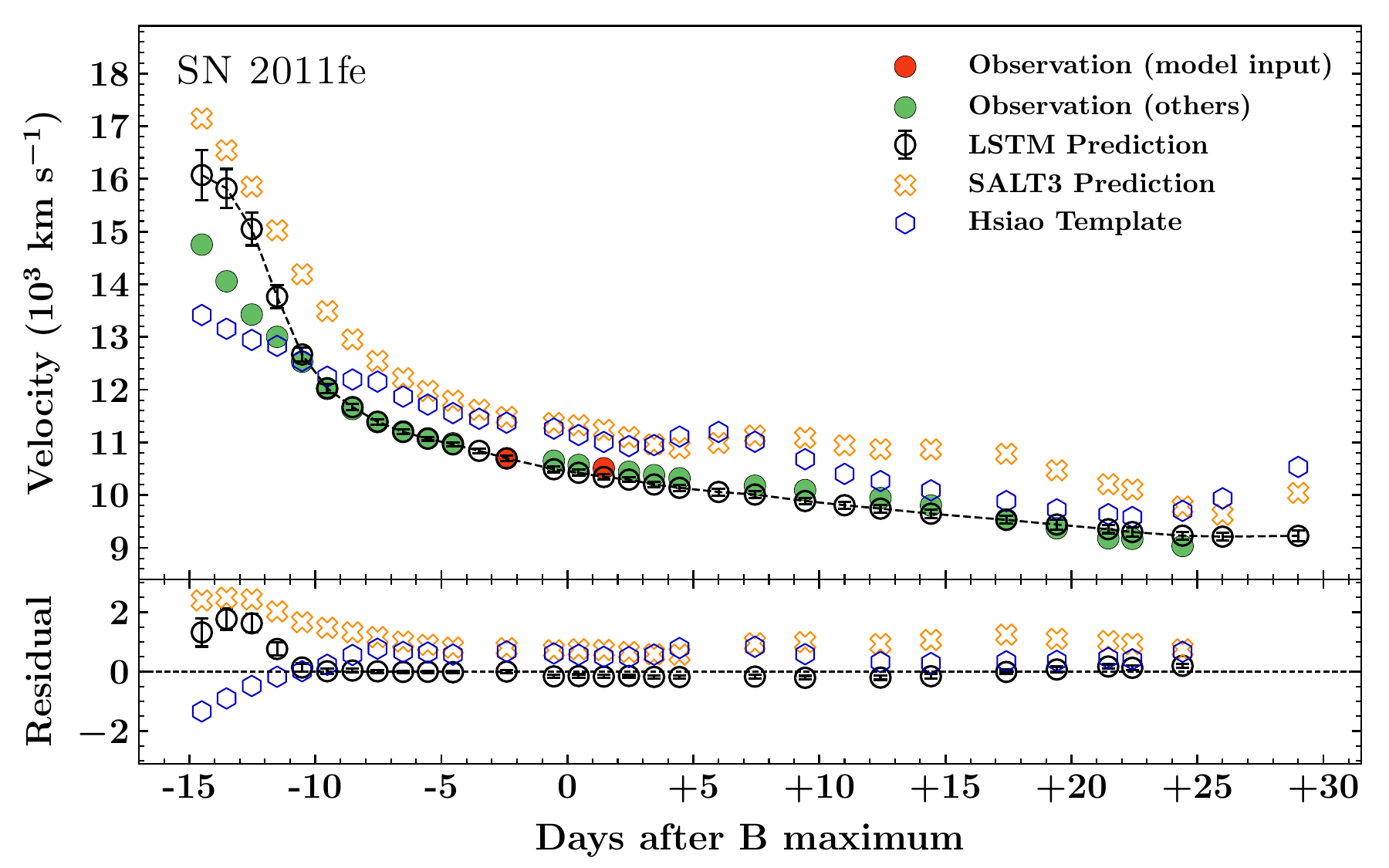}{0.37\textwidth}{(b) SN~2011fe, $\Delta{P}=4\text{d}$}
        \fig{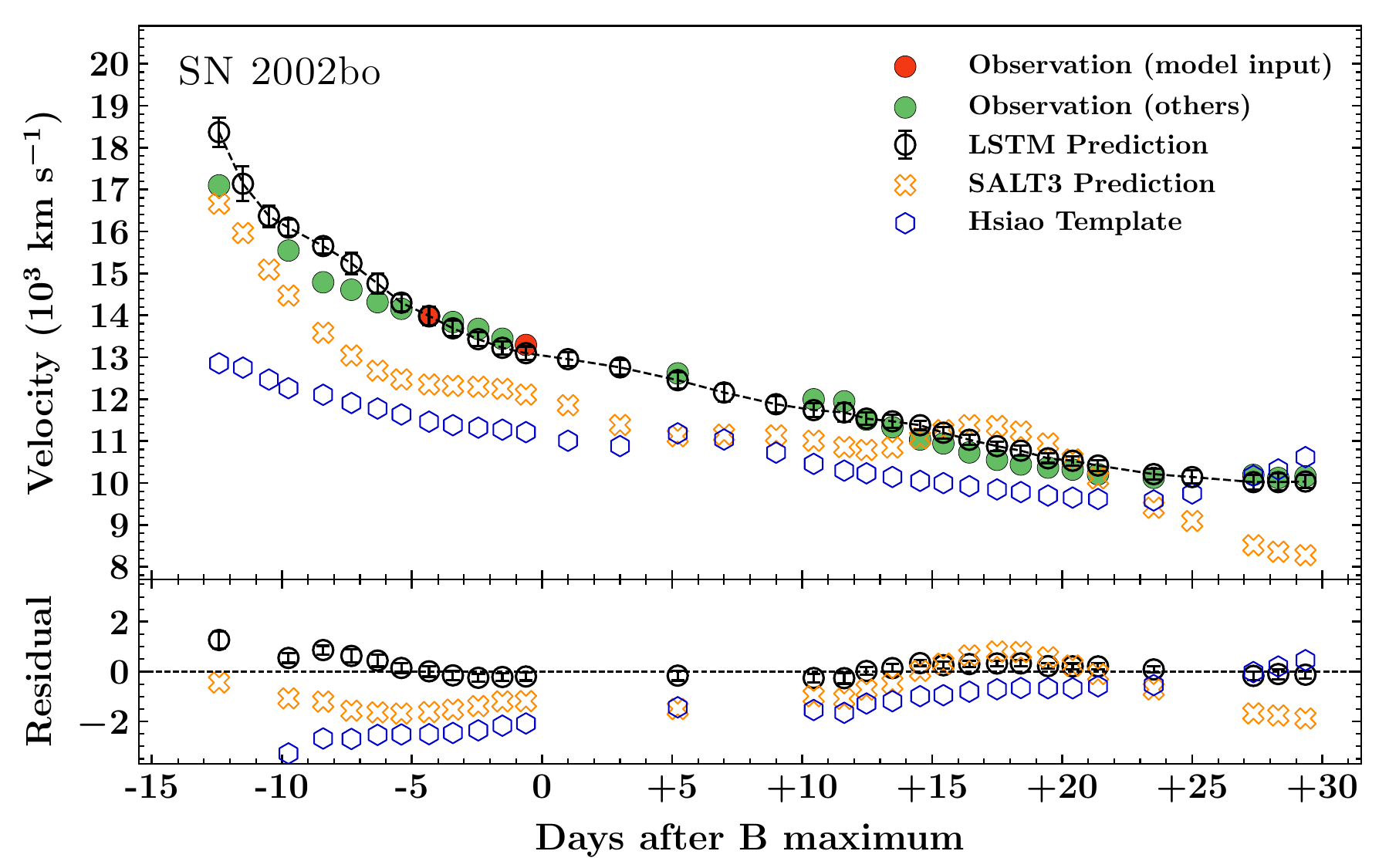}{0.37\textwidth}{(f) SN~2002bo, $\Delta{P}=5\text{d}$}
        }
    \gridline{
        \fig{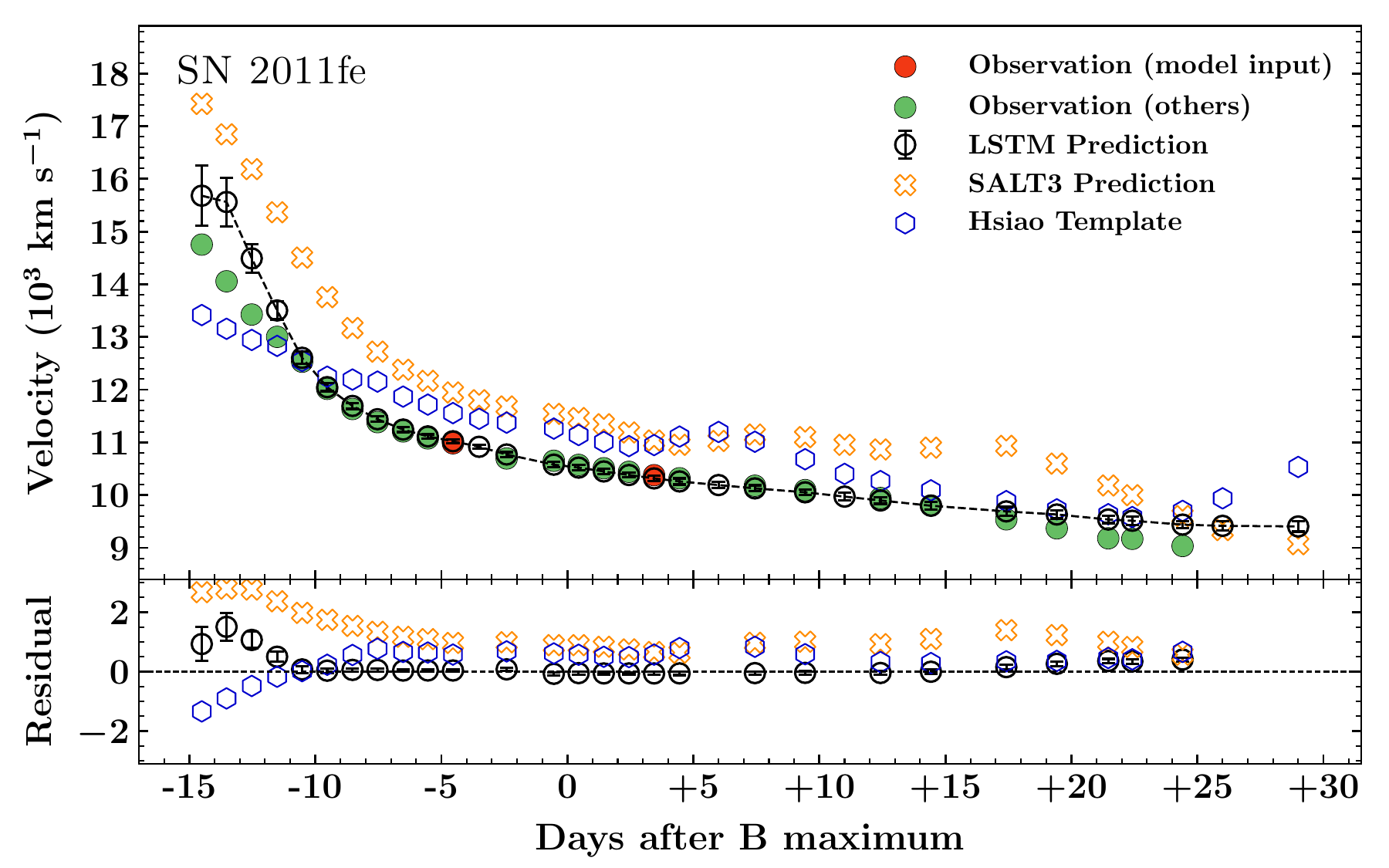}{0.37\textwidth}{(c) SN~2011fe, $\Delta{P}=8\text{d}$}
        \fig{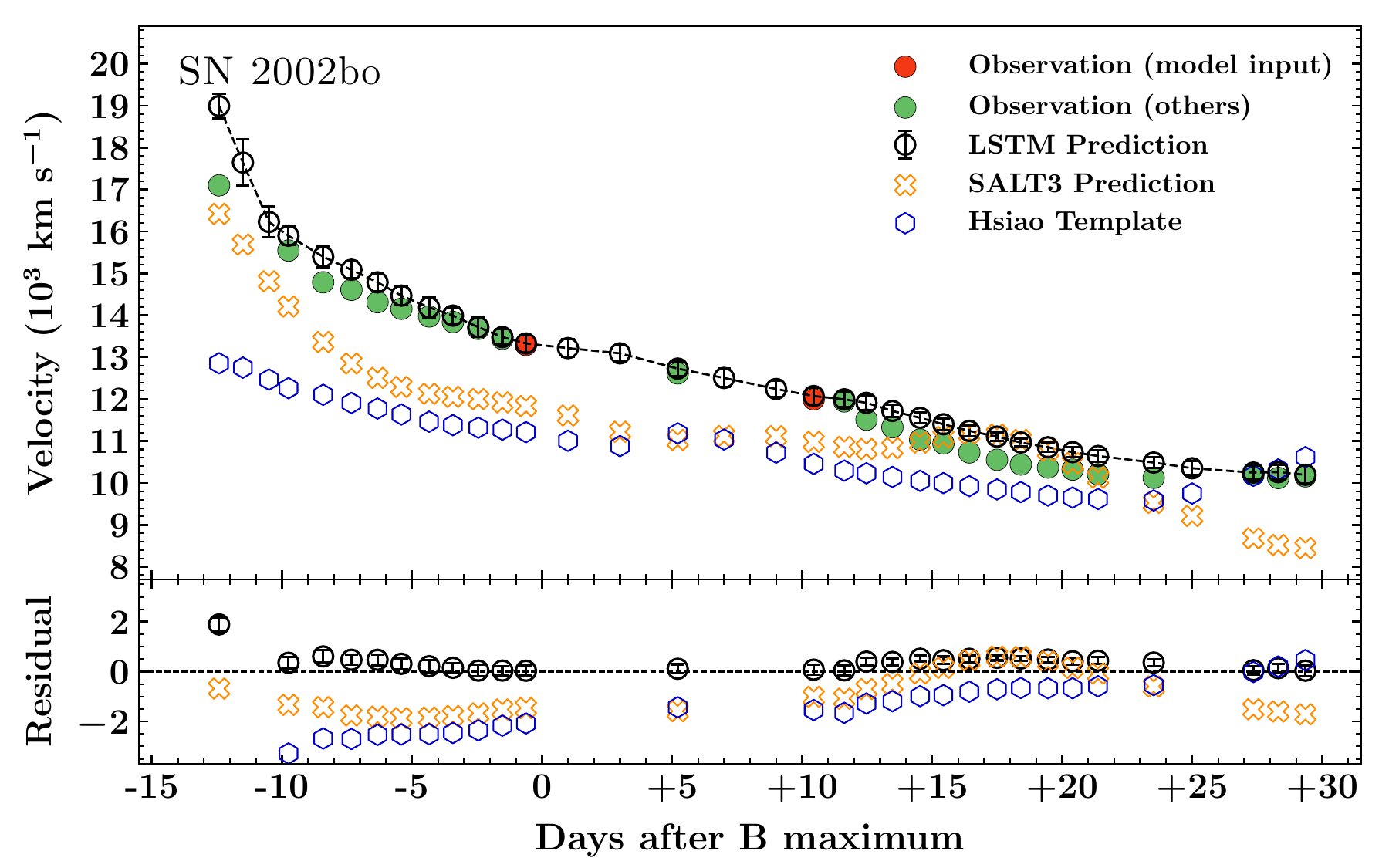}{0.37\textwidth}{(g) SN~2002bo, $\Delta{P}=11\text{d}$}
        }
    \gridline{
        \fig{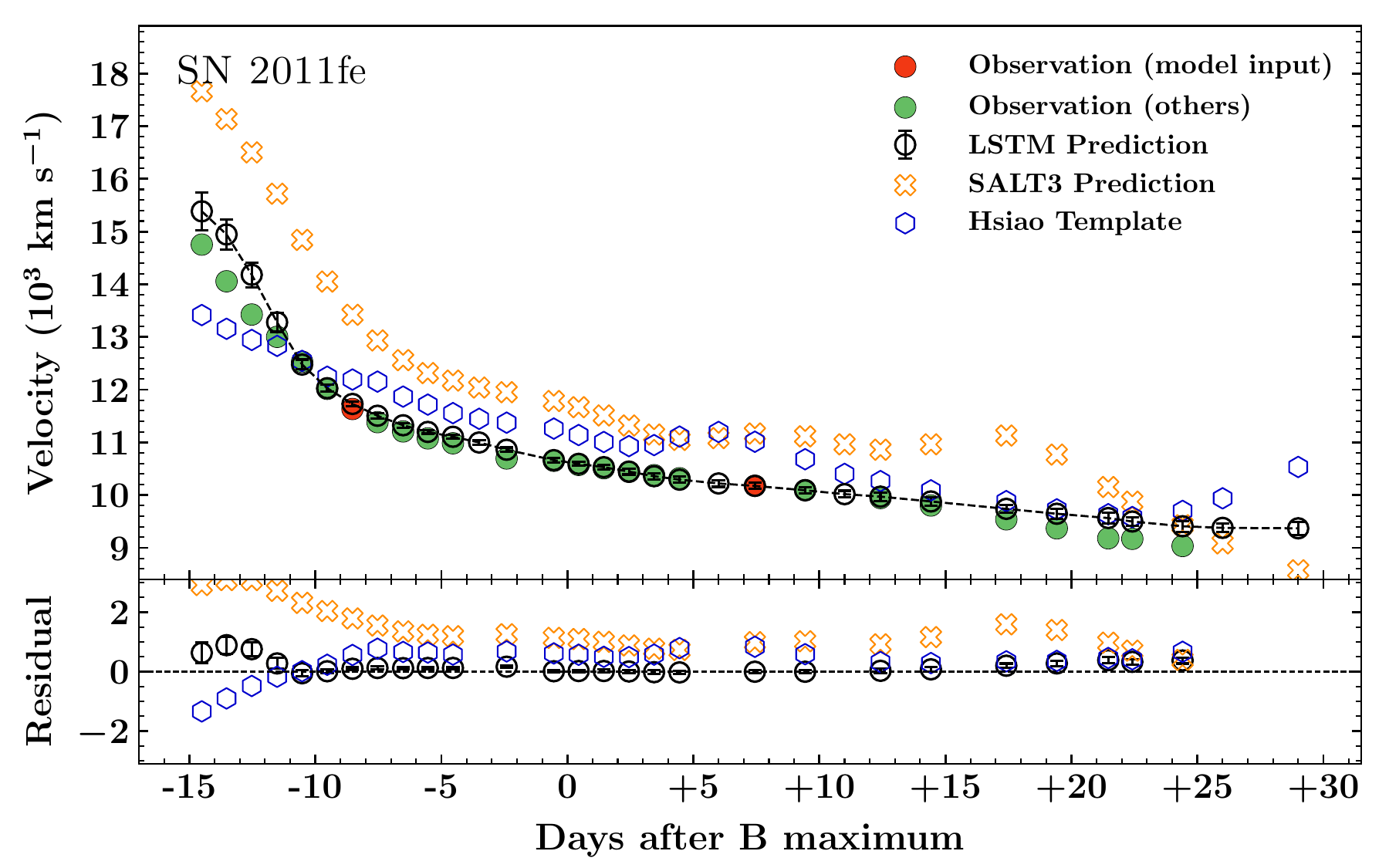}{0.37\textwidth}{(d) SN~2011fe, $\Delta{P}=16\text{d}$}
        \fig{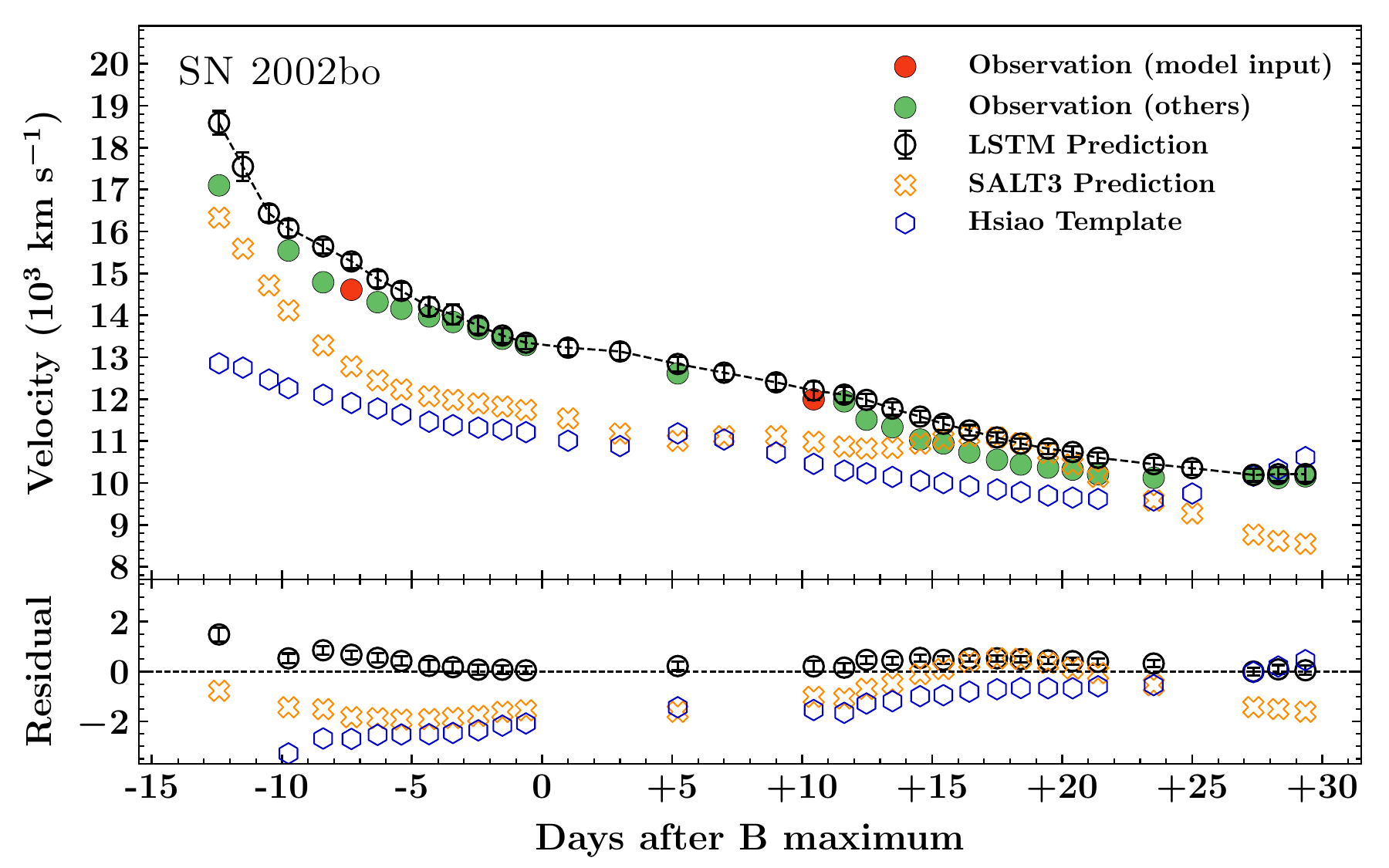}{0.37\textwidth}{(h) SN~2002bo, $\Delta{P}=18\text{d}$}
        }
    \caption{\label{fig:TwoSpec-BaselineValidate-11fe02bo-velocity} The Si~II $\lambda 6355$\AA\ velocity measured on the spectral sequence predicted from two spectra with phase difference $\Delta{P}$ using LSTM neural networks (black) and the corresponding observed spectra (green) for SN~2011fe (\textit{left column}) and SN~2002bo (\textit{right column}). The panel format of each panel is the same as in Figure~\ref{fig:OneSpec-Validate-11fe02bo-velocity}.}
\end{figure*}

\begin{figure*}[ht!]
    \centering
    \gridline{
        \fig{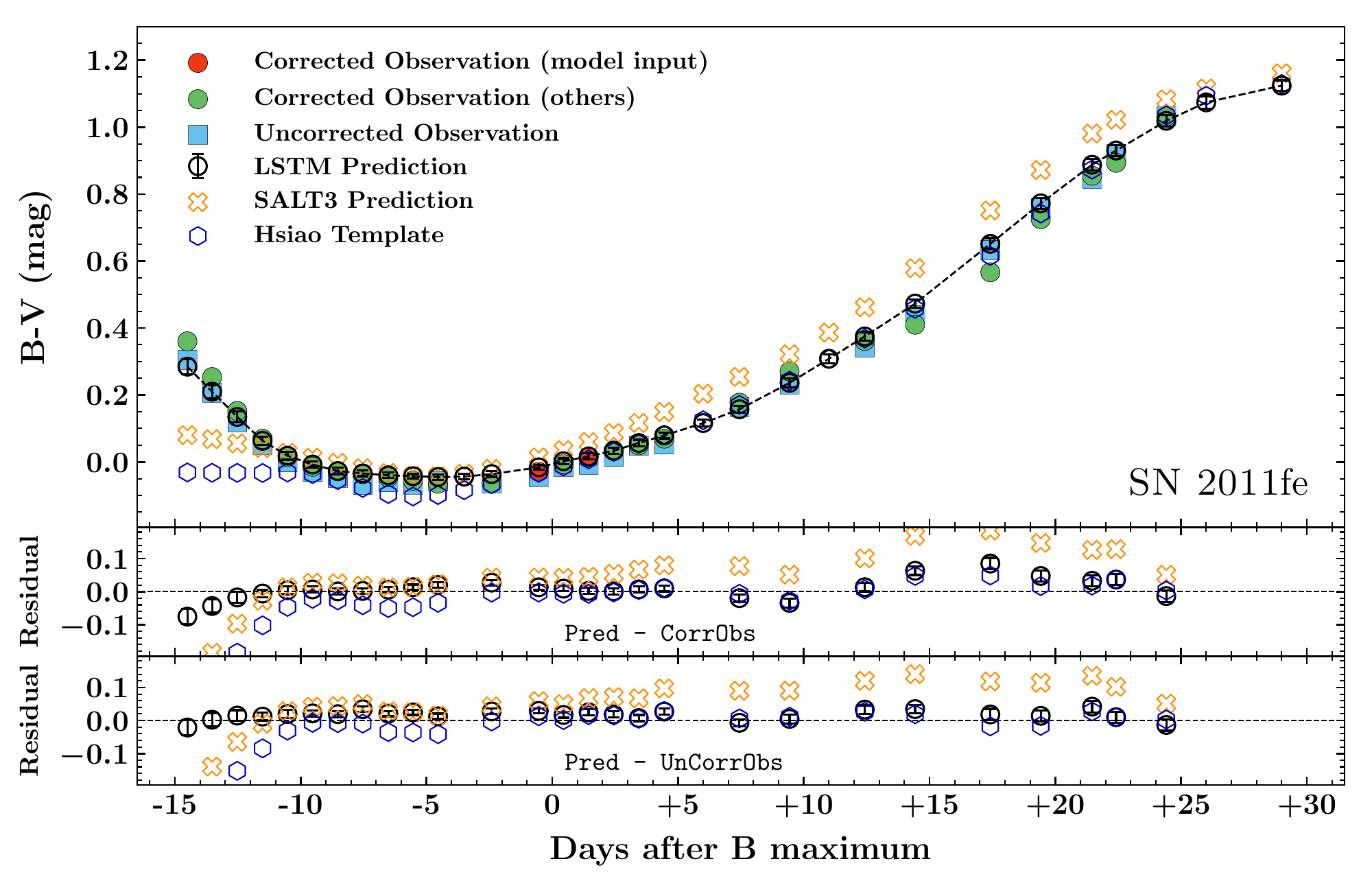}{0.37\textwidth}{(a) SN~2011fe, $\Delta{P}=2\text{d}$}
        \fig{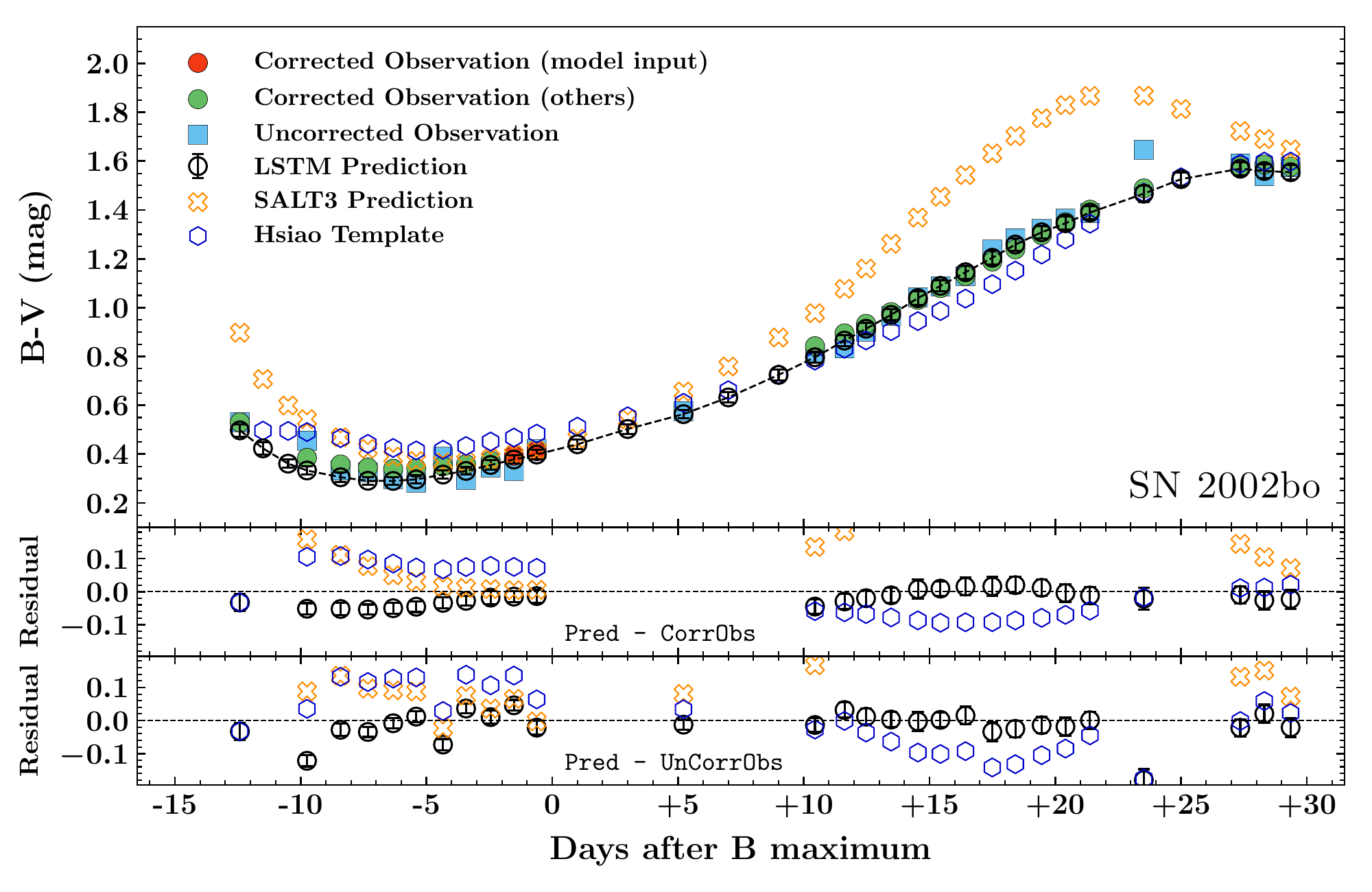}{0.37\textwidth}{(e) SN~2002bo, $\Delta{P}=2\text{d}$}
        }
    \gridline{
        \fig{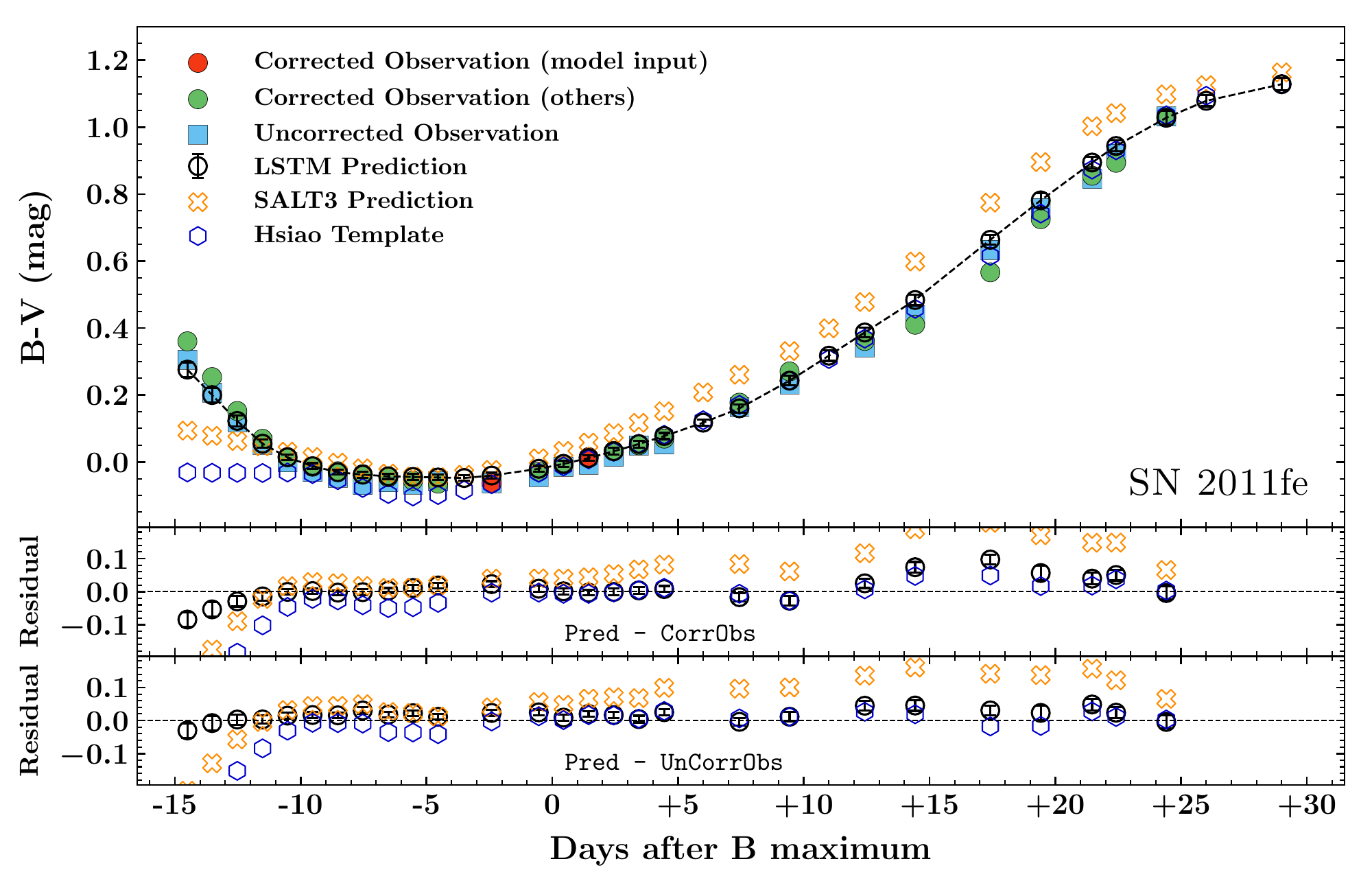}{0.37\textwidth}{(b) SN~2011fe, $\Delta{P}=4\text{d}$}
        \fig{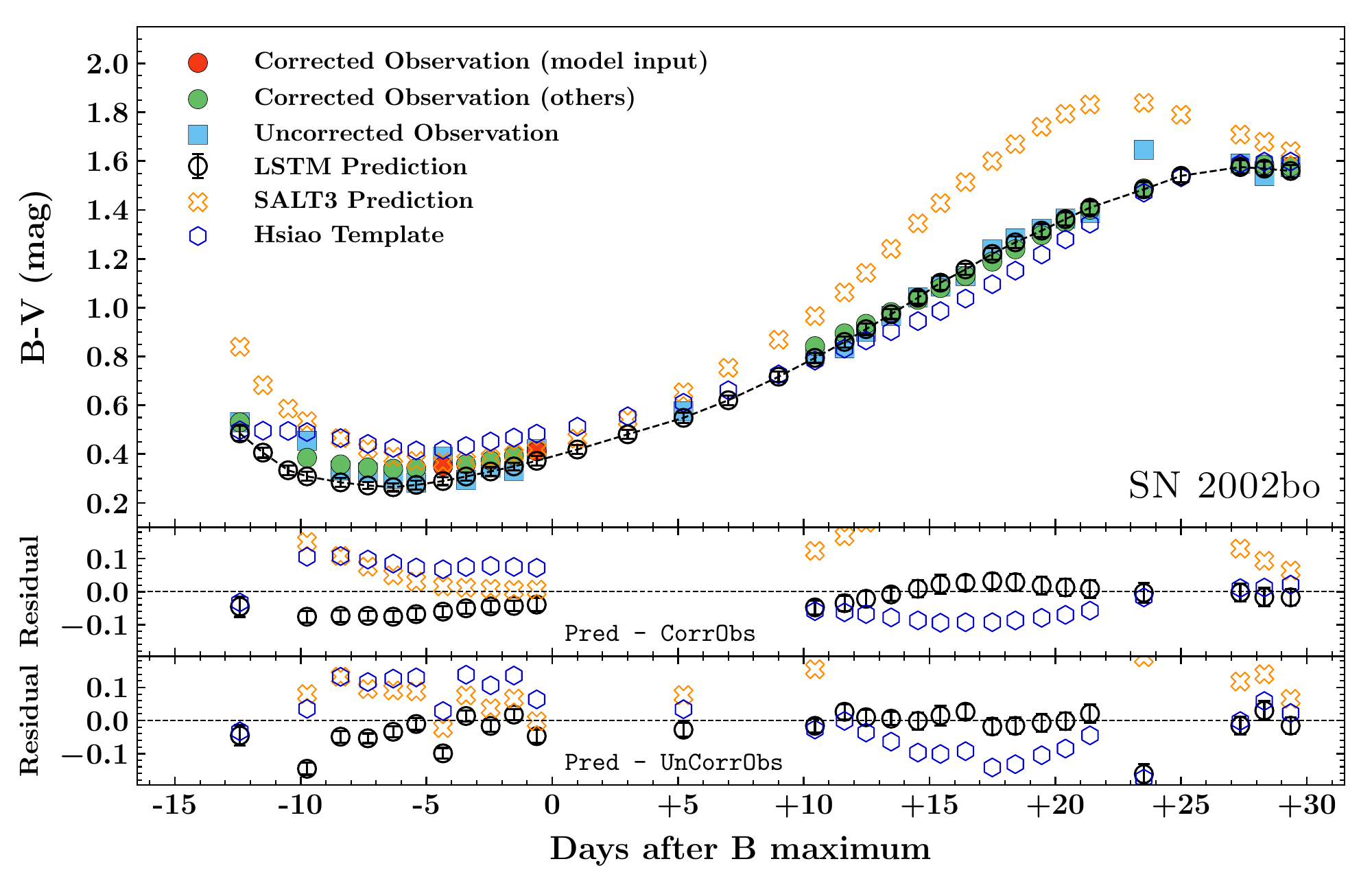}{0.37\textwidth}{(f) SN~2002bo, $\Delta{P}=5\text{d}$}
        }
    \gridline{
        \fig{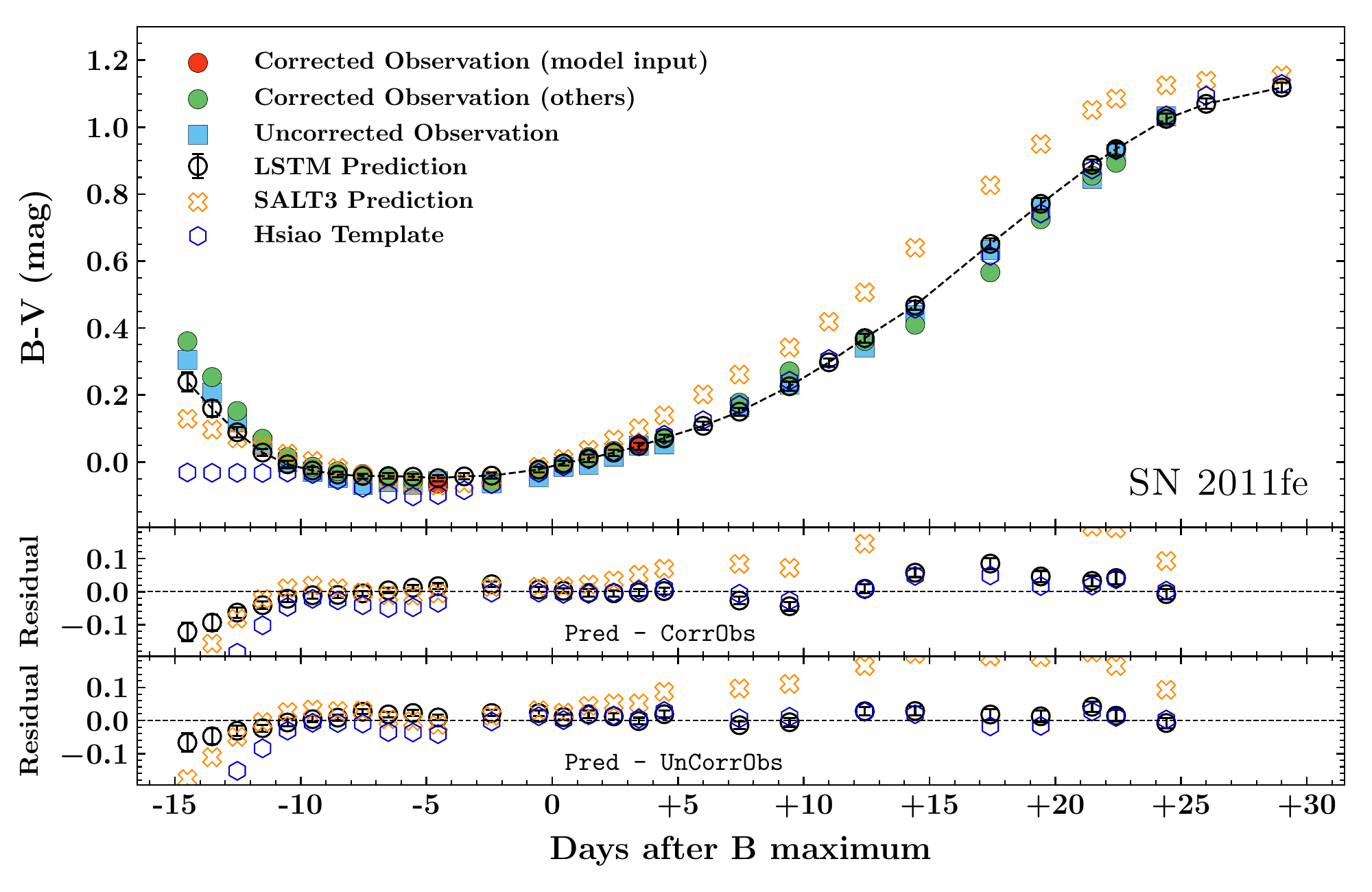}{0.37\textwidth}{(c) SN~2011fe, $\Delta{P}=8\text{d}$}
        \fig{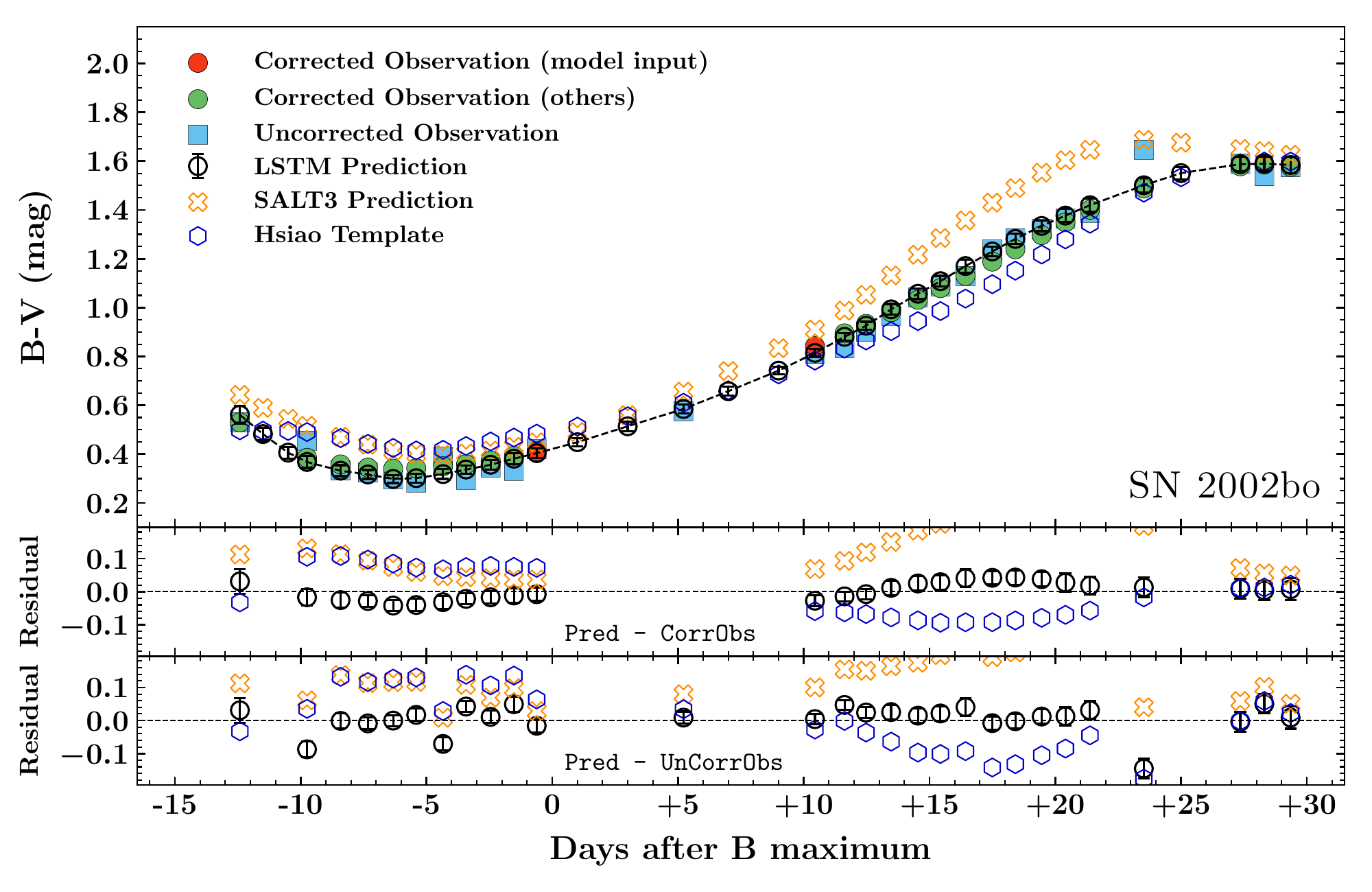}{0.37\textwidth}{(g) SN~2002bo, $\Delta{P}=11\text{d}$}
        }
    \gridline{
        \fig{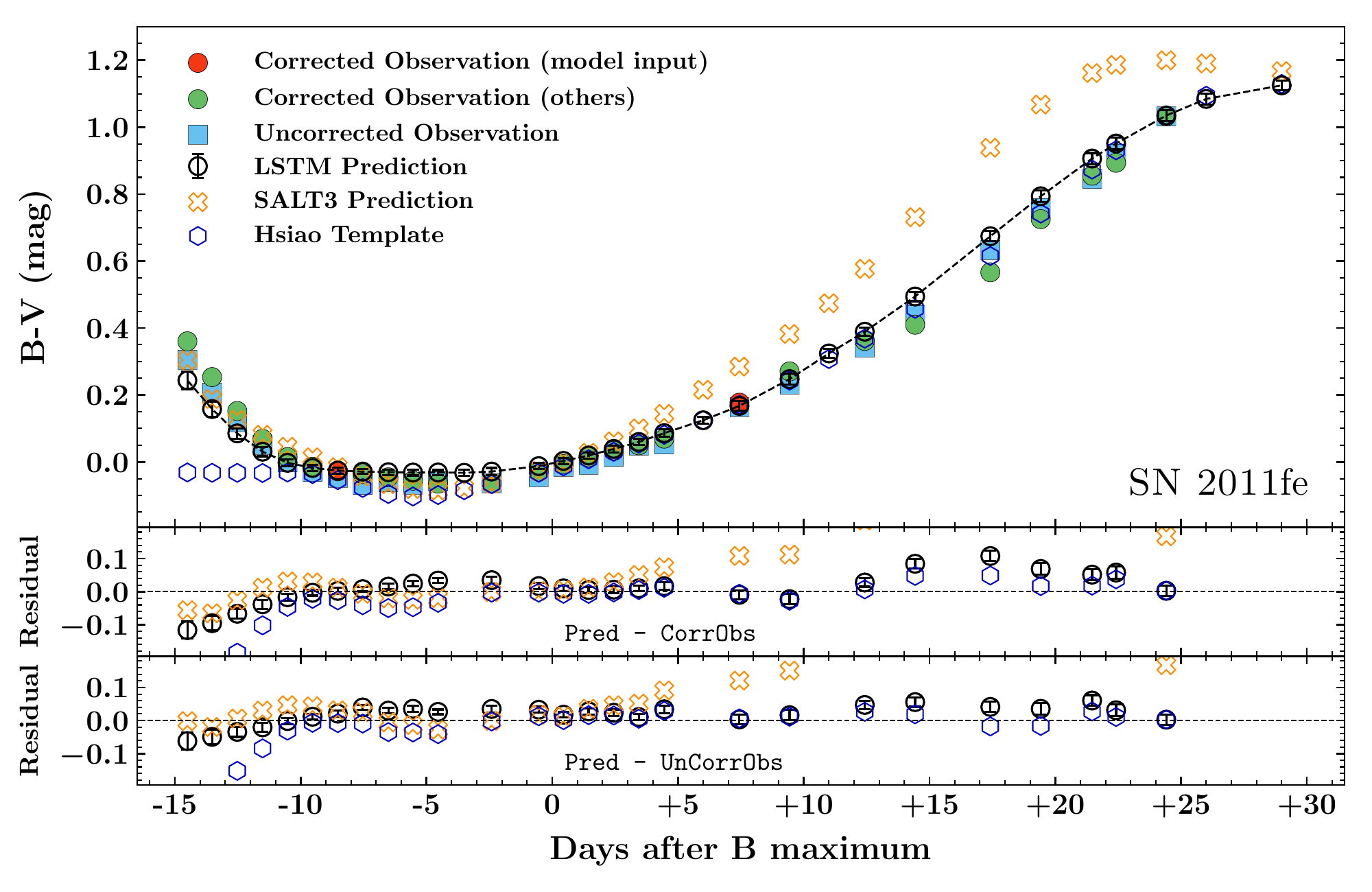}{0.37\textwidth}{(d) SN~2011fe, $\Delta{P}=16\text{d}$}
        \fig{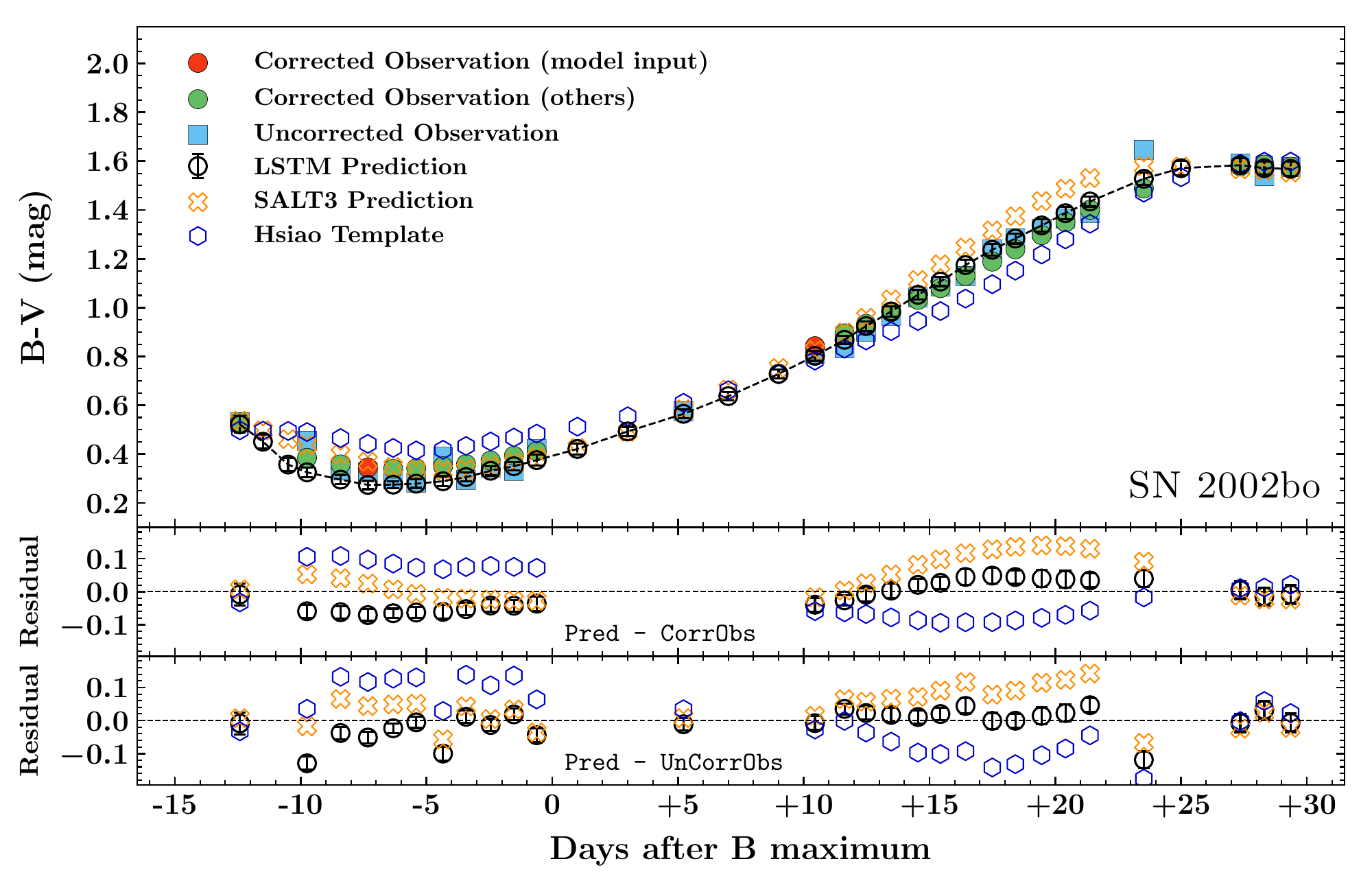}{0.37\textwidth}{(h) SN~2002bo, $\Delta{P}=18\text{d}$}
        }
    \caption{\label{fig:TwoSpec-BaselineValidate-11fe02bo-color} Synthetic $B-V$ color measured from the spectral sequence predicted from two spectra with phase difference $\Delta{P}$ using LSTM (black) (black circles) and the corresponding observational data (green circles and blue squares) for SN~2011fe (\textit{left column}) and SN~2002bo (\textit{right column}). The panel format of each panel is the same as in Figure~\ref{fig:OneSpec-Validate-11fe02bo-color}.}
\end{figure*}

\subsection{The Diverse SN~Ia Spectral Family} \label{ssec:onespec-subtypes}

SNe~Ia demonstrate a wide range of spectral diversities. A critical question is whether the neural networks can capture such diversities. 
In this section, we use four representative SNe (LSQ~12gdj, SN~2008Z, SN~2005ke, and SN~2012Z) to examine the performance of the neural networks on the diverse subtypes of SN~Ia. 
For each representative SN, a separate LSTM model is trained over the samples generated from all of the SNe~Ia excluding the representative SN under study, as in Section~\ref{ssec:onesepc-NVHV}. The spectral sequence is constructed similarly following Algorithm~\ref{algo:makeseq-12spec}.

Figure~\ref{fig:OneSpec-Validate-pec-direct}a shows the spectral sequence predicted from one spectrum at maximum light for LSQ~12gdj \citep{2014ApJ...795..142G} which is a Ia-91T SN \citep{1992ApJ...384L..15F,1992AJ....103.1632P} with a very shallow Si~II 6355 \AA\ line before optical maximum. 
The spectral sequences constructed by using a spectrum taken at +0.7 days after $B$ maximum are shown together with the observed spectra. At the time of optical maximum, the Si~II 6355 \AA\ line is well developed. The spectral features match well throughout the period covered by the observations. 
In particular, the neural network is able to reproduce the extremely shallow Si 6355 \AA\ feature at epochs around 1 week before maximum. This suggests that the spectra around the optical maximum carry enough information to define a Ia-91T event. 
Identifying Ia-91T events from normal SNe~Ia is important for supernova cosmology as it is shown by recent studies that Ia-91T SNe are potential sources of systematic errors of supernova cosmology (Jiawen Yang et al., in prep.).

Object SN~1999aa represents another subtype of peculiar SN~Ia, which is similar to Ia-91T SNe but with the subtle differences of having weak signatures of Ca II H \& K and Si~II absorption prior to maximum light \citep{2004AJ....128..387G}. Figure~\ref{fig:OneSpec-Validate-pec-direct}b shows the model sequence calculated with a single spectrum +0.3 days from maximum together with the observations for SN~2008Z \citep{2012MNRAS.425.1789S} which is a Ia-99aa SN. In this case, the Si~II feature is stronger a week before maximum than Ia-91T and again well reproduced by the neural network. This demonstrates that the neural network can effectively distinguish Ia-91T and Ia-99aa SNe based on spectroscopy around optical maximum. 

Object SN~2005ke \citep{2017AJ....154..211K} is a Ia-91bg \citep{1997ARA&A..35..309F} subluminous supernova. The SN shows rapid spectral evolution and much redder overall spectral color. Figure~\ref{fig:OneSpec-Validate-pec-direct}c shows the spectral sequence computed from a spectrum of SN~2005ke at +0.3 days from optical maximum and the observed spectra. The rapid evolution of the spectral features is well produced from more than a week before maximum to about a month after maximum. 

Object SN~2012Z \citep{2019MNRAS.490.3882S} belongs to the peculiar type Iax \citep{2013ApJ...767...57F} SNe. The spectral data of this subtype are sparse, but nonetheless, the neural network captures all of the major spectral features of the observations, as shown in Figure~\ref{fig:OneSpec-Validate-pec-direct}d. 

Like in Section~\ref{ssec:baseline4ctemp}, we compared the results of the LSTM neural networks with the other two template models, as shown in Figure~\ref{fig:OneSpec-BaselineValidate-91T99aa-direct} and Figure~\ref{fig:OneSpec-BaselineValidate-91bgIax-direct}. 
We found again that the \cite{Hsiao2007Kcorrection} model and the SALT3 model using only one spectrum around maximum can not accurately reconstruct the spectral evolution of any of these objects. The LSTM models can robustly reproduce spectral sequences of these SNe from the date of explosion to about a month past optical maximum, except for the Iax SN~2012Z which shows increasing deviations at around 2 weeks past maximum.

\subsection{Spectral Sequences from Two Spectra} \label{ssec:twospec}

%% ***** Section: Phase Prediction
\begin{figure*}[!ht]
    \centering
    \includegraphics[trim=0cm 0cm 0cm 0cm,clip=true,width=11cm]{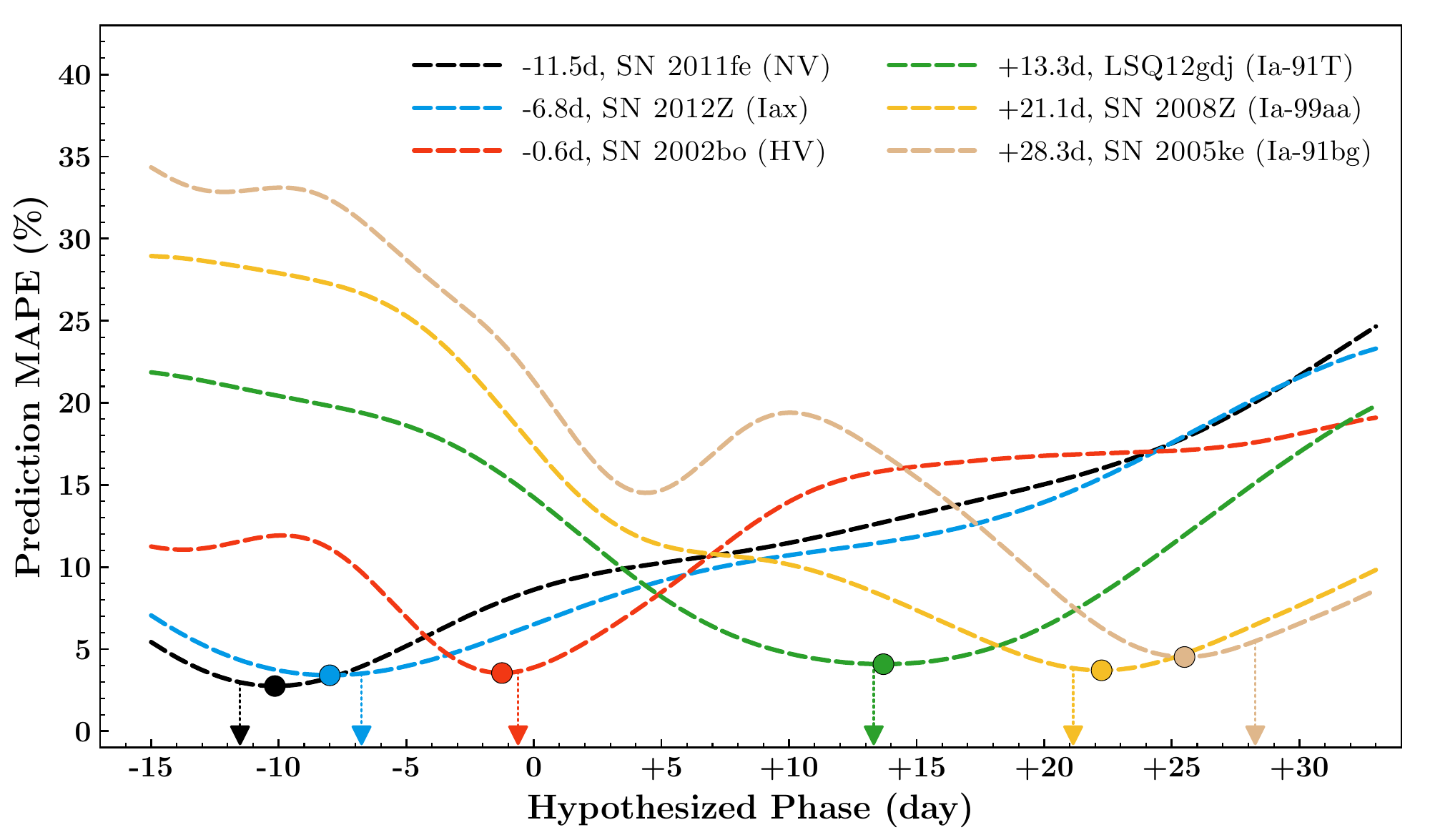}
    \caption{\label{fig:PhaseEstimate-Example} The MAPEs between the observed spectra and the predicted spectra using LSTM neural networks as a function of the hypothesized phase. This figure presents the results for six representative spectra with different subtypes and phases. The curves show the MAPEs for SN~2011fe (black), SN~2012Z (light blue), SN~2020bo (red), LSQ~12gdj (green), SN~2008Z (yellow), and SN~2005ke (brown). The filled circles show the locations of the minimum MAPEs, and the arrows point to the true phases as derived from photometric light curves.}
\end{figure*}

\begin{figure*}[!ht]
    \centering
    \includegraphics[trim=0cm 0cm 0cm 0cm,clip=true,width=11cm]{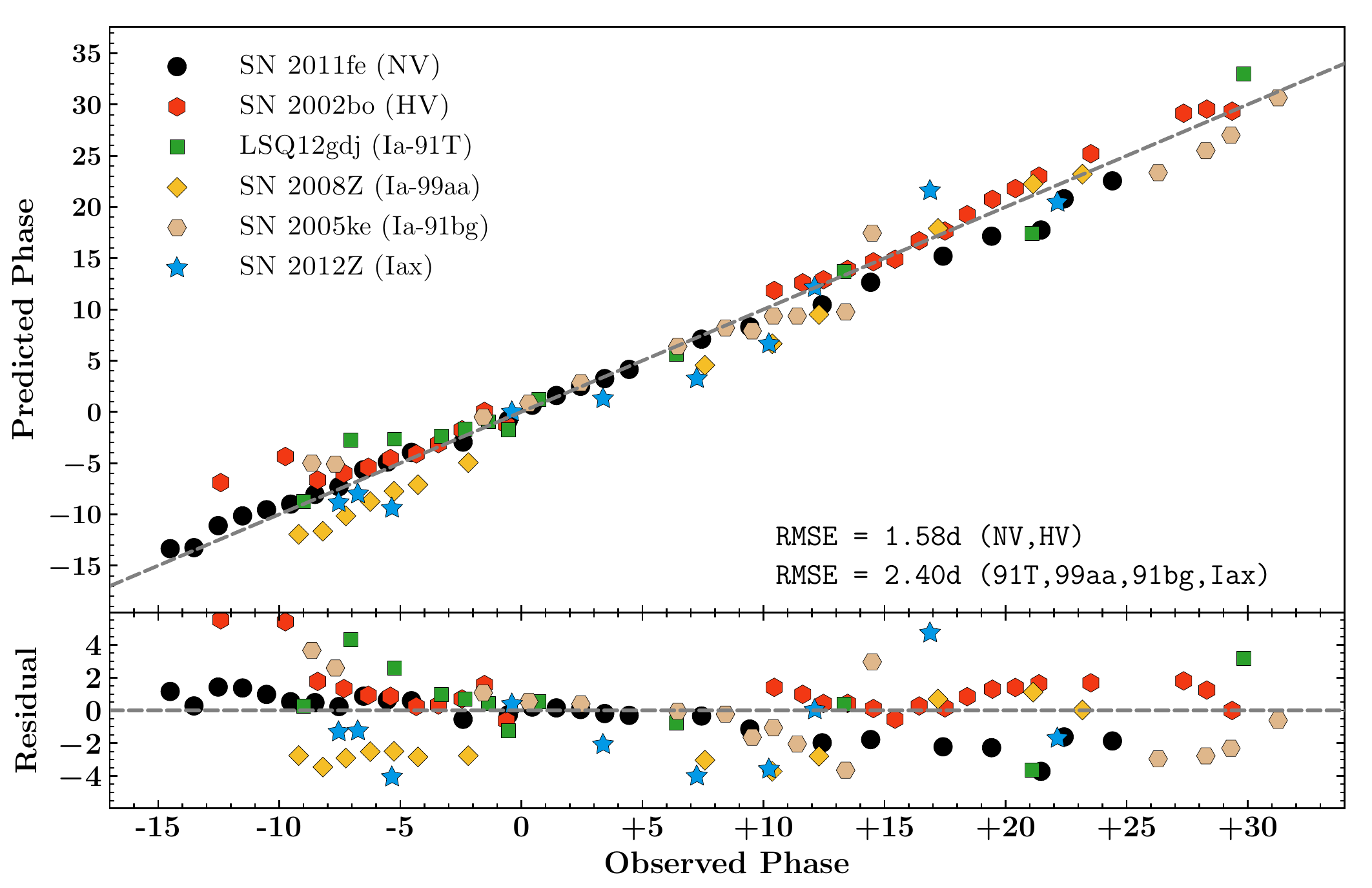}
    \caption{\label{fig:PhaseEstimate-Stat} Predicted phase versus the observed phase for six representative SNe~Ia discussed in Section~\ref{sec:lstm-12spec}. The predicted phase is determined by minimizing the MAPE, as demonstrated in Figure~\ref{fig:PhaseEstimate-Example}. The legend shows the symbols used for each SN. The upper panel shows the correlation between the predicted and observed phases. The dashed gray line shows the relation when the predicted and observed phases are equal. The residuals of the predictions are shown in the lower panel. The RMSEs of the phases predictions are given in the bottom right corner of the upper panel for the SNe representing NV, HV, Ia-91T, Ia-99aa, Ia-91bg, and Iax.}
\end{figure*}

One important question in scheduling spectroscopic observations of SNe~Ia is what the optimal time gaps between observations would be if multiple observations could be acquired. With the neural networks, such a question is related to the spectral sequence prediction using multiple spectra as input data. 
Figure~\ref{fig:TwoSpec-Validate-NVHV-direct}(a-d) show the predictions with two spectra of SN~2011fe separated by $\sim$ 2, 4, 8 and 16 days in spectral phase, each at [-0.5, +0.4], [-2.4, 1.4], [-4.5, +3.4], and [-8.5, +7.4] days, respectively. 
The corresponding measurements of Si~II $\lambda$6355 \AA\ velocity and spectral $B-V$ colors are given by Figures~\ref{fig:TwoSpec-BaselineValidate-11fe02bo-velocity}(a-d) and \ref{fig:TwoSpec-BaselineValidate-11fe02bo-color}(a-d).
The overall spectral fits at early phases improve significantly, as a spectrum at phases well before maximum is used as the input. The fits to the data around optical maximum do not show any obvious deterioration, even when both input spectra are more than 1 week from optical maximum.
The same is true for HV SN~2002bo, as shown in Figures~\ref{fig:TwoSpec-BaselineValidate-11fe02bo-velocity}(e-h) and \ref{fig:TwoSpec-BaselineValidate-11fe02bo-color}(e-h).

Based on these neural network predictions, we may conclude that spectral data separated by more than 8-16 days around optical maximum can provide the maximum constraining power on the intrinsic properties of an SN Ia.

\subsection{When the Spectral Phase Is Unknown} \label{ssec:unknownphase}

In our framework, the spectral phase is set as prior knowledge, whereas the time of maximum light may not always be available, especially when a newly discovered SN is still being actively monitored. A CNN can be used to provide a phase estimate based on an SN spectrum {\tt\string deepSIP} \citep{Stahl2020deepSIP}). The success of the LSTM neural networks allows for an alternative approach to derive the phase of an SN spectrum without light-curve data.
The basic concept is that a wrong spectral phase fed into the LSTM neural networks is likely to degrade the resulting predictions; that is to say, the correct spectral phase stands the highest chance of maximizing the predictive performance.

We tested this concept with six representative SNe~Ia (SN~2011fe, SN~2002bo, LSQ~12gdj, SN~2008Z, SN~2005ke, and SN~2012Z). Their spectra are shown in  Figures~\ref{fig:OneSpec-Validate-NVHV-direct} and \ref{fig:OneSpec-Validate-pec-direct} (except the WHT spectrum of SN~2002bo with data missing). Each spectrum is assigned a sequence of phases from -15 to +33 days and fed into the LSTM model (trained without the same SN). 
The predictive MAPEs at the time of the input spectrum are used as the loss function to estimate the optimal phase. As shown in Figure~\ref{fig:PhaseEstimate-Example} for these six SNe, the MAPEs are at their minimum only when the input phases are close to the true phases. Figure~\ref{fig:PhaseEstimate-Stat} compares the predicted and observed phases. 
The root mean squared error (RMSE) of this method is $\sim$1.6 days for normal SNe~Ia and $\sim$2.4 days for other subtypes. In addition, our publicly available software also accepts two spectra with unknown phases as input.

\section{Discussions and Conclusions} \label{sec:discussions}

We constructed a neural network-based algorithm to predict the spectral time series of SNe~Ia from sparsely time-sampled spectroscopy. 
In order to train and test the models, we have compiled a spectral database of 3091 optical spectra from 361 SNe~Ia. 
Given the heterogeneous nature of the spectroscopic observations, we homogenized the spectra to obtain uniform wavelength sampling, then performed flux recalibration through photometric observations. The spectral data were reprojected into a lower-dimensional space by FPCA parameterization. 
At the heart of the proposed method is the multilayer LSTM network. It allows predictions of spectra at any specific phase by using a sequence of observed spectra of an SN as input. The model thus enables the construction of spectral sequences from observations with limited time coverage. 

With this method, we have constructed 361 spectral templates for the 361 SNe~Ia in our dataset, where each template is a spectral time series from -15 to 33 days relative to maximum light covering 3800 to 7200 \AA\ in the rest frame. 
Running the testing procedures of the method with the test set, which was not involved in the template construction, has confirmed that our model can reliably build spectral sequences up to a median MAPE error of 4.4\%. 
No obvious bias appears in the distribution of the reconstruction accuracy over the evolution stages of the SNe. Although normal SNe~Ia dominate the spectral dataset, the neural network seems to be able to show reasonable performance on other less represented subtypes, such as Ia-91T, Ia-91bg, Ia-99aa, and Iax. 

We further verified that the method can work well when only one observed spectrum is available. We used SN~2011fe and SN~2002bo as the representative cases of NV and HV objects, respectively, and found that their spectral sequences can be accurately predicted by using a single spectrum around maximum light. We also confirmed the model accuracy by measuring the Si~II $\lambda$6355 \AA\ velocity and spectral $B-V$ colors. 

The difficulties in acquiring spectroscopic data have been the biggest challenge in SN studies and for future time-domain surveys. In SN cosmology, spectroscopy of SNe~Ia is normally only attempted for a single epoch around the optical maximum. 
In the upcoming era of LSST/Rubin and the WFIRST/Roman survey, a large number of transients will be discovered whereas detailed spectroscopy will be impossible for the majority of the transients. The spectral follow-ups of the transients will need to be built with the knowledge of the existing dataset.
The trend motivated us to develop this data-driven method for spectral inference. This is more than an interpolation tool. It allows reconstruction of complete spectral time series from limited available observations, as has been applied to the analyses of the Kepler-observed SN~2018agk in a recent study \citep{Qinan_SN2018agk}.

Our method can be used to investigate the spectral properties and reveal the intrinsic diversities among SNe~Ia. Therefore, it may give a new insight into the error budget of cosmological parameters. 
The direct applications of the proposed method in SN cosmology also include K-correction and searching for spectroscopic twin SNe. With spectroscopy from a single epoch, one may reconstruct the entire spectral time sequence to derive more reliable K-corrections. The method may also project SN observations from different epochs to the same epoch for direct comparisons to search for spectral twins \citep{2015ApJ...815...58F}.
For SN surveys, our method might be useful to optimize the spectroscopic follow-up strategies. The LSTM neural network allows for the phases of SNe to be estimated during an observing campaign and spectral follow-ups to be compared with LSTM predictions in real time.

For future work, the number of input spectra $K$, which is fixed to 2 in this paper, may be increased to allow for direct spectral reconstruction with more input spectra.   
Our choice of wavelength coverage from 3800 to 7200 \AA, which was chosen to include as many spectra as possible, may be extended when more data become available.
The loss function MSE is a generic metric and prone to spectra with incorrect colors. We may develop more sophisticated loss functions that weigh more heavily on specific spectral features. 
The neural network uncertainties used in this work are still oversimplified. We certainly need to better understand the predictive uncertainty.
This may include by dividing it into epistemic uncertainty and aleatoric uncertainty \citep{Kiureghian2009,gal2016uncertainty,2020PhRvD}. 
The epistemic uncertainty is reducible by increasing the number of observations as the limited training set can be insufficient for the entire feature space, while the aleatoric uncertainty captures the noise of intrinsic randomness (such as photon noise) and cannot be reduced by collecting more data.

\acknowledgments

L.H. acknowledges support by the National Natural Science Foundation of China (11761141001) and the Major Science and Technology Project of Qinghai Province (2019-ZJ-A10). L.W. and X.C. acknowledge support from NSF grant AST-1817099. We acknowledge the Supernova Polarimetry Project for providing the high-S/N VLT dataset. We thank Jiawen Yang for helpful discussions on the package developments. The authors are also grateful to an anonymous referee whose report has significantly improved this manuscript.

\bibliography{SpecLSTM.bib}
\end{document}